\documentclass[aps,prc,preprint,times,graphicx,tighten,showpacs]{revtex4}
\usepackage[dvips]{graphicx}
\usepackage[dvips]{xcolor}
\usepackage{ulem}

\newcommand{\bea}{\begin{eqnarray}}
\newcommand{\eea}{\end{eqnarray}}
\newcommand{\be}{\begin{equation}}
\newcommand{\ee}{\end{equation}}

\newcommand{\carbon}{$^{12}$C }
\newcommand{\oxygen}{$^{16}$O }
\newcommand{\argon}{$^{40}$Ar }

\begin{document}

\preprint{JLAB-THY-17-2586}

\title{%SuSAv2-MEC predictions for the $CC0\pi$ neutrino and antineutrino cross section on Water //
  Neutrino-Oxygen CC0$\pi$ scattering in the SuSAv2-MEC model 
}

%--------------------
\author{
G.D. Megias$^a$,
M.B. Barbaro$^b$,
J.A. Caballero$^a$,
J.E. Amaro$^c$,
T.W. Donnelly$^d$,
I. Ruiz Simo$^c$,
J. W. Van Orden$^{e,f}$
}
%--------------------

\affiliation{$^a$Departamento de F\'{i}sica At\'omica, Molecular y Nuclear, Universidad de Sevilla, 41080 Sevilla, Spain}

\affiliation{$^b$Dipartimento di Fisica, Universit\`{a} di Torino and INFN, Sezione di Torino, Via P. Giuria 1, 10125 Torino, Italy}

\affiliation{$^c$Departamento de F\'{\i}sica At\'omica, Molecular y Nuclear,
and Instituto de F\'{\i}sica Te\'orica y Computacional Carlos I,
Universidad de Granada, Granada 18071, Spain}

\affiliation{$^d$Center for Theoretical Physics, Laboratory for Nuclear Science and Department of Physics, Massachusetts Institute of Technology, Cambridge, Massachusetts 02139, USA}

\affiliation{$^e$Department of Physics, Old Dominion University, Norfolk, Virginia 23529, USA}
\affiliation{$^f$Jefferson Laboratory, 12000 Jefferson Avenue, Newport News, Virginia 23606, USA\footnote{Notice: Authored by Jefferson Science Associates, LLC under U.S. DOE Contract No. DE-AC05-06OR23177.
		The U.S. Government retains a non-exclusive, paid-up, irrevocable, world-wide license to publish or reproduce this manuscript for U.S. Government purposes}}

\date{\today}
\begin{abstract}
  We present the predictions of the joint SuSAv2-MEC (SuperScaling Approach version 2 - Meson Exchange Currents) model for the double
  differential charged-current muonic neutrino (antineutrino) cross section on water
  for the T2K neutrino (antineutrino) beam. We validate our model by comparing with the
  available inclusive electron scattering data on oxygen and compare our predictions with the recent T2K $\nu_\mu$-\oxygen data~\cite{T2Kwater}, finding good agreement at all kinematics. We show that
  the results are very similar to those obtained for $\nu_\mu-^{12}$C
  scattering, except at low energies, and we comment on the origin of
  this difference. A factorized spectral function model of \oxygen is also included for purposes of comparison.
\end{abstract}


\maketitle

\section{Introduction}

The accurate understanding of medium effects in neutrino-nucleus
scattering has become a major challenge in recent years due to the
essential role played by nuclear physics in the analysis of neutrino
oscillation experiments.  In fact, nuclear modeling uncertainties for
this process represent the main source of systematic error for present
(T2K, NOvA) and future (HyperK, DUNE) long baseline neutrino
experiments, aiming at precision measuremens of neutrino oscillation
parameters and searching for leptonic CP violation.  This has
triggered intense activity in the nuclear theory community with the goal of describing neutrino-nucleus observables with high accuracy~\cite{Martini:2009aa,Amaro:2010sd,Nieves:2011yp,Nieves:2013nubar,Mosel:MEC2016,Lalakulich:2012ac,Lalakulich12PRC,Mosel,Megias:2016ee,Simo:2014wka,Pandey:2016ee,RuizSimo:2016ikw,Meucci,Meucci1,Amaro:2004bs,Gonzalez-Jimenez:2014eqa,Megias:2014qva}.  For a detailed and comprehensive study of neutrino-nucleus cross sections and their impact on the measurement of neutrino properties through oscillation experiments, the reader is referred to the NuSTEC White Paper~\cite{White-paper} (see also~\cite{KatoriMartini}). 

Most of the past work was focused on scattering of neutrinos and
antineutrinos on mineral oil, CH$_2$, which has been the most commonly
used target up to now.  However, there is increasing interest in
theoretical predictions for cross sections on different targets,
specifically \argon and \oxygen. 
In particular, in the T2K experiment
the near and far detectors are made of different nuclear targets, mineral oil and
water, respectively; it is then crucial to explore the differences between $\nu$-C and $\nu$-O observables and to understand how to extrapolate the results from one target to another.

The aim of this paper is, within the framework of the
SuSAv2-MEC nuclear model~\cite{Gonzalez-Jimenez:2014eqa,Megias:2016ee,Megias:2016nu}, to explore the similarities and differences between
charged current (CC) (anti)neutrino scattering with no pions in the final
state (the so-called CC0$\pi$ process) on \oxygen and \carbon. This
process receives contributions mainly 
%(I think we should not include any reference to pion absorption, if not the referees could ask why we are not considering that) (OK, m.)
from two different reaction mechanisms:
quasielastic (QE) scattering, where the probe couples to the one-body
current of a single bound nucleon, and the process where scattering
occurs on a pair of nucleons interacting through the exchange of a
meson, giving rise to two-body meson exchange currents (MEC). These
two mechanisms in general have different dependences on the nuclear
species, namely they scale differently with the nuclear density~\cite{kfdep}. Therefore a careful investigation of this behavior for the two contributions must be performed before extrapolating the results
from one nucleus to another. 

A further difficulty arises from the fact that in oscillation
experiments the neutrino energy is not known precisely, but broadly
distributed around a maximum value: as a consequence each kinematic situation for a given outgoing lepton corresponds to a range of different neutrino
energies and the one-and two-body responses cannot be disentangled in
the experimental data. The situation is different in electron
scattering, where the very precise knowledge of energy and momentum transfer
allows one to identify clearly the different reaction mechanisms. At this point, it is worth mentioning that a consistent evaluation of ($e,e'$) cross-section data
in the same kinematical regime is a key input for a proper analysis of
neutrino-nucleus interactions as it provides a decisive benchmark for
assessing the validity of the theoretical nuclear models, not only in the
QE regime, but also for the 2p-2h MEC contributions as well as at
higher energy transfers (nucleonic resonances, inelastic
spectrum). This has been studied in detail
in~\cite{Megias:2016ee}, where good agreement with ($e,e'$) data is
reached in the framework of the SuSAv2-MEC model for a
wide range of kinematics, covering from the QE regime to the deep
inelastic spectrum.

The paper is organized as follows: in Sec.~\ref{sec:formalism} we
introduce the basic formalism and briefly review the SuSAv2-MEC
model. In Sec.~\ref{sec:results} we present our results: we validate
the model by comparing with inclusive electron scattering data on
\oxygen (Sec.~\ref{sec:electron}), we show our predictions for the T2K
CC0$\pi$ cross section on \oxygen and compare with recent data~\cite{T2Kwater} 
(Sect.~\ref{sec:T2K}), 
intercompare the results on \carbon and \oxygen (Sec.~\ref{sec:T2Kcomp}) and present predictions for antineutrinos on oxygen and water in Sec.~\ref{sec:anti}. A factorized spectral function model \cite{Benhar:1994hw,Benhar:2005dj,VanOrden:2017uyy} for \oxygen is shown for purposes of comparison.
        Finally, in Sec.~\ref{sec:conclusions} we draw our conclusions.

\section{Theoretical formalism: The model}
\label{sec:formalism}

The general formalism describing electron and charged-current
neutrino-nucleus scattering processes has already been presented in
detail in previous works \cite{KatoriMartini,Amaro:2004bs,Gonzalez-Jimenez:2014eqa,Caballero:2006wi,Martini:ee,Nieves13PRD,Gil:1997bm}. Here we summarize the basic
expressions involved in the differential cross sections for the discussion that follows. 
We work in the laboratory frame where the initial nucleus is at rest.
In the case of electron
scattering, the double differential $(e,e')$ inclusive cross section
is given in terms of two response functions that account for all of the
information on the nuclear effects involved in the process,
\begin{equation}
\frac{d^2\sigma}{d\Omega_e d\omega}=\sigma_{Mott}\left[v_LR^L(q,\omega)+v_TR^T(q,\omega ) \right] \, ,
\label{equation:1}
\end{equation}
where $\sigma_{Mott}$ is the Mott cross section and the $v$'s are
kinematical factors that only depend on the leptonic variables (see
\cite{Day:1990} for their explicit expressions). The response functions are
given by $R^{L,T}$ with $L$ ($T$) referring to the longitudinal
(transverse) components with respect to the direction of the transferred momentum, $q$. Notice that
both responses contain isoscalar and isovector contributions.

In the case of CC neutrino-nucleus scattering, the double differential
cross section is also decomposed in a sum of responses, which are now purely isovector and are expressed in terms of the charge/longitudinal ($CC, CL, LL$) and transverse ($T, T'$) responses with respect to the direction of $q$. The CC, CL, LL and T responses are composed of pure vector (VV) and axial (AA) components, while the T' response contains only the interference (VA) component.
The general expression for the cross section is given by
\begin{equation}
  \frac{d\sigma}{dk' d\Omega}=\sigma_0\left[\hat V_{CC}\hat R^{CC}+2\hat V_{CL}\hat R^{CL}+\hat V_{LL}\hat R^{LL}
  +\hat V_T\hat R^T\pm2\hat V_{T'}\hat R^{T'}\right] \, ,
\label{CS}
\end{equation}
with $\hat R^K$ the weak nuclear response functions, and 
\begin{equation}
\sigma_0=
\frac{G_F^2\cos^2\theta_c}{2\pi^2}
\left(k^\prime \cos\frac{\tilde\theta}{2}\right)^2 \, ,
\end{equation}
that depends on the Fermi
constant $G_F$, the Cabibbo angle $\theta_c$, the outgoing lepton
momentum $k^\prime$, and the generalized scattering angle
$\tilde\theta$, whose explicit expression is given by (see~\cite{Amaro:2004bs})
\begin{equation}
\tan^2\frac{\tilde\theta}{2} = \frac {|Q^2|} {4 \varepsilon \varepsilon'-|Q^2|}                            
\end{equation}
with $\varepsilon$ ($\varepsilon'$) the neutrino (muon) energy. Notice that $\tilde{\theta}$ coincides with the leptonic scattering angle in the limit of the lepton masses being zero, which is not the case for CC neutrino scattering.
Finally, the terms $\hat V_K$ in eq.~(\ref{CS}) are kinematical factors whose explicit expressions can be found in \cite{Amaro:2004bs,Amaro:2005sr}. Note that the transverse channel contains an
interference vector-axial (VA) response that is constructive (+) for
neutrino scattering and destructive (--) for antineutrinos. 
%{\ttblue Do you think we should add more details here? (No, m.)}
%\begin{eqnarray}
%\hat V_{CC}
%&=&
%1-\delta^2\tan^2\frac{\tilde{\theta}}{2}
%\label{vcc}\\
%\hat{V}_{CL}
%&=&
%\frac{\omega}{q}+\frac{\delta^2}{\rho'}\tan^2\frac{\tilde{\theta}}{2}
%\\
%\hat{V}_{LL}
%&=&
%\frac{\omega^2}{q^2}+
%\left(1+\frac{2\omega}{q\rho'}+\rho\delta^2\right)\delta^2
%\tan^2\frac{\tilde{\theta}}{2}
%\\
%\hat{V}_{T}
%&=&
%\tan^2\frac{\tilde{\theta}}{2}+\frac{\rho}{2}-
%\frac{\delta^2}{\rho'}
%\left(\frac{\omega}{q}+\frac12\rho\rho'\delta^2\right)
%\tan^2\frac{\tilde{\theta}}{2}
%\\
%\hat{V}_{T'}
%&=&
%\frac{1}{\rho'}
%\left(1-\frac{\omega\rho'}{q}\delta^2\right)
%\tan^2\frac{\tilde{\theta}}{2}\ .
%\label{vtp}
%\end{eqnarray}
%In Eqs.~(\ref{vcc}--\ref{vtp}) we have defined
%\begin{eqnarray}
%\delta &=& \frac{m_l}{\sqrt{|Q^2|}}\\
%\rho &=& \frac{|Q^2|}{q^2}\\
%\rho' &=& \frac{q}{\epsilon+\epsilon'}\ ,
%\end{eqnarray}
%where $m_l$ is the final charged lepton mass.
%}

\subsection{SuSAv2: brief summary}
 
In this work all the electromagnetic and weak nuclear responses have
been evaluated within the framework of the SuSAv2 (SuperScaling Approach version 2)
model~\cite{Gonzalez-Jimenez:2014eqa,Megias:2016ee}. This approach, based on the
scaling and superscaling properties~\cite{Day:1990,super,super1,Amaro:2004bs} exhibited by electron scattering
data, also takes into account the behavior of the responses provided
by the Relativistic Mean Field (RMF) theory~\cite{Caballero:2005sj,Caballero:2006wi,Caballero:2007tz}.

The SuperScaling approach (SuSA)~\cite{Day:1990,super,super1,Amaro:2004bs} is a semiphenomenological model that assumes
the existence of universal scaling functions for both electromagnetic
and weak interactions in such a way that nuclear effects can be analyzed through a scaling function extracted from the ratio between the experimental QE cross section and the appropriate single-nucleon one. Analyses of inclusive ($e,e'$) data have shown
that at energy transfers below the QE peak superscaling is fulfilled
with very good accuracy  \cite{super,super1,Chiara1}:
this implies that the reduced cross section (scaling function) exhibits an independence of the momentum transfer (first-kind scaling) and of the nuclear
target (second-kind scaling) when expressed as a function of the
appropriate scaling variable ($\psi$), see Eq.~\ref{scaling}, itself a function of the energy
($\omega$) and momentum transfer ($q$) to the nucleus.
\begin{eqnarray}\label{scaling}
 \psi &\equiv&
\frac{1}{\sqrt{\xi_F}}\frac{\lambda-\tau}
             {\sqrt{(1+\lambda)\tau+\kappa\sqrt{\tau(\tau+1)}}};\\ \lambda&=&\omega/2m_N, \kappa=q/2m_N, \tau=\kappa^2-\lambda^2, \xi_F=\sqrt{1+(k_F/m_N)^2}-1, m_N: \mbox{nucleon mass}\nonumber
\end{eqnarray}
Accordingly, the previous nuclear responses (see Eqs.~\ref{equation:1}-\ref{CS}) can be described in this context with the general structure
\begin{eqnarray}\displaystyle
R^K=\frac{1}{k_F}f^K_{model}(\psi)G^K \quad ; \qquad K= L,T,CC,CL,LL,T,T'
\end{eqnarray}
where $G^K$ are the single-nucleon responses and $f^K_{model}(\psi)$ are the corresponding scaling functions for a particular model and for each longitudinal or transverse channel (see~\cite{Megias:2017PhD,Amaro:2004bs,Amaro:2005sr} for details).

Nevertheless, it is worth remarking that at energies
above the QE peak both kinds of scaling are violated, which is
associated with effects beyond the impulse approximation (IA), such as 2p-2h MEC or with inelastic
contributions. An extension of this formalism, originally introduced
to describe the QE regime, to the $\Delta$-resonance domain and the
complete inelastic spectrum -- resonant, non-resonant and deep
inelastic scattering (DIS) -- has also been proposed in recent
works~\cite{Barbaro:2003ie,Maieron:2009an,Ivanov:2016Delta}.

Contrary to the original SuSA
model where a unique phenomenological scaling function, extracted from
the analysis of the longitudinal response data for electron
scattering, is used for both longitudinal and transverse
electromagnetic responses as well as for all the weak neutrino
responses, SuSAv2 is constructed by considering RMF effects. The RMF theory has the merit of accounting for the difference betweeen isoscalar and isovector channels as well as for the separate vector-vector, axial-axial and vector axial channels.
% and a separate analysis of the isovector and isoscalar channels. 
Within the RMF model [96] the bound and scattered nucleon wave functions are solutions of the Dirac-Hartree equation in the presence of energy-independent real scalar (attractive) and vector (repulsive) potentials. Since the same relativistic potential is used to describe the initial and final nucleon states, the model is shown to preserve the continuity equation [97]; hence the results are almost independent of the particular gauge selected [93,94]. In the RMF model the nucleons are dynamically and strongly {\it off-shell} and, as a consequence, the cross section is not factorized into a spectral funcion and an elementary lepton-nucleus cross section. The RMF has achieved significant success in describing ($e,e'$) data and their superscaling properties. On the one hand, its validity has been widely proved through comparisons with QE ($e,e'$) data [93], also reproducing surprisingly well the magnitude and shape of the experimental longitudinal scaling function and thus the nuclear dynamics. On the other hand, the model predicts a natural enhancement of the transverse nuclear responses as a genuine relativistic effect related with the treatment of the final state interactions (FSI) between the outgoing nucleon and the residual nucleus.

Thus, all these ingredients are incorporated in the SuSAv2 model~\cite{Gonzalez-Jimenez:2014eqa,Megias:2016ee} to produce fully theoretical scaling functions that provide a good agreement with ($e,e'$) data, also fulfilling the superscaling behavior exhibited by electron scattering data. All this constitutes a solid proof of the validity and consistency of the RMF theory and, subsequently, of the SuSAv2 model for the analysis of nuclear dynamics in electron-nucleus reactions and its extension to neutrino scattering processes, producing a more realistic and accurate model than the original SuSA one.

However,
in spite of the undeniable merits of the RMF description, {\it i.e.}, it
provides a scaling function with the right asymmetry (tail extended to
high values of the energy transferred $\omega$) and a transverse
scaling function that exceeds by $\sim 20\%$ the longitudinal one, RMF
predictions do not behave properly at high values of the momentum
transfer $q$. In particular, the RMF peak position and the asymmetry
of the scaling function keep growing with $q$. This is a consequence
of the strong energy-independent scalar and vector potentials involved
in the RMF. Hence, while the RMF approach works properly at low to intermediate
values of $q$, where the effects linked to the treatment of FSI are significant, it clearly fails at higher
$q$ where FSI become less and less important and the relativistic
plane-wave impulse approximation (RPWIA) is much more appropriate. Both
approaches, RMF and RPWIA, are incorporated in the SuSAv2 model by
using scaling functions that are given as linear combinations of the
RMF and RPWIA predictions with a $q$-dependent ``blending'' function that
allows a smooth transition from low to intermediate $q$-values
(validity of RMF) to high $q$ (RPWIA-based region).

Although SuSAv2 was originally applied to the analysis of data within
the quasielastic (QE) domain, {\it i.e.}, based on the validity of the
IA, in \cite{Megias:2016ee} the model was extended also
to the inelastic region by employing phenomenological fits to the
single-nucleon inelastic electromagnetic structure functions. Notice
that in both regimes, QE and inelastic, the general structure of the
``blending'' scaling functions is similar, and the difference in the
nuclear responses comes essentially from the single-nucleon structure
functions used, elastic versus inelastic, as well as from the
different region ($q,\omega$-values) explored. The sensitivity of the
model to several choices of the parameters involved in the ``blending''
function as well as a detailed comparison between the SuSAv2
predictions and inclusive $(e,e')$ data on $^{12}$C for very different
kinematical situations was presented in \cite{Megias:2016ee}.
In the case of \oxygen the available electron scattering data cover only
a limited kinematic region (see \cite{Anghinolfi:1996vm} and \cite{O'Connell:1987ag}) and can be well represented using constant parameters, specifically a Fermi momentum $k_F$=230 MeV/c and an energy shift $E_{shift}$=16 MeV, as discussed below. In order to apply the model to the wider kinematic range of interest in neutrino experiments, we assume the same $q$-dependence of the parameters found by fitting the carbon data, with a global rescaling of the Fermi momentum and energy shift to the values above specified. This choice is motivated by the validity of second-kind scaling, which is fulfilled very well by electron scattering data on different nuclei.

The SuSAv2 model, with the separate analysis of the isoscalar and
isovector channels, makes it very well suited for describing
charged-current (CC) neutrino-nucleus scattering processes. This has
been clearly illustrated in \cite{Gonzalez-Jimenez:2014eqa} where the model was applied to
CC neutrino reactions within the QE domain. Furthermore, its extension
to the inelastic region was introduced in \cite{Megias:2016nu}, but restricting
the analysis to the $\Delta$ resonance that in most of the cases plays
a major role. The addition of higher inelasticities is %still
in progress and the results will be presented in a forthcoming
publication.

\subsection{2p-2h MEC responses}

Ingredients beyond the IA, namely 2p-2h MEC effects, have been shown
to play a very significant role in the ``dip'' region between the QE and
$\Delta$ peaks. This has clearly been illustrated in \cite{Megias:2016ee} where
the 2p-2h MEC effects added to the SuSAv2 predictions (denoted as
SuSAv2-MEC) provide a very good description of inclusive $(e,e')$ data
on $^{12}$C covering the entire energy spectrum at very different
kinematics. Contrary to the SuSAv2 approach, that is based on the RMF predictions in the QE domain and also on the phenomenology of electron scattering data, the 2p-2h MEC calculations
are entirely performed within the Relativistic Fermi Gas (RFG)
model. This is due to the
technical difficulties inherent to the calculation of relativistic two-body
contributions even in the simple RFG model. However, it is noteworthy
to point out that the present 2p-2h MEC contributions correspond to a
fully relativistic calculation, needed for the extended kinematics
involved in neutrino reactions. Moreover, all the electromagnetic
(longitudinal and transverse) as well as the whole set of weak
responses, including the vector, axial and axial-vector interference
contributions in all channels have been
evaluated exactly, that is, no particular assumption on the behavior
and/or magnitude of any response has been considered. 
{Following
previous work\cite{Simo:2016aa,Simo:2016ab,RuizSimo:2016ikw,RuizSimo:2017hlc}, here we consider only the real part of the $\Delta$-propagator. The missing
MEC terms --- that is, contributions linked to the imaginary part of the propagator (on-shell intermediate delta) --- are included in the scaling functions that were fitted to the data. 
As a consequence the 2p-2h MEC model contains a suppression of the $\Delta$ peak to be consistent with our parametrization of the inelastic scaling function.  The full SuSAv2-MEC model has been validated by its excellent description of $(e,e')$ data for a very wide range of kinematics.
In order to apply the present calculation to the analysis of current and
future neutrino oscillation experiments, with its possible
implementation into the Monte Carlo generators, we have developed a
parametrization of the MEC responses in the range of
momentum transfer $q= 50, \ldots, 2000$ MeV/c, that significantly reduces the computational time. Its functional form is given in
detail in \cite{Megias:2016ee,Megias:2016nu} for the case of $^{12}$C, and here it is extended
to the analysis of $^{16}$O data.

\begin{figure}%[h]
  \begin{minipage}{\textwidth}
    \begin{flushleft}
%\vspace{-0.2cm}
\includegraphics[width=6.cm, angle=270]{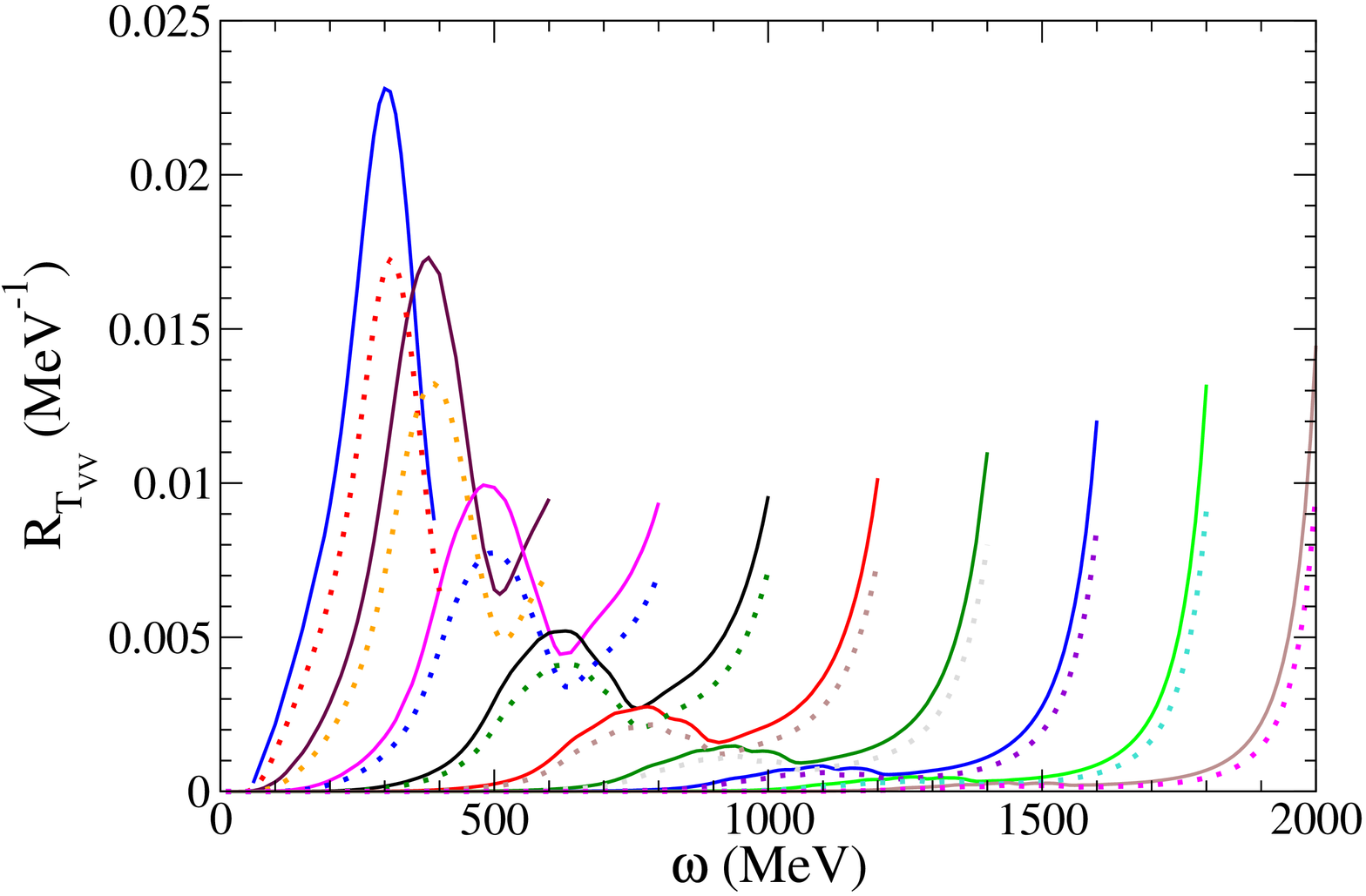}
\includegraphics[width=6.cm, angle=270]{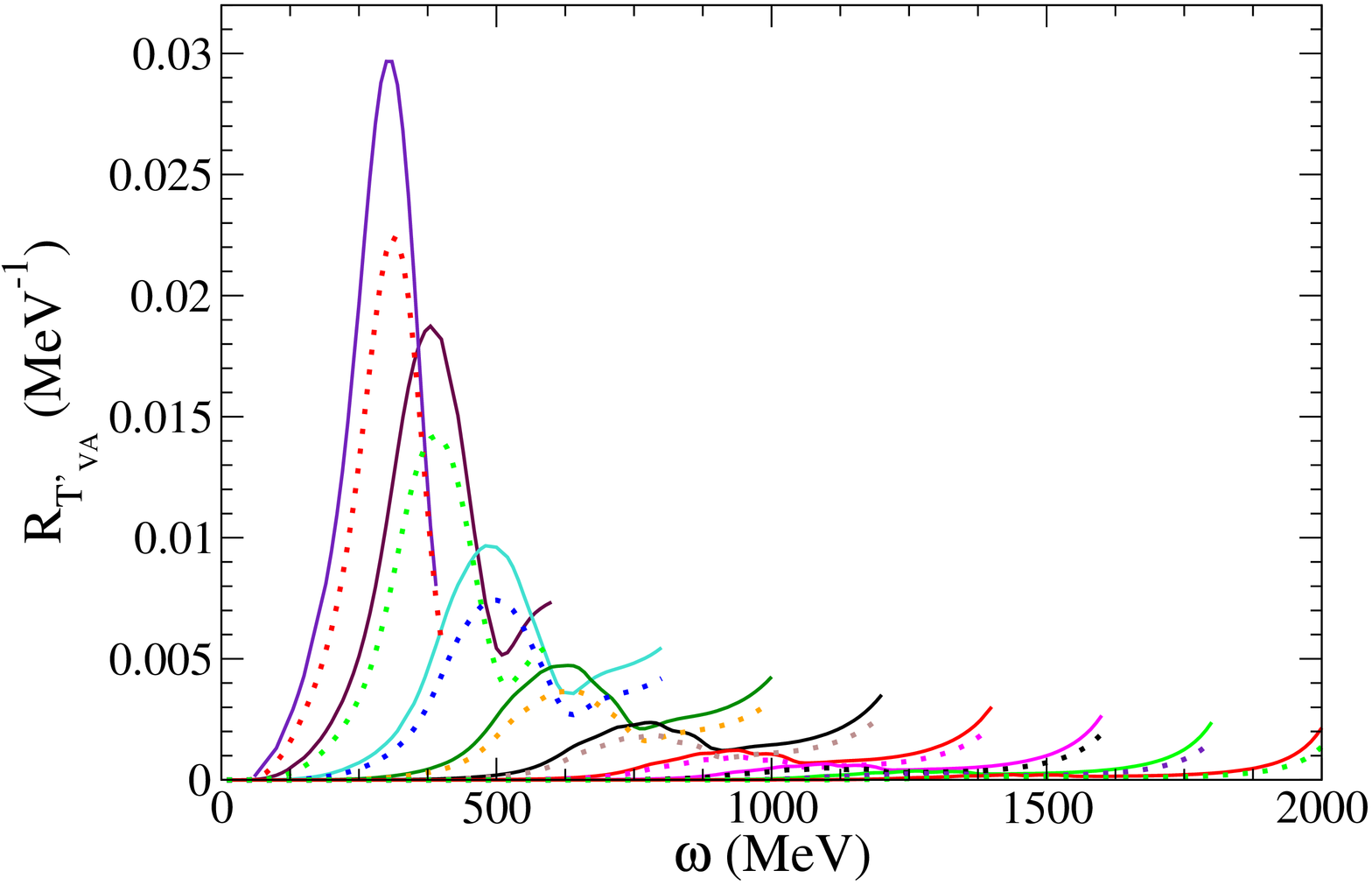} 
\\
\includegraphics[width=6.cm, angle=270]{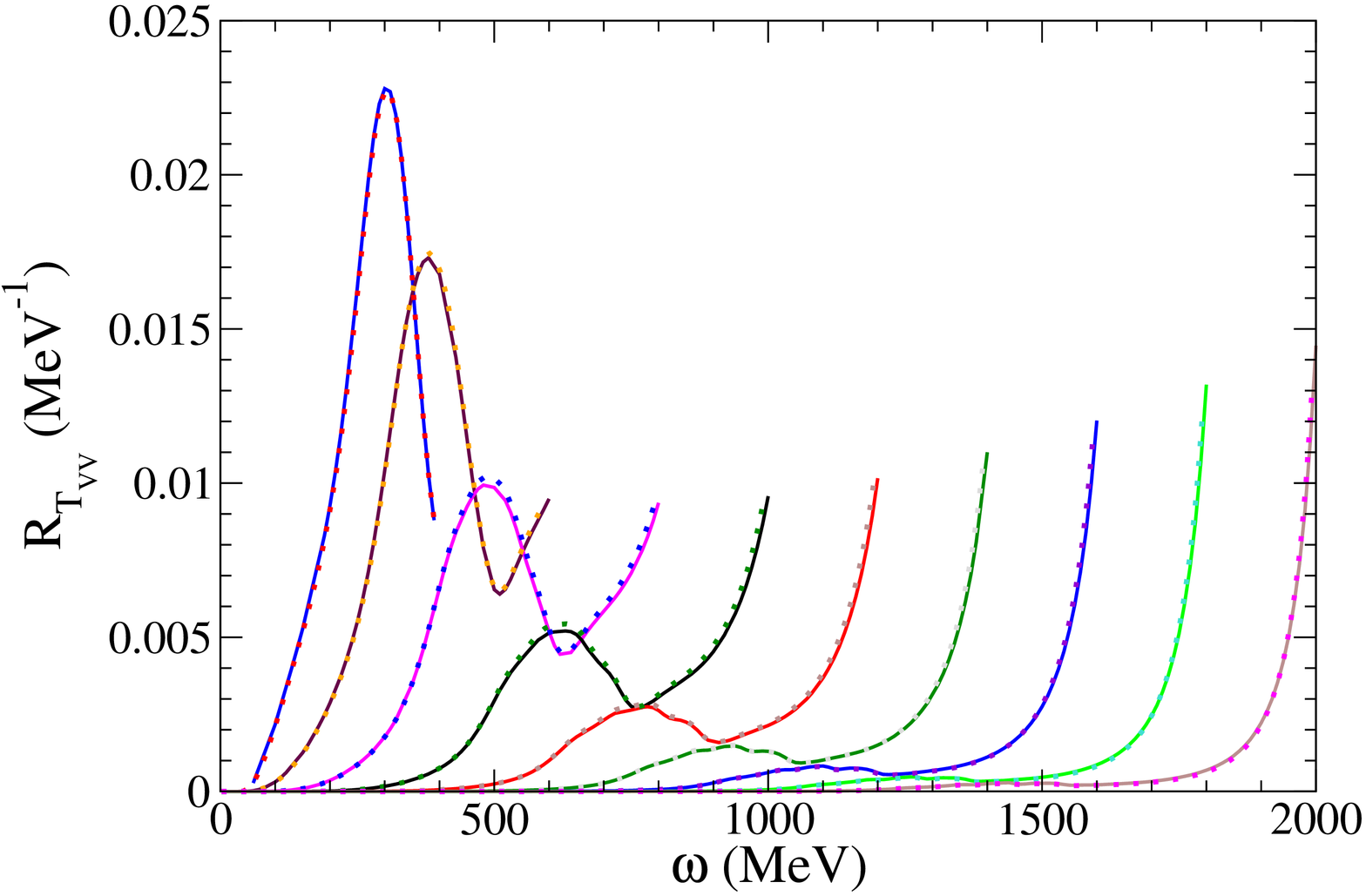}
\includegraphics[width=6.cm, angle=270]{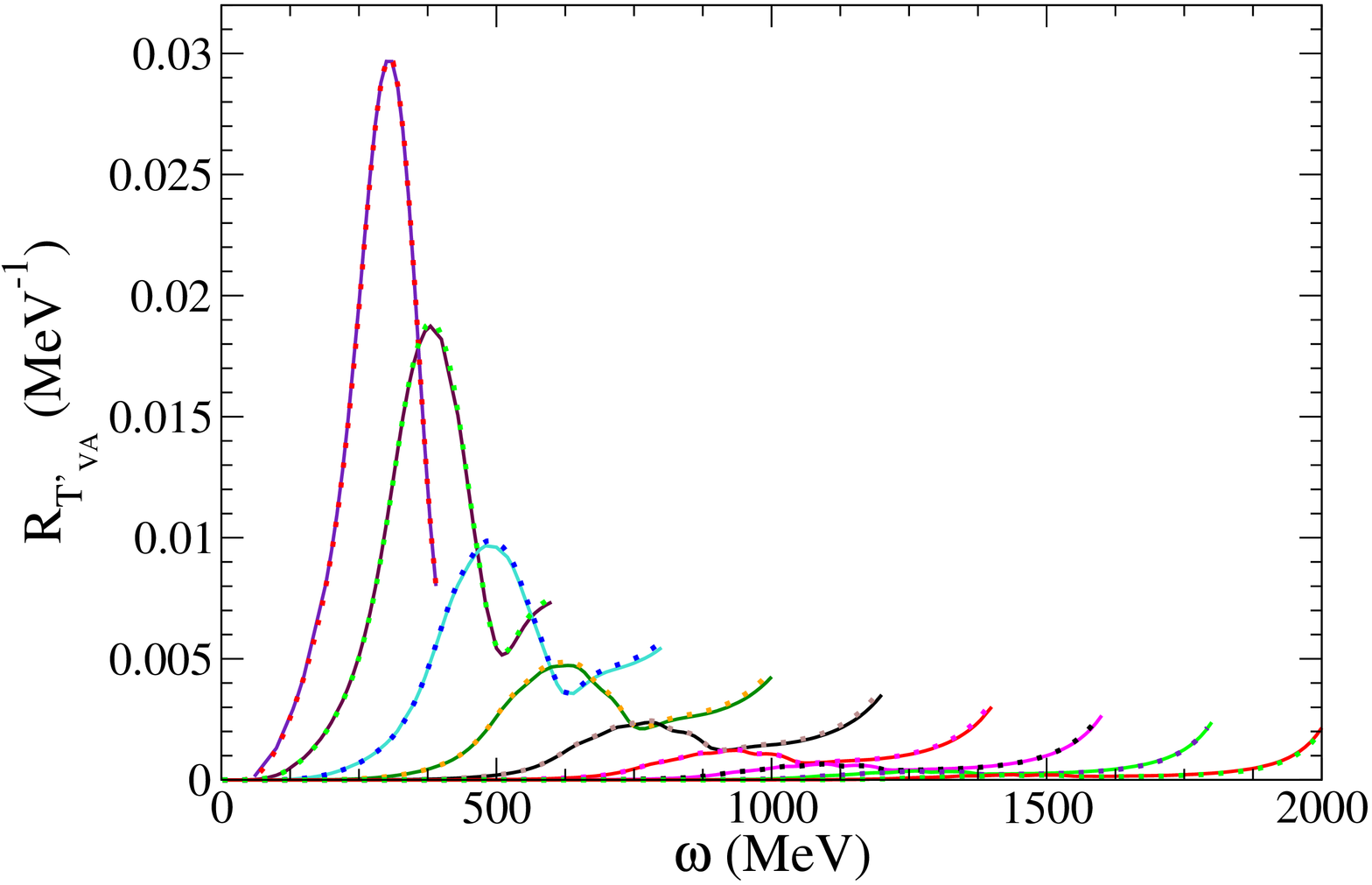}
%\vspace{-1cm}
\end{flushleft}

\caption{(Color online) 2p-2h MEC vector-vector transverse ($T_{VV}$)
  response and the axial-vector interference ($T'_{VA}$) one.
  Comparison between the results for $^{12}$C (dots) and $^{16}$O
  (solid). 
 Bottom panels: comparison by re-scaling the $^{12}$C
  results with a factor 1.35 (see text). The curves are displayed from
  left to right in steps of $q=200$ MeV/c, starting at $q=400$ MeV/c
\label{fig:fig0} }

\end{minipage}
\end{figure}

\subsection{Oxygen versus Carbon predictions}

Some comments are in order concerning the present results
for $^{16}$O compared with the previous ones for $^{12}$C
(see~\cite{Megias:2016ee,Megias:2016nu}). In the particular case of the SuSAv2 model, no significant
differences in the scaling functions are assumed for the two nuclei,
except for the values used for the Fermi momentum and energy shift: $k_F=228$ MeV/c, $E_{shift}=20$
MeV for $^{12}$C and $k_F=230$ MeV/c, $E_{shift}=16$ MeV for $^{16}$O. These values are in accordance with the ones considered in \cite{Megias:2016ee,Megias:2016nu,Chiara1} for \carbon. In the case of $^{16}$O, the $k_F$ and $E_{shift}$-values selected are also consistent with the general trend observed in \cite{Chiara1}. This is at variance with some previous works~\cite{Caballero:2005sj,Caballero:2006wi,Amaro:2005sr,Amaro:2006if} where \oxygen was described by using $k_F=216$ MeV/c and $E_{shift}=25$ MeV. Although the two sets of values only lead to small differences in the cross sections (see below), the present choice does provide a more consistent analysis, and more importantly, it also improves the comparison with electron scattering data. The use of the same scaling functions for both nuclear systems is
consistent with the property of scaling of second kind, {\it i.e.}
independence of the scaling function with the nucleus, and it also
follows from the theoretical predictions provided by the RMF and RPWIA
models on which SuSAv2 relies. This has been studied in detail in
previous works (see \cite{Gonzalez-Jimenez:2013plb,Caballero:2007tz,Meucci1,Caballero:2005sj,Megias:2017PhD}) where the electromagnetic and weak
scaling functions evaluated with the RMF and RPWIA approaches have
been compared for $^{12}$C and $^{16}$O.
Although the
two models lead to significant differences, with the asymmetry (long
tail extended at high $\omega$-values) only emerging when FSI are
accounted for through the strong energy-independent scalar and vector
potentials that are present in the RMF model, only very minor
differences appear in the inter-comparisons for the two nuclei. It is also worth mentioning that the assumption of a scaling relationship between the $^{12}$C and $^{16}$O scaling functions could be inaccurate at very low kinematics (few MeV). These possible differences could be computed and quantified within the RMF theory. Nevertheless, this is beyond the scope of this work that is focused on the analysis of T2K results with a mean neutrino energy of around 0.8 GeV.

Regarding the 2p-2h MEC contributions, our calculations show that they
approximately scale as $k_F^2$. This result, that is consistent with
some analyses presented in the past~\cite{VanOrden:1980tg,DePace:2004xu}, also matches the detailed study we have recently pursued in \cite{kfdep}.
However, it is
important to point out that, although the $k_F^2$-scale rule for the
MEC responses works remarkably well at the peak of the MEC response (see \cite{kfdep}),
the degree of its validity depends on the particular region explored. In the present analysis, we have checked that the same
parametrization already considered for $^{12}$C can be extended to
$^{16}$O but re-scaled with a factor 1.35, that is close to the ratio $8/6 \left[{k_F(O)}/{k_F(C)}\right]^2$
between the nucleon numbers and the squares of the Fermi momenta for
the two nuclei, and taking into account the different energy shifts. This 
provides the best fit of the fully relativistic
results for $^{16}$O.  These results are presented in
  Fig.~\ref{fig:fig0} where two of the MEC responses are shown for the
  two nuclear systems (top panels), and their comparison when using the re-scale
  factor (bottom panels).
  Note the degree of accuracy between the results for both
  nuclei. Similar comments also apply to all the remaining
  electromagnetic and weak responses (not shown for brevity).
 
\section{Results}
\label{sec:results}

In what follows we apply our SuSAv2-MEC model to electron and CC
neutrino scattering reactions on $^{16}$O and compare the theoretical
predictions with data taken at different kinematics and, in the
case of neutrinos, 
given by the T2K collaboration~\cite{T2Kwater}. 
The
discussion follows closely the analysis already presented in the case
of $^{12}$C for electron~\cite{Megias:2016ee} and neutrino (antineutrino)~\cite{Megias:2016nu} processes, where data are given in \cite{T2Kincl,T2Kinclelectron,T2Kcc0pi}.

\subsection{Electron scattering}
\label{sec:electron}

Any theoretical model that aspires to describe neutrino-nucleus
scattering processes should be first tested against electron
scattering data. Thus a consistent description of electron scattering
cross sections including not only the QE regime but also higher energy
transfer regions (nucleon resonances and inelastic spectrum) is
essential for the analysis of current neutrino oscillation
experiments. Following our previous study on $^{12}$C~\cite{Megias:2016ee},
here we apply the SuSAv2-MEC model to $^{16}$O for which the amount of
available $(e,e')$ experimental data is, unfortunately, much smaller (see http://faculty.virginia.edu/qes-archive/ and \cite{QESarchive}). 
We employ the Gari-Krumpelmann (GKex) model for the
elastic electromagnetic form factors~\cite{Gari2}, whereas the inelastic
structure functions are described making use of the Bosted and Christy
parametrization~\cite{Bosted1}. The contribution of the 2p-2h MEC is also
included in both the longitudinal and
transverse channels. In accordance with previous comments, the
value of the Fermi momentum is fixed to $k_F=230$ MeV/c.

In Fig.~\ref{fig:fig1} our predictions are compared with data for six
different kinematical situations, corresponding to all the available %(e,e')$ data on
$^{16}$O data. In all the cases we present the
separate contributions for the QE, 2p-2h MEC and inelastic
regimes. The inclusive cross sections are given versus the transferred
energy ($\omega$), and each panel corresponds to fixed values of the
incident electron energy ($E_i$) and the scattering angle
($\theta$). Whereas the latter is fixed to $32^0$
\cite{Anghinolfi:1996vm} except for one case (center panel on the top,
{\it i.e.}, $\theta=37.1^0$)~\cite{O'Connell:1987ag}, the electron energy
values run from $700$ MeV (left-top panel), where the QE peak
dominates, to $1500$ MeV (right-bottom) with the inelastic channel
providing a very significant contribution. This is due to the values of
the transferred momentum $q$ involved in each situation. Although $q$ is
not fixed in each of the panels, {\it i.e.}, it varies as $\omega$ also
varies, the range of $q$-values allowed by the kinematics increases
very significantly as the electron energy grows (for fixed
scattering angles). Thus, for higher $E_i$ the two regimes, QE and
inelastic, overlap strongly, the inelastic processes being responsible
for the large cross sections at increasing values of $\omega$. This
different range of $q$-values spanned in each panel also explains the
relative role played by the RMF versus the RPWIA approaches. Although
not shown in the figure for sake of clarity, whereas the RMF response
dominates at lower $E_i$-values (panels from left to right on the
top), the reverse occurs, that is, the scaling function is essentially
given by the RPWIA prediction, as $E_i$ increases (panels on the
bottom).

As observed, the SuSAv2-MEC predictions are in very good accordance
with data for all kinematical situations. Although the relative role
of the 2p2h-MEC effects is rather modest compared with the QE and
inelastic contributions,
its maximum is located in the dip region between the QE
and inelastic peaks. This makes 2p2h-MEC essential in order to
describe successfully the behavior of $(e,e')$ data against the
transferred energy $\omega$. This is clearly illustrated for all the
panels in Fig.~\ref{fig:fig1}. Data in the dip region can only be
reproduced by adding MEC effects to the tails of the QE and inelastic
curves. Indeed, at the peak of the 2p2h response the three contributions
  are comparable in size.

The spectral function model, as used here, is described in more detail in \cite{VanOrden:2017uyy} for semi-inclusive CC$\nu$ scattering on \oxygen. It is factorized with the relativistic single-nucleon cross section folded with a non-relativistic spectral function \cite{Benhar:1994hw,Benhar:2005dj}. It contains the correct relativistic kinematics, but since it is essentially rooted in PWIA it contains no transverse enhancement as in SuSAv2 approach and has no two-body MEC or meson production contributions. Its magnitude is therefore generally somewhat smaller than the SuSAv2 QE contribution and differs slightly in the position of the QE peak. This said, it is encouraging that the SF and SuSAv2 results for the QE contributions are not dramatically different. This result is in line with what was shown in \cite{Ankowski:2015lma} for the case of neutral current Neutrino-Oxygen scattering as well as introduced in~\cite{Ivanov14} for CC Neutrino-Carbon reactions where NN correlations and FSI effects were also considered in a spectral function model.

\begin{figure}
  \begin{minipage}{\textwidth}
    \begin{flushleft}
      %\begin{center}
\includegraphics[width=4.cm,angle=270]{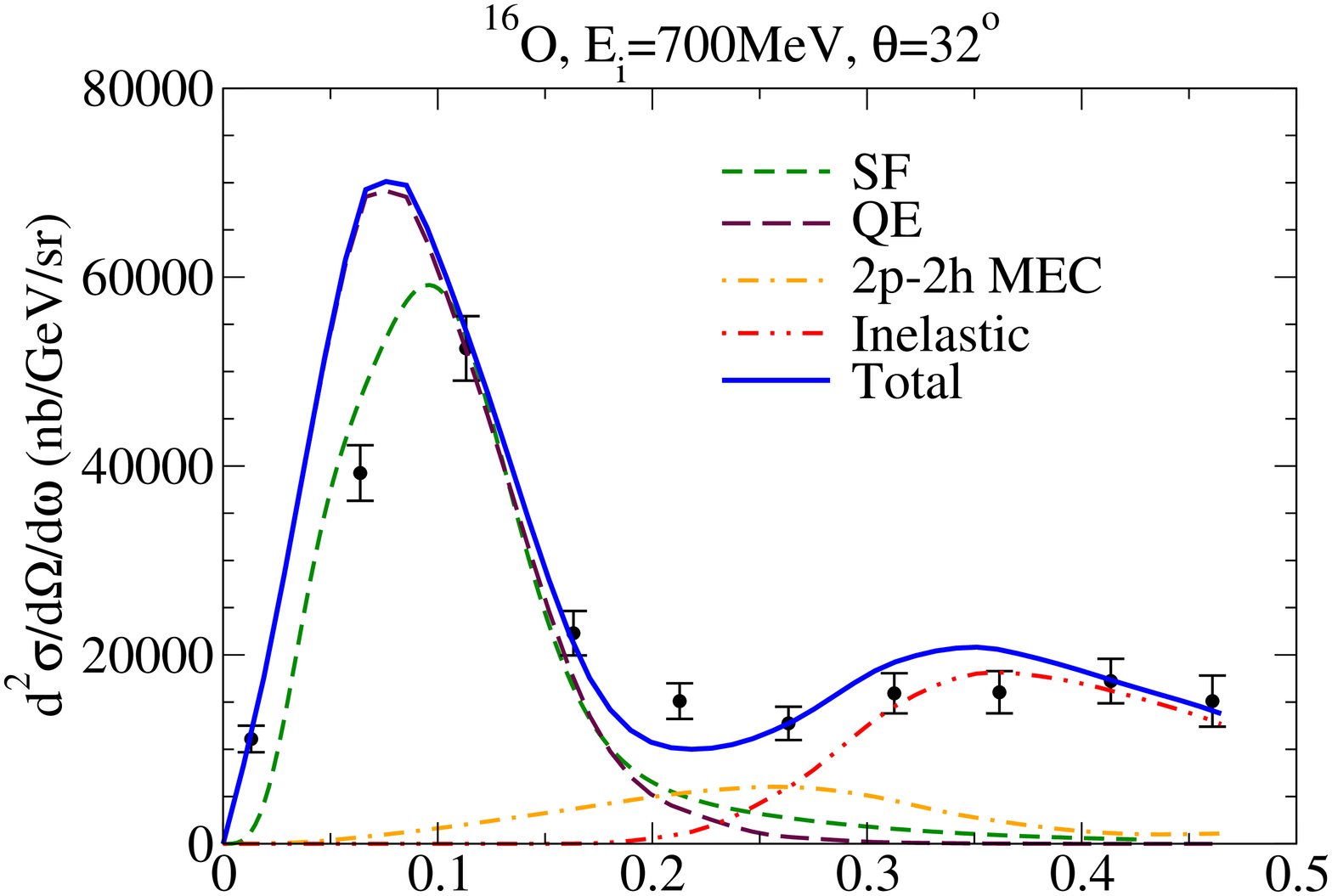}
\includegraphics[width=4.cm,angle=270]{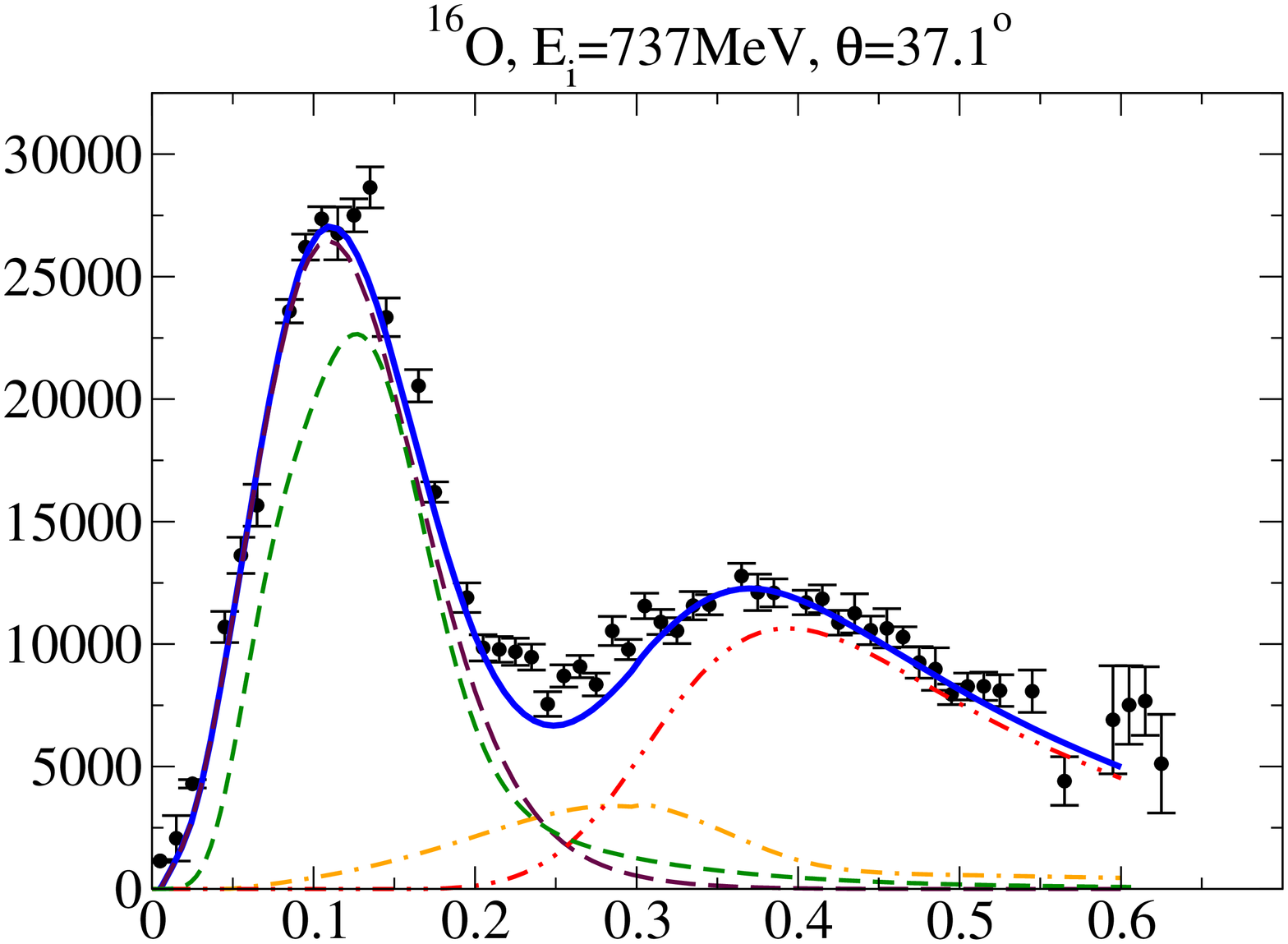}
\includegraphics[width=4.cm,angle=270]{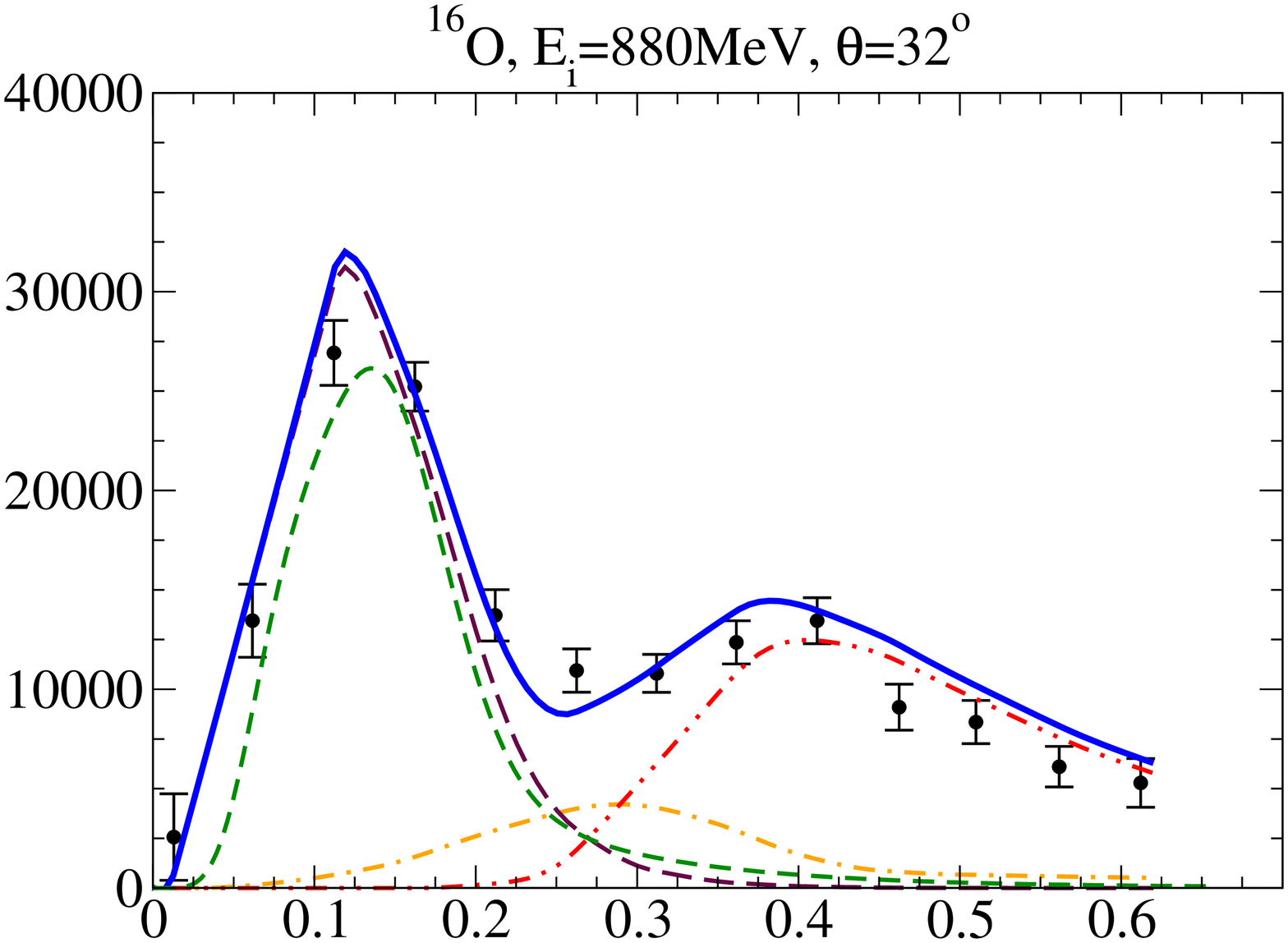}
\\
\includegraphics[width=4.cm,angle=270]{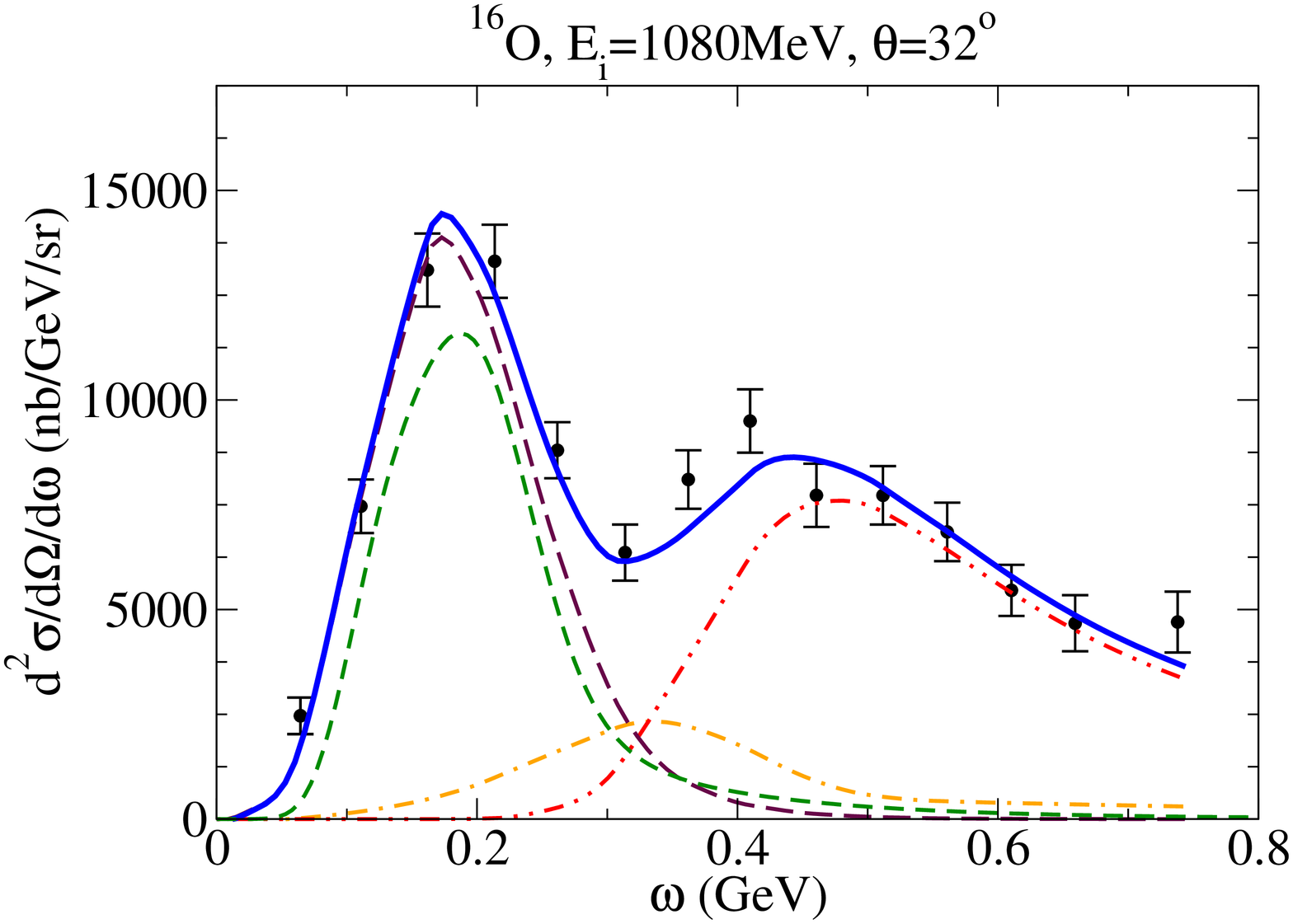}
\includegraphics[width=4.cm,angle=270]{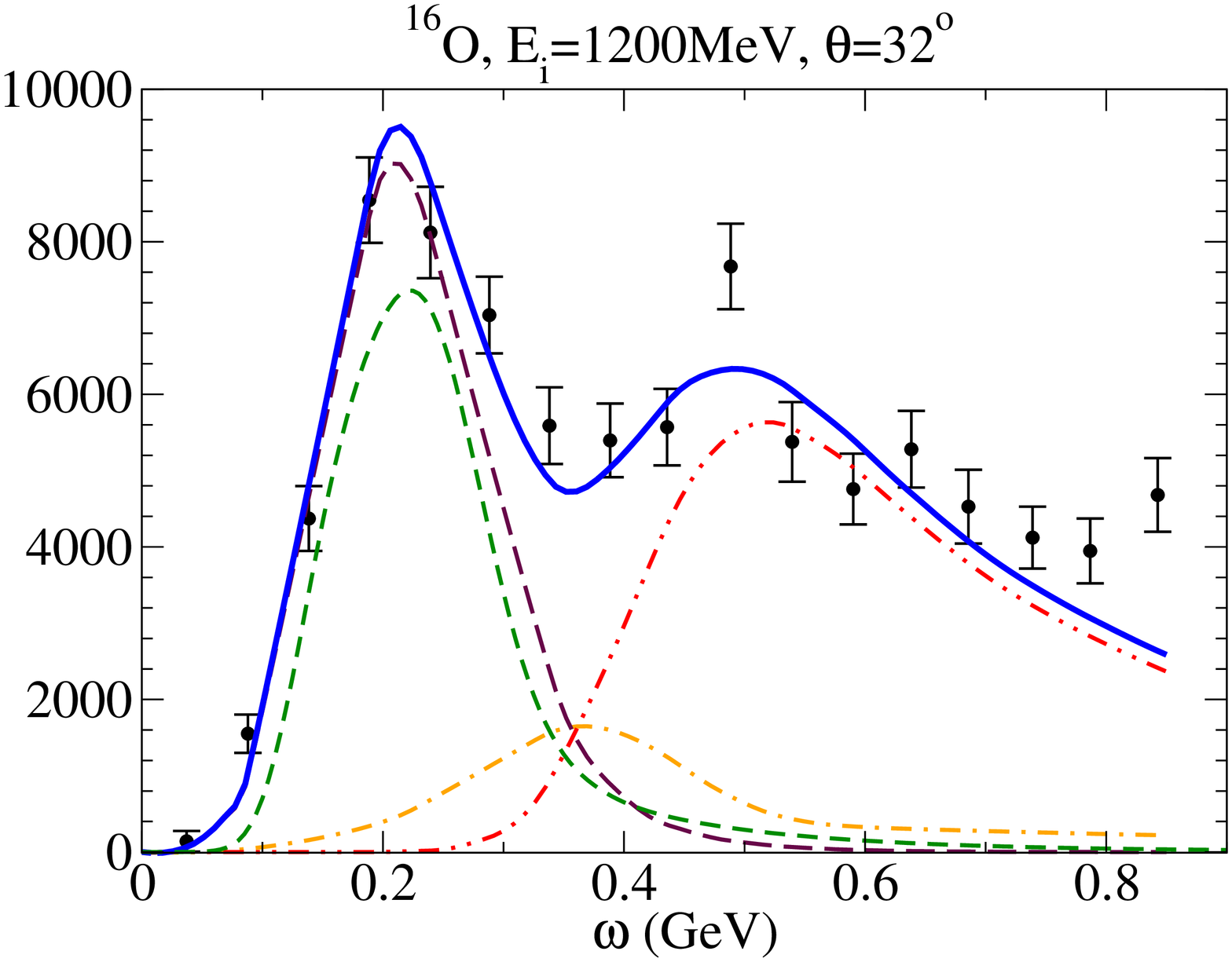}
\includegraphics[width=4.cm,angle=270]{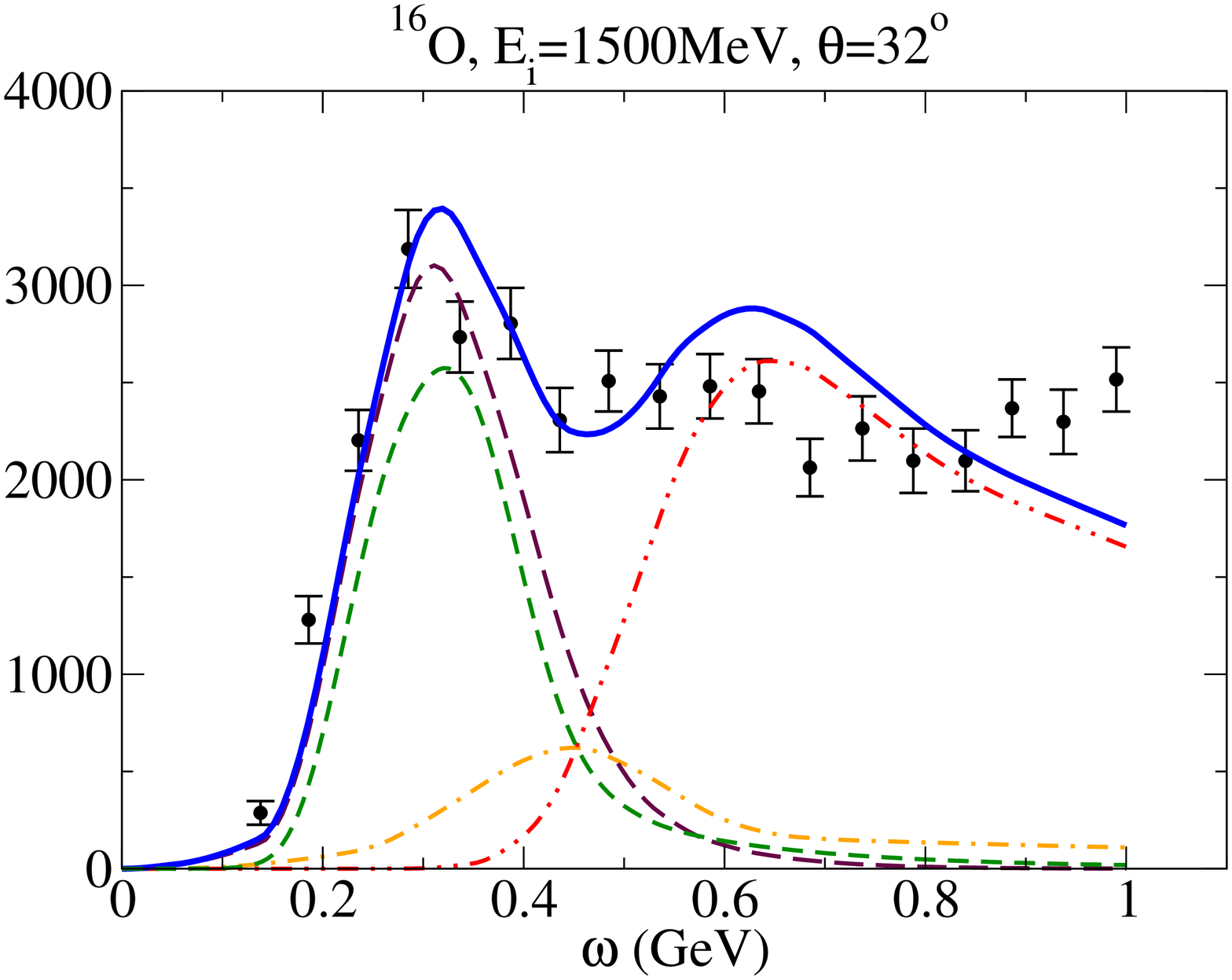}
\end{flushleft}
%\end{center}
\caption{\label{fig:fig1} (Color online) 
  Comparison of inclusive $^{16}$O$(e,e')$ cross sections and predictions of the SuSAv2-MEC model. The separate contributions of the pure QE response (dashed violet line), the 2p-2h MEC (dot-dashed), inelastic (double-dot dashed) are displayed. The sum of the three contributions is represented with a solid blue line. The spectral function (SF) result for the QE
cross section is also shown for comparison (dashed green curve).
  %The $y$ axis represents $d^2\sigma/d\Omega/d\omega$ in nb/GeV/sr.
  The data are from \cite{Anghinolfi:1996vm} and \cite{O'Connell:1987ag}.
}
  \end{minipage}
\end{figure}

\subsection{T2K neutrino --$^{16}$O scattering}
\label{sec:T2K}

Results for CC neutrino reactions on $^{16}$O are shown in
Fig.~\ref{fig:fig2}. Each panel presents the double differential cross
section averaged over the T2K muonic neutrino flux versus the muon
momentum for fixed bins of the muon scattering angle. These kinematics
correspond to the T2K experiment~\cite{T2Kwater}. SuSAv2-MEC
  predictions are compared with data. Contrary to the $(e,e')$ cross
sections shown in the previous section, here only the QE and 2p-2h MEC
contributions are taken into account, as this is consistent with the
analysis of T2K-$^{16}$O data that is restricted to charged-current
processes with no pions in the final state (CC0$\pi$).
We show the separate contributions of the pure QE, the 2p-2h MEC and the sum of
both. Notice the role of the MEC effects compared with the pure
QE ones --- of the order of $\sim$15$\%$ at the maximum
of the peak, except for forward angles, where they represent about 20\% of the total cross section. Furthermore, the MEC peak compared with the QE one is
%slightly
shifted to smaller $p_\mu$-values. These results, which have already been
observed in the case of T2K-$^{12}$C (see \cite{Megias:2016nu}), are
in contrast with the analysis of other experiments, namely, MiniBooNE
and MINERvA, that show 2p-2h MEC relative effects to be larger and the
peak location more in accordance with the QE maximum. This can be
connected with the much narrower distribution presented by the T2K
neutrino flux that explains the smaller 2p-2h MEC contribution and the
location of its peak.

The SuSAv2-MEC approach provides predictions in good agreement with T2K data in
most of the situations, although here 2p-2h MEC effects do not seem to
improve in a significant way the comparison with data. This is at
variance with other experiments, MiniBooNE and MINERvA, and it is
connected with the minor role played by MEC. Notice that in most of
the situations, both the pure QE and the total QE+MEC predictions
describe data with equal success. A similar discussion was already
presented in \cite{Megias:2016nu} for $^{12}$C. It is also important to remark that the agreement with both data sets is widely affected by the current experimental uncertainties, where the low statistics and the large error bars prevent to draw further and clearer conclusions on the goodness of the comparison.

Figure \ref{fig:fig2Wally} compares the SuSAv2 CCQE to the SF calculation for the weighted cross section averaged over bins in scattering angle. Two versions of the SF calculation are shown, one that integrates over all possible values of the neutrino momentum and another that integrates over values for which $\omega\ge 50 \mathrm{MeV}$ (see also the discussions concerning this strategy for exploring the sensitivities to the near-threshold region in the following section). Note that the fully integrated calculation is much larger than the SuSAv2 result at forward angles, but with the difference decreasing with increasing angle, becoming in reasonable agreement with SuSAv2 in the largest first angular bin. The cutoff SF result is smaller than SuSAv2 at forward angles, but comes into reasonable agreement as the angle increases. This is reason for some caution, since none of the CCQE models contains a complete description of inelastic scattering for ejected protons with kinetic energies below 50 MeV. Certainly this is the case for the PWIA SF model where plane-waves are involved and the near-threshold region cannot be successfully represented. 
 A similar comment also applies to a fully relativistic plane-wave impulse approximation (RPWIA) calculation that shows at forward scattering angles cross sections which have much larger than the ones obtained when final-state interactions (FSI) are included. Hence the significant discrepancy introduced by the SF prediction is mostly due to the plane-wave limit approach. Authors in \cite{Ankowski} show that the description of data improves when the hole spectral function is complemented by the particle spectral function and Pauli blocking.
Importantly, a large amount of the data collected in the T2K experiment shown here falls into this region. The SuSAv2 approach involves an assumption which is discussed more fully in previous work where the ideas were developed about how so-called Pauli Blocking can be generalized from the only model where the concept is well-founded, namely, the extreme RFG model. The results obtained within the SuSAv2 approach are not in disagreement with the data, even at forward angles. However, one should still exercise some caution in drawing any final conclusions about how well one can claim to understand this region, {\it i.e.,} in any existing model. This problem deserves to be given greater attention in the future.

\begin{figure}
  \begin{minipage}{\textwidth}
    \begin{flushleft}
      %\begin{center}
\includegraphics[width=4.cm,angle=270]{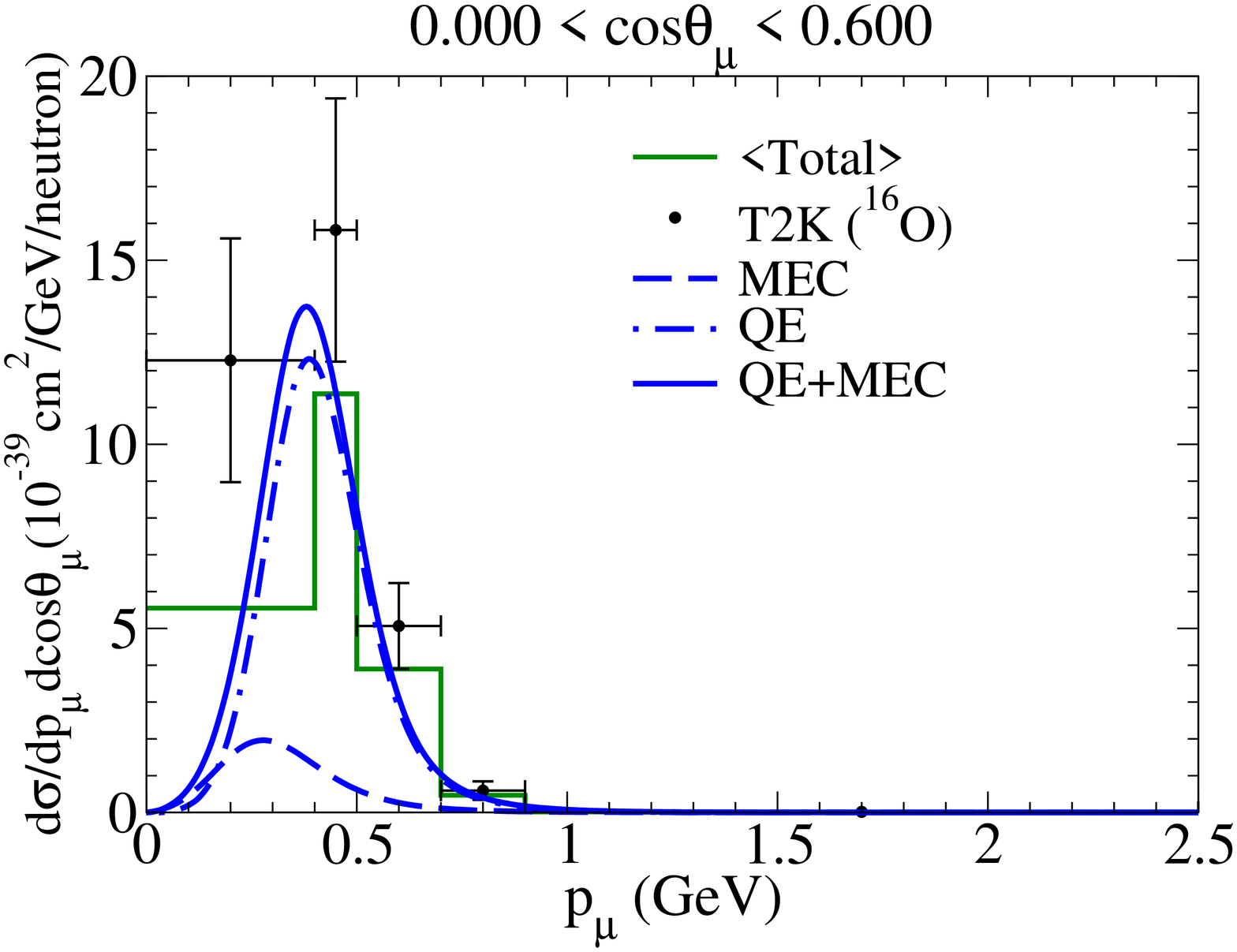}
\includegraphics[width=4.cm,angle=270]{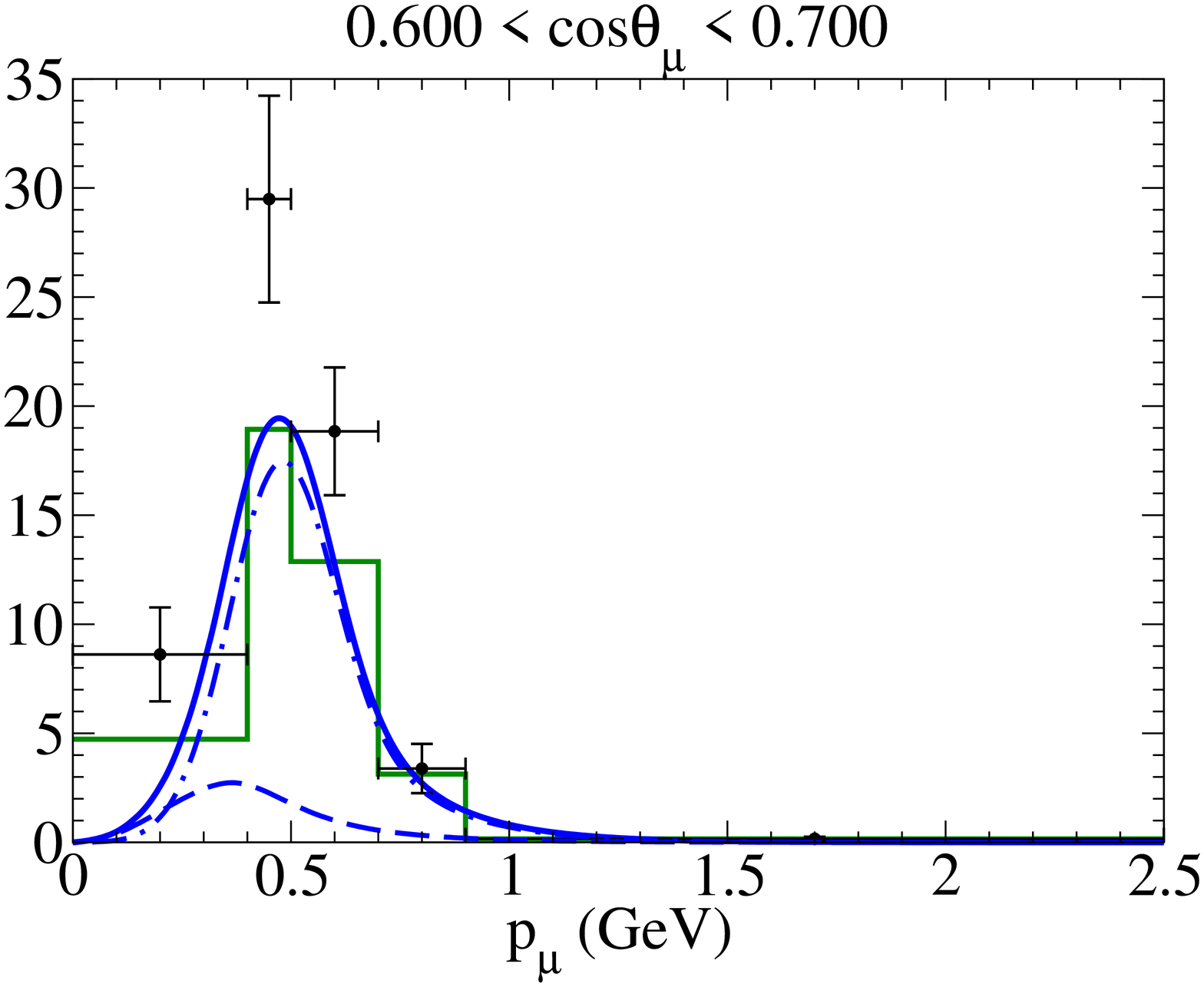}
\includegraphics[width=4.cm,angle=270]{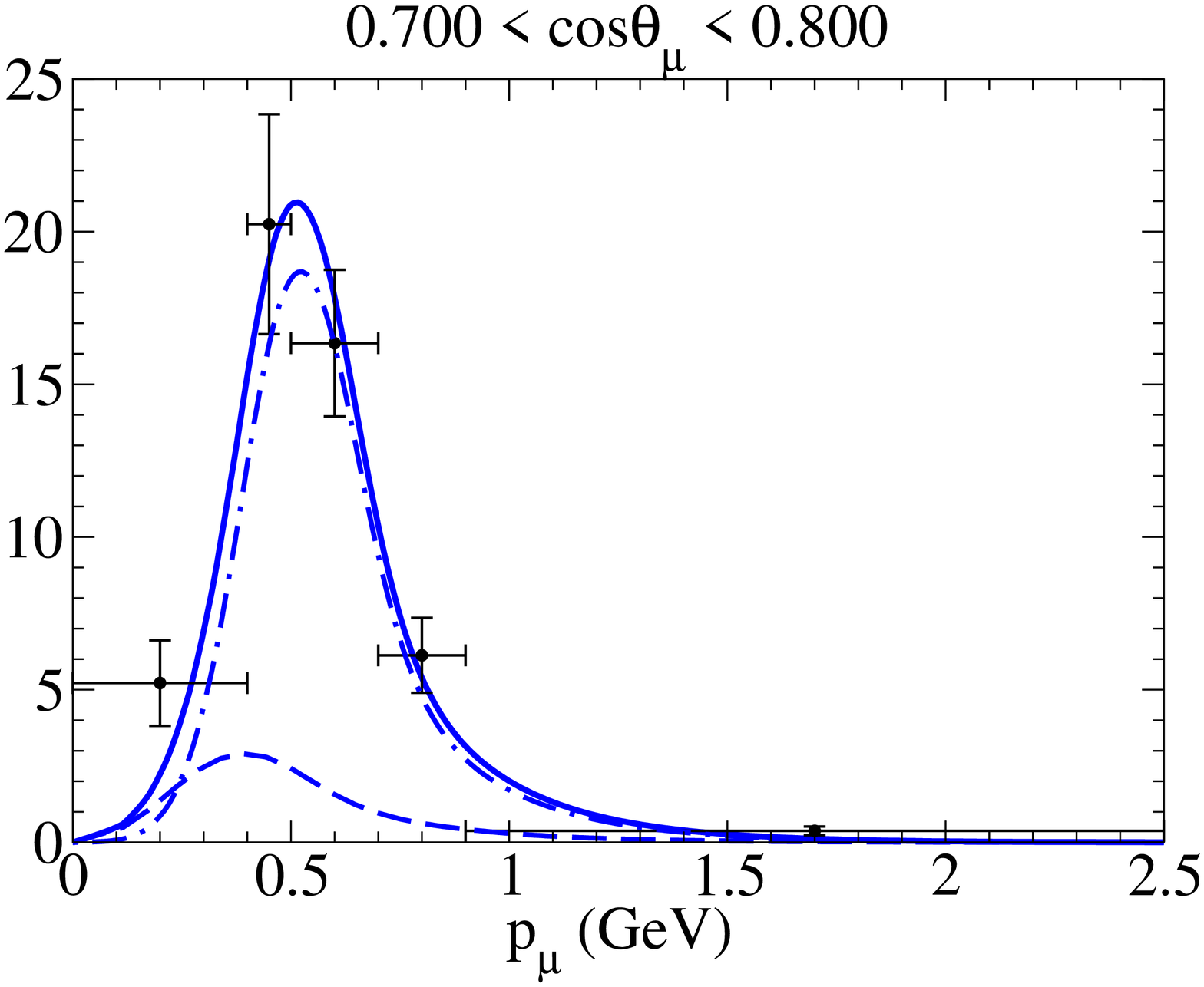}
\\
\includegraphics[width=4.cm,angle=270]{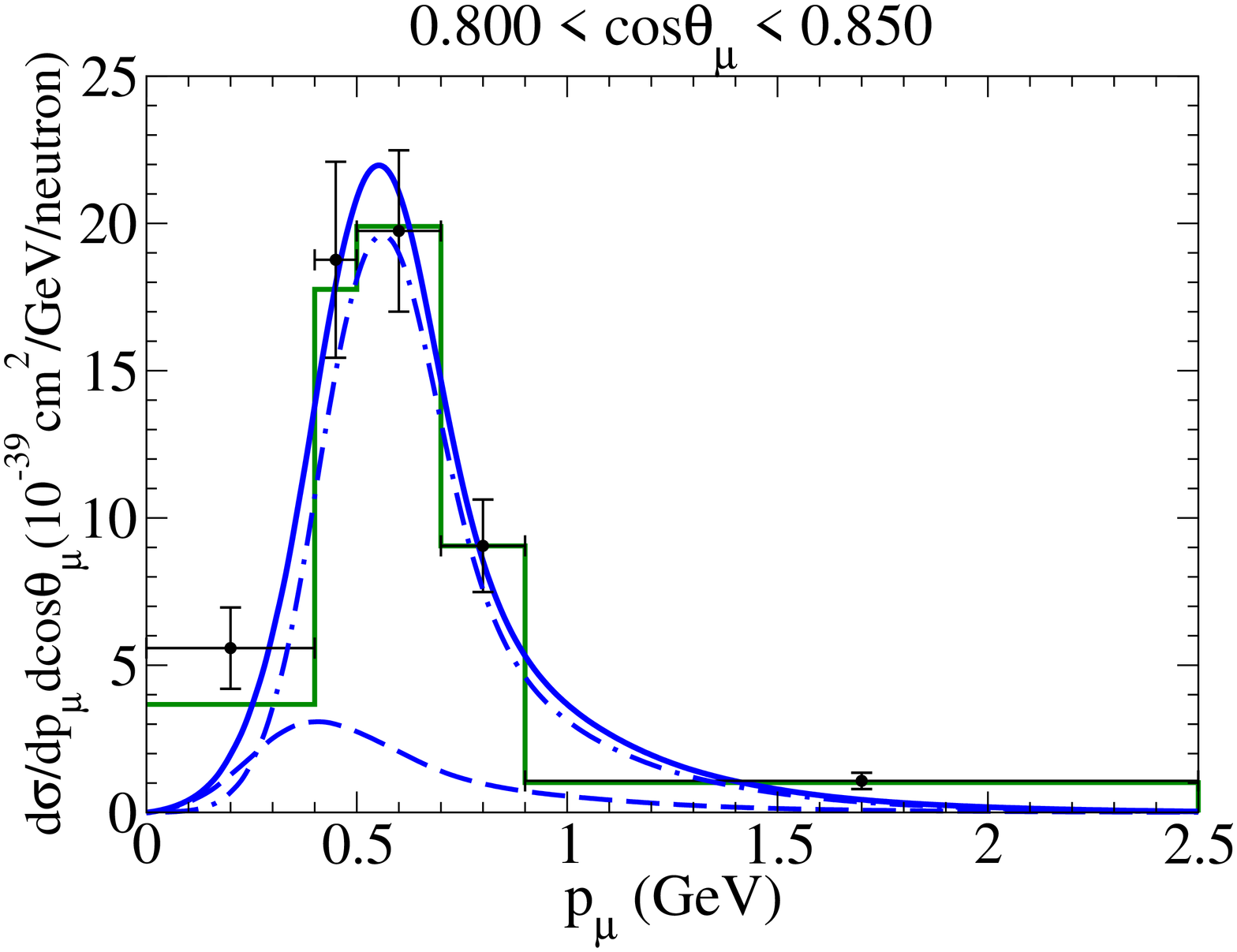}
\includegraphics[width=4.cm,angle=270]{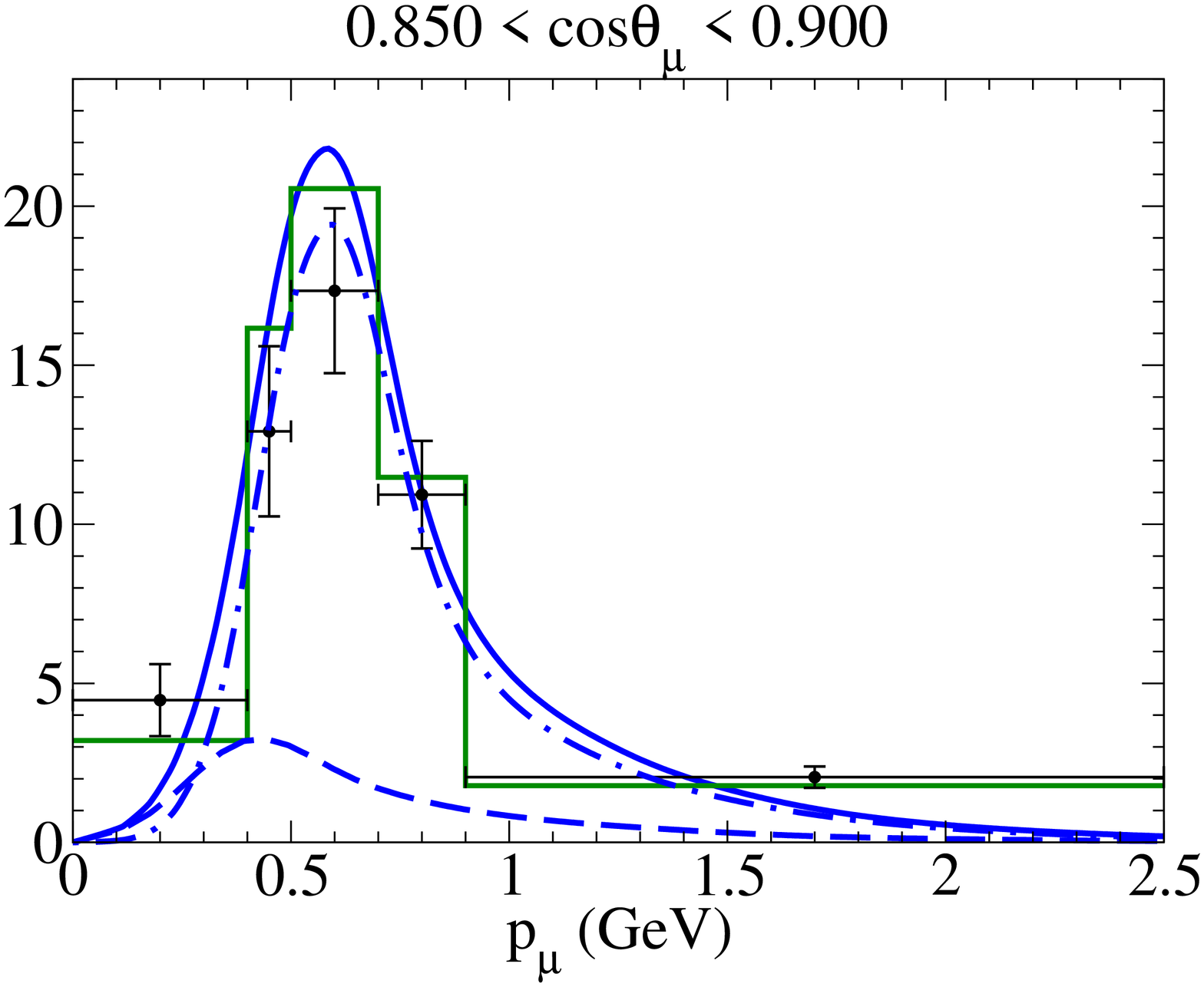}
\includegraphics[width=4.cm,angle=270]{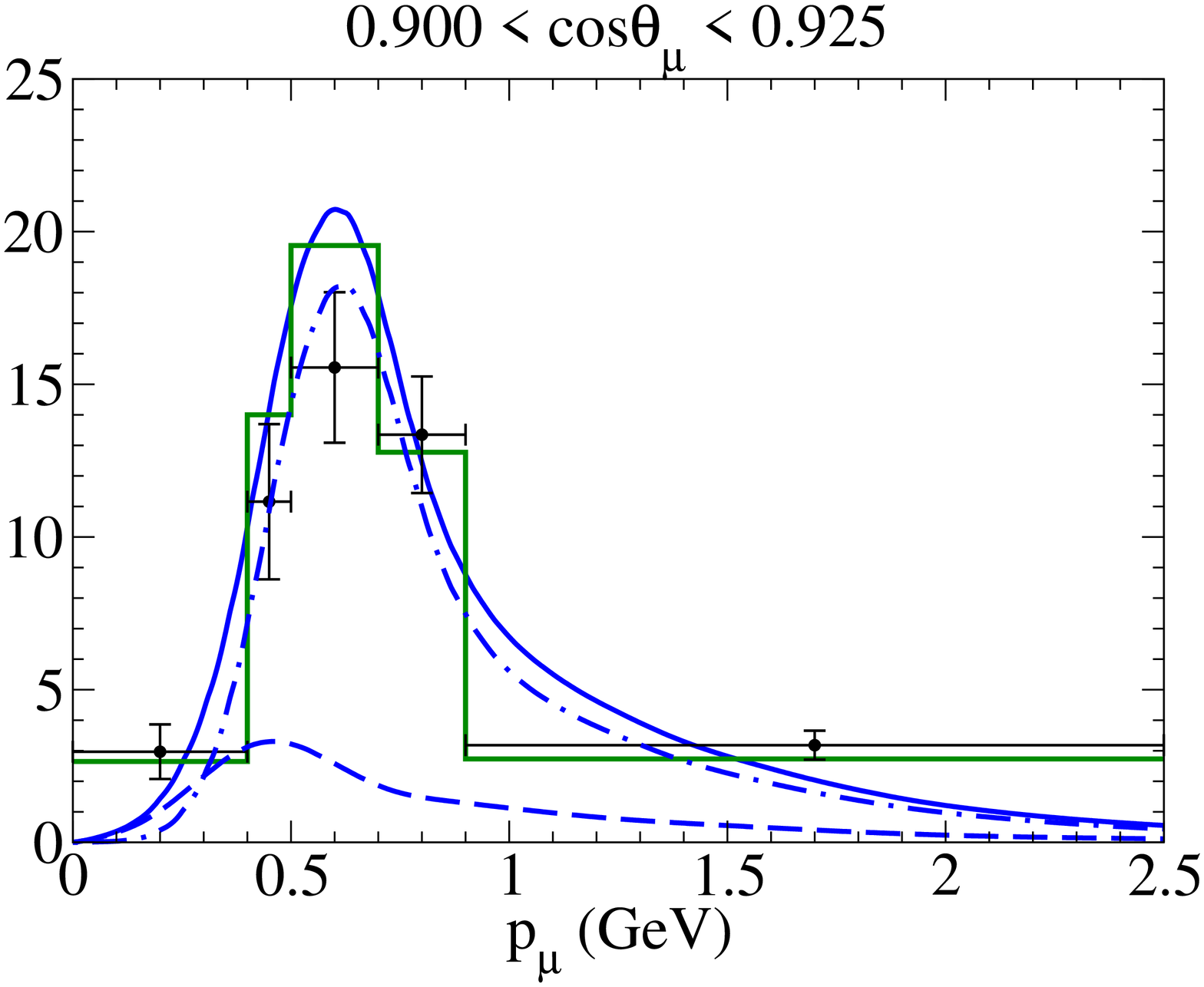}
\\
\includegraphics[width=4.cm,angle=270]{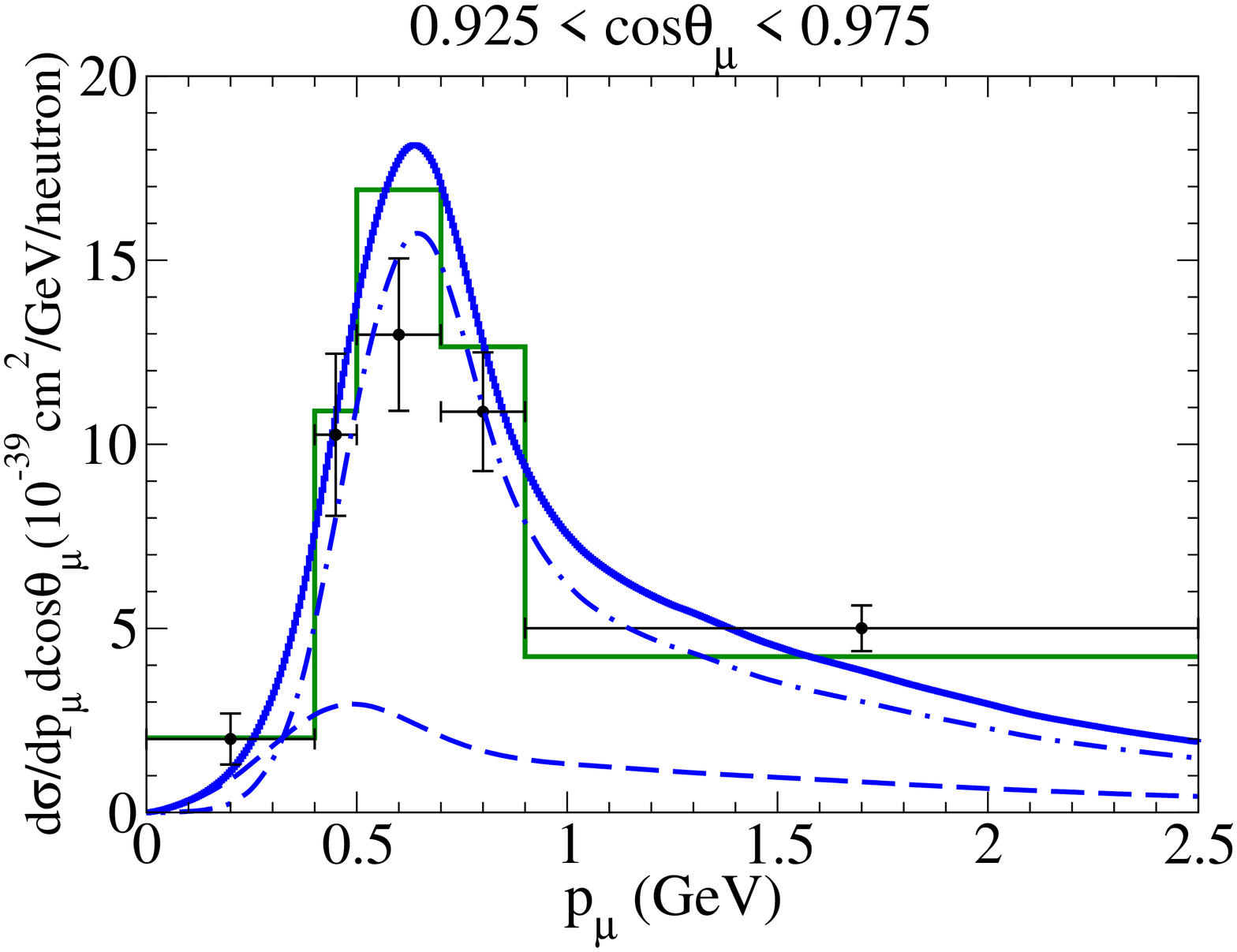}
\includegraphics[width=4.cm,angle=270]{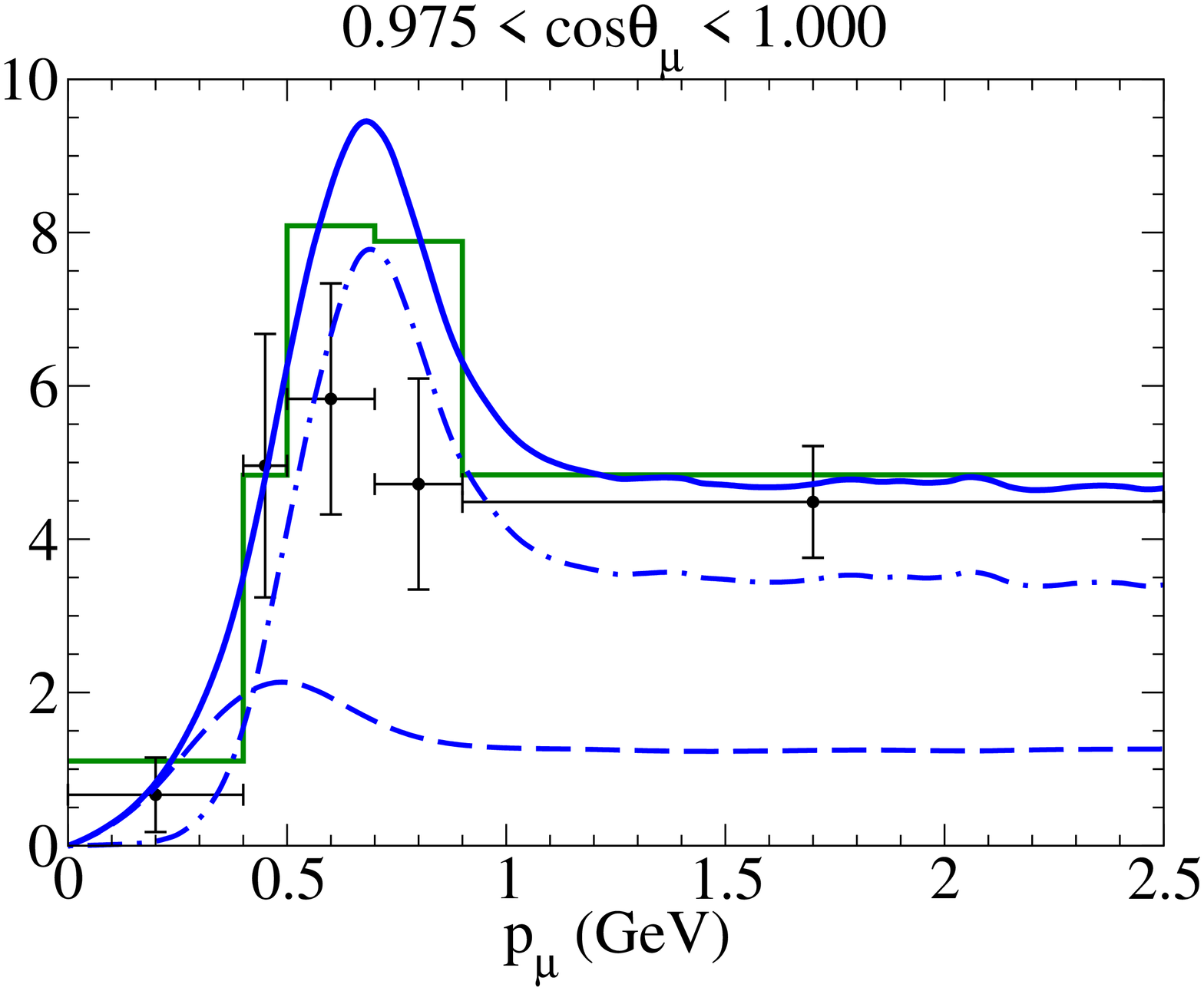}
\end{flushleft}
%\end{center}
\caption{\label{fig:fig2} (Color online)
T2K flux-folded double differential cross section per target neutron for the $\nu_\mu$ CCQE process on \oxygen displayed versus the
muon momentum $p_\mu$ for various bins of $\cos\theta_\mu$ obtained within the SuSAv2-MEC approach. QE and 2p-2h MEC results
are shown separately. The histogram represents the theoretical average of the total result over each bin of $p_\mu$. The data are from~\cite{T2Kwater}.
}
  \end{minipage}
\end{figure}

\begin{figure}
  \begin{minipage}{\textwidth}
    \begin{flushleft}
      %\begin{center}
\includegraphics[width=4.cm,angle=270]{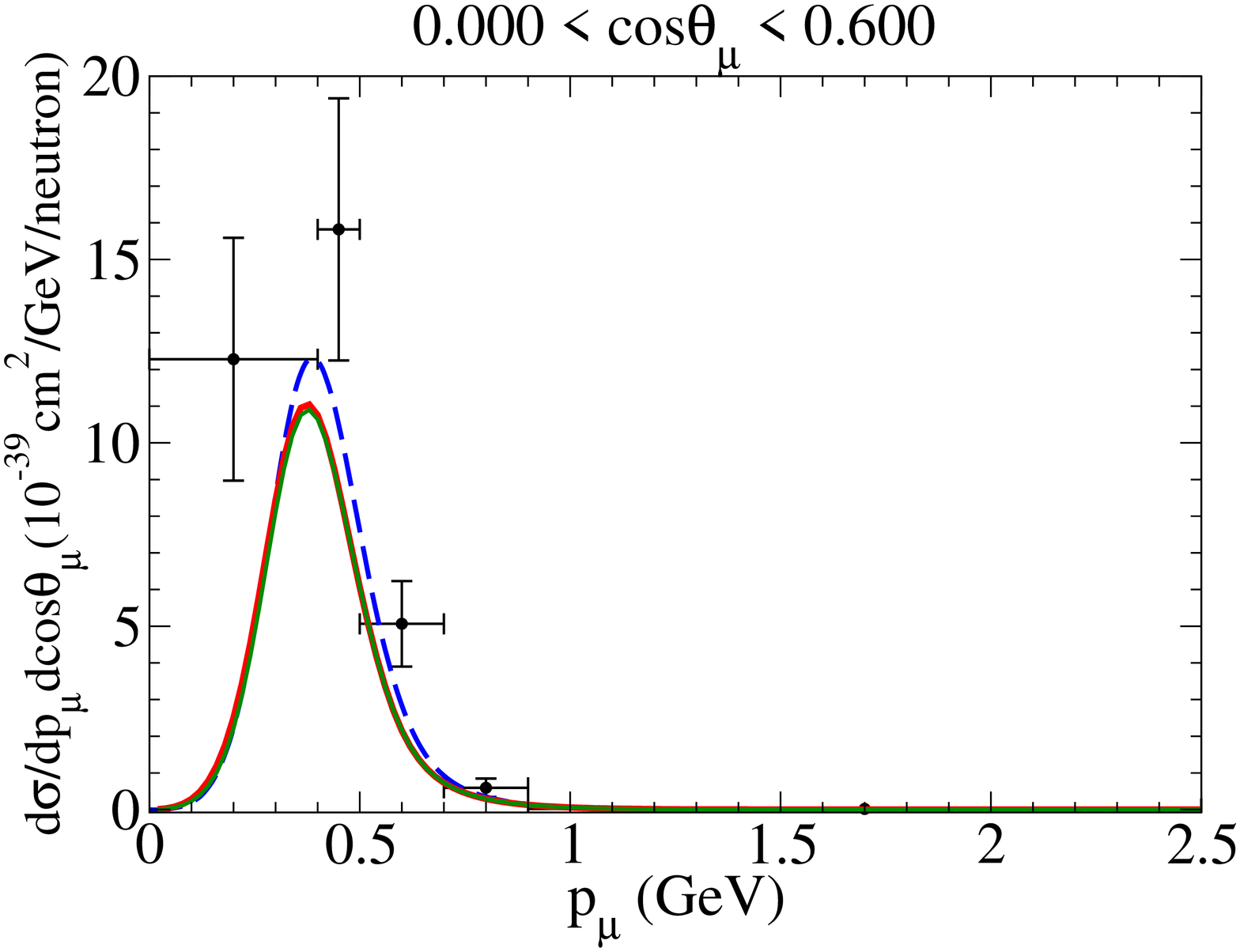}
\includegraphics[width=4.cm,angle=270]{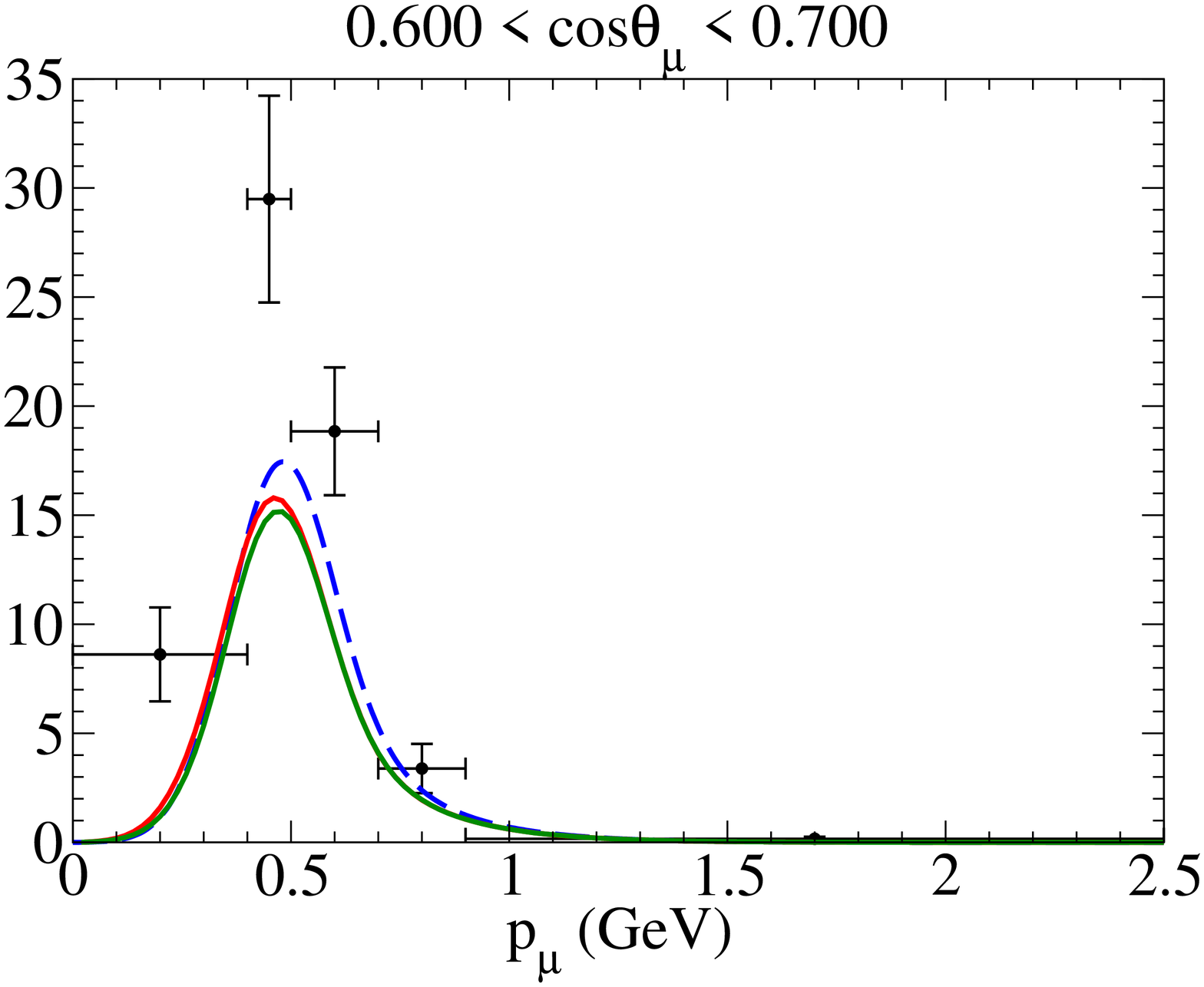}
\includegraphics[width=4.cm,angle=270]{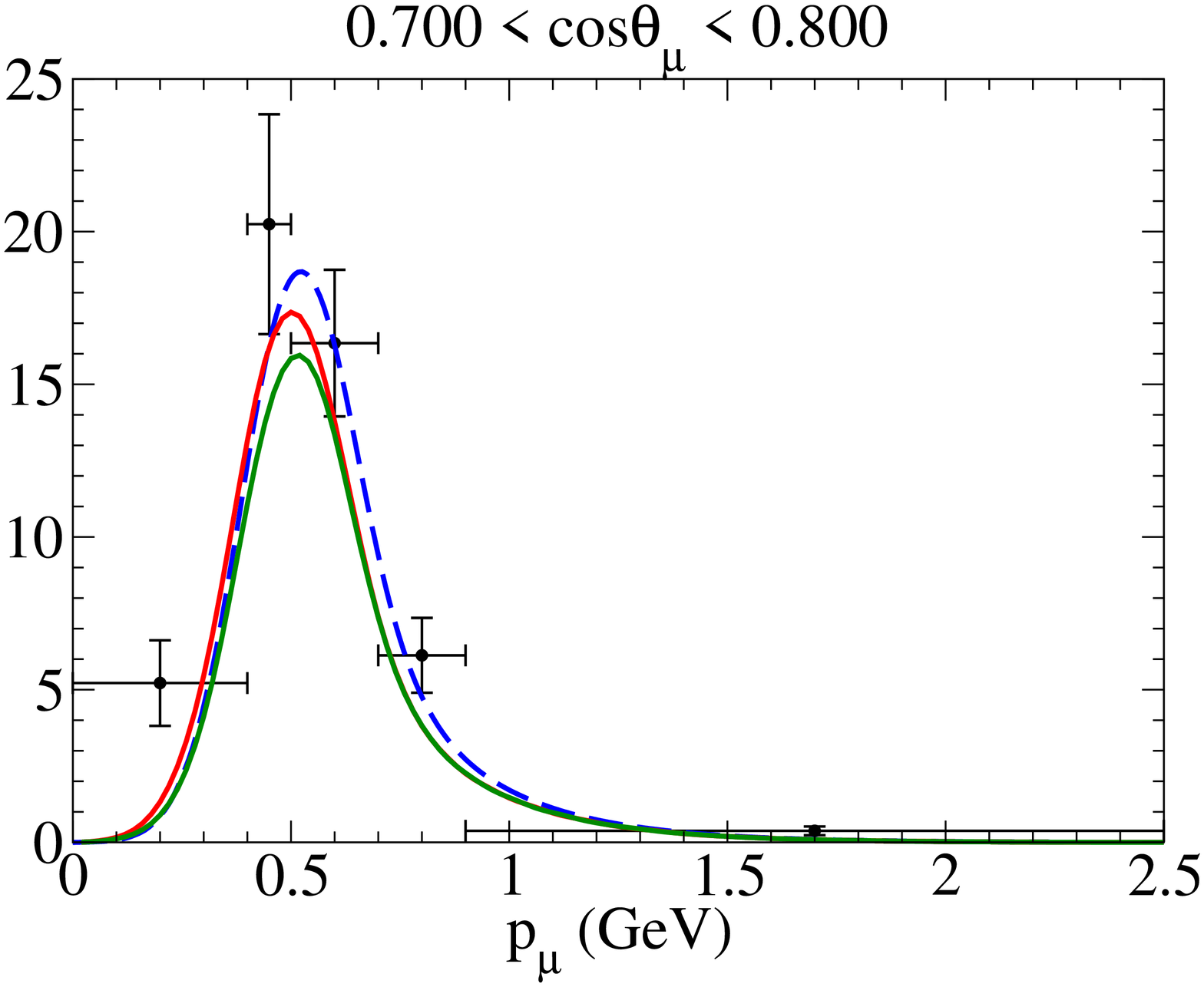}
\\
\includegraphics[width=4.cm,angle=270]{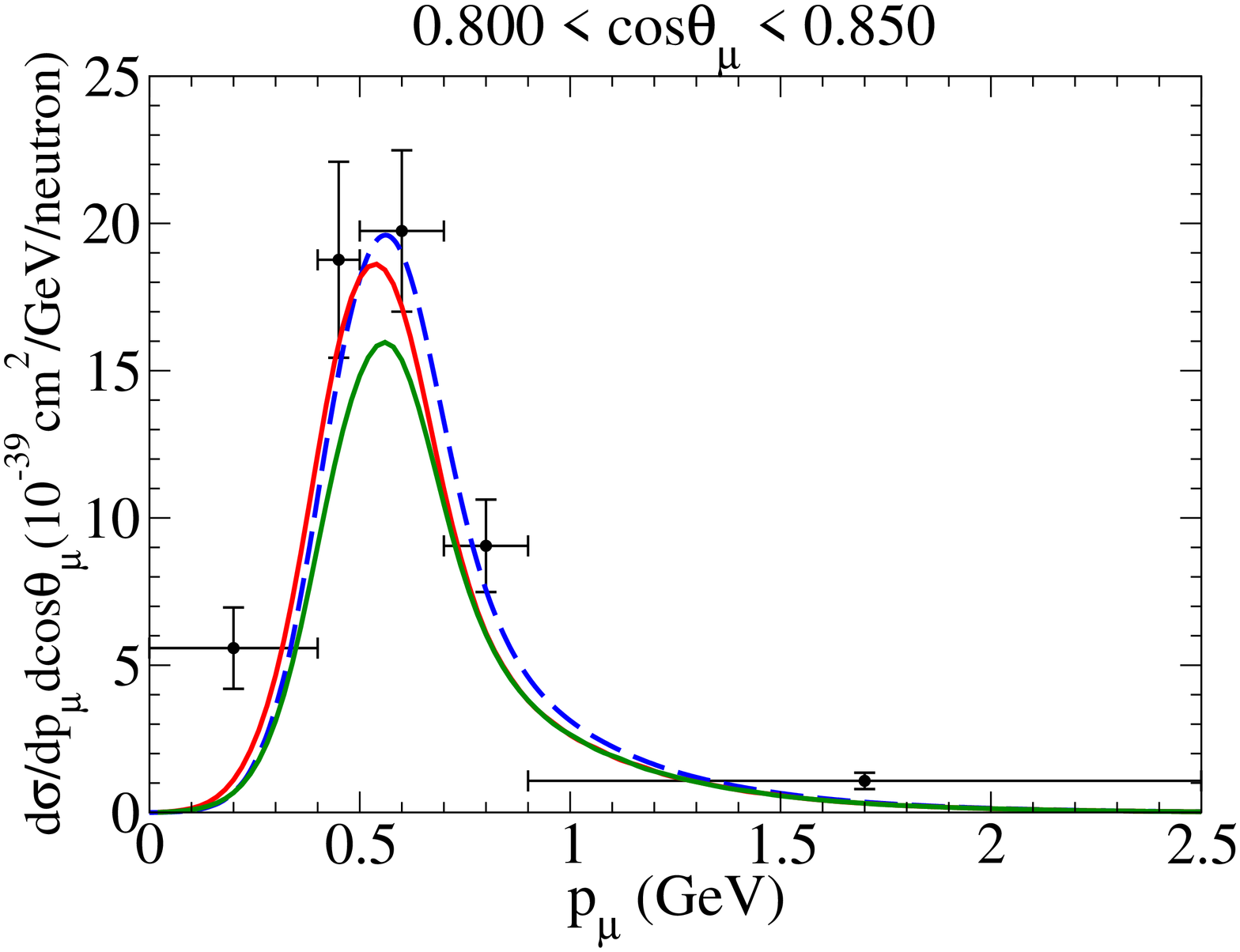}
\includegraphics[width=4.cm,angle=270]{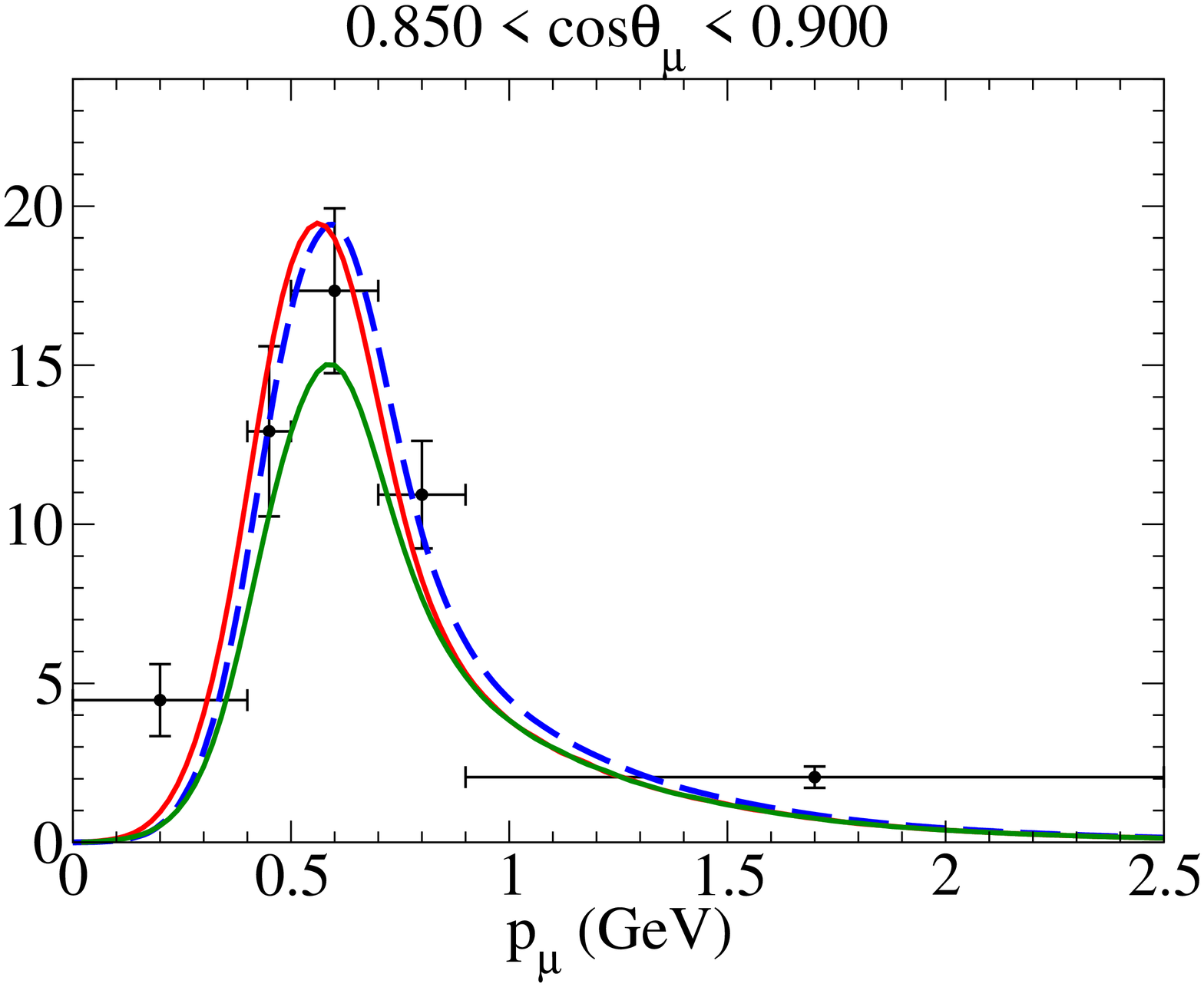}
\includegraphics[width=4.cm,angle=270]{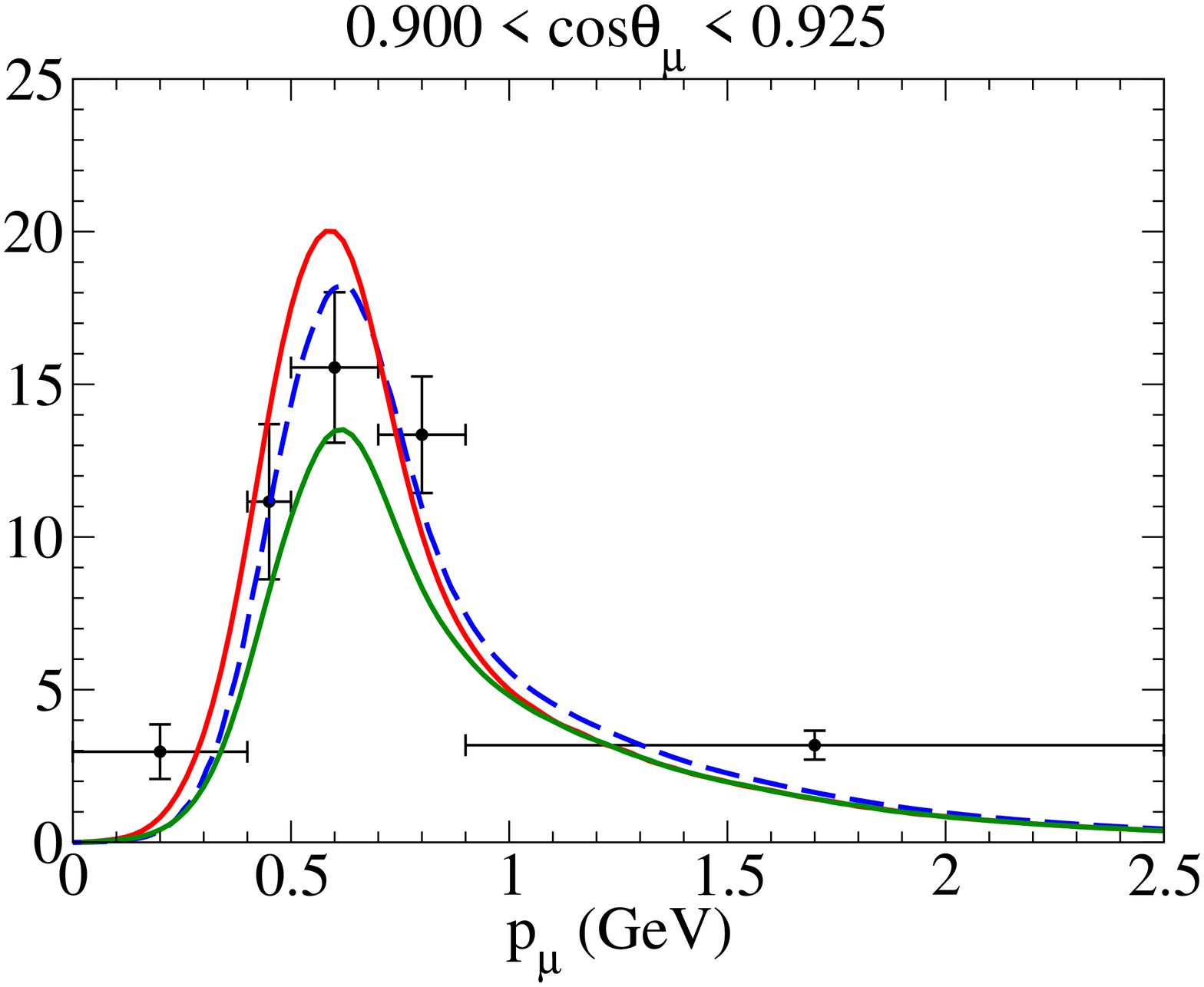}
\\
\includegraphics[width=4.cm,angle=270]{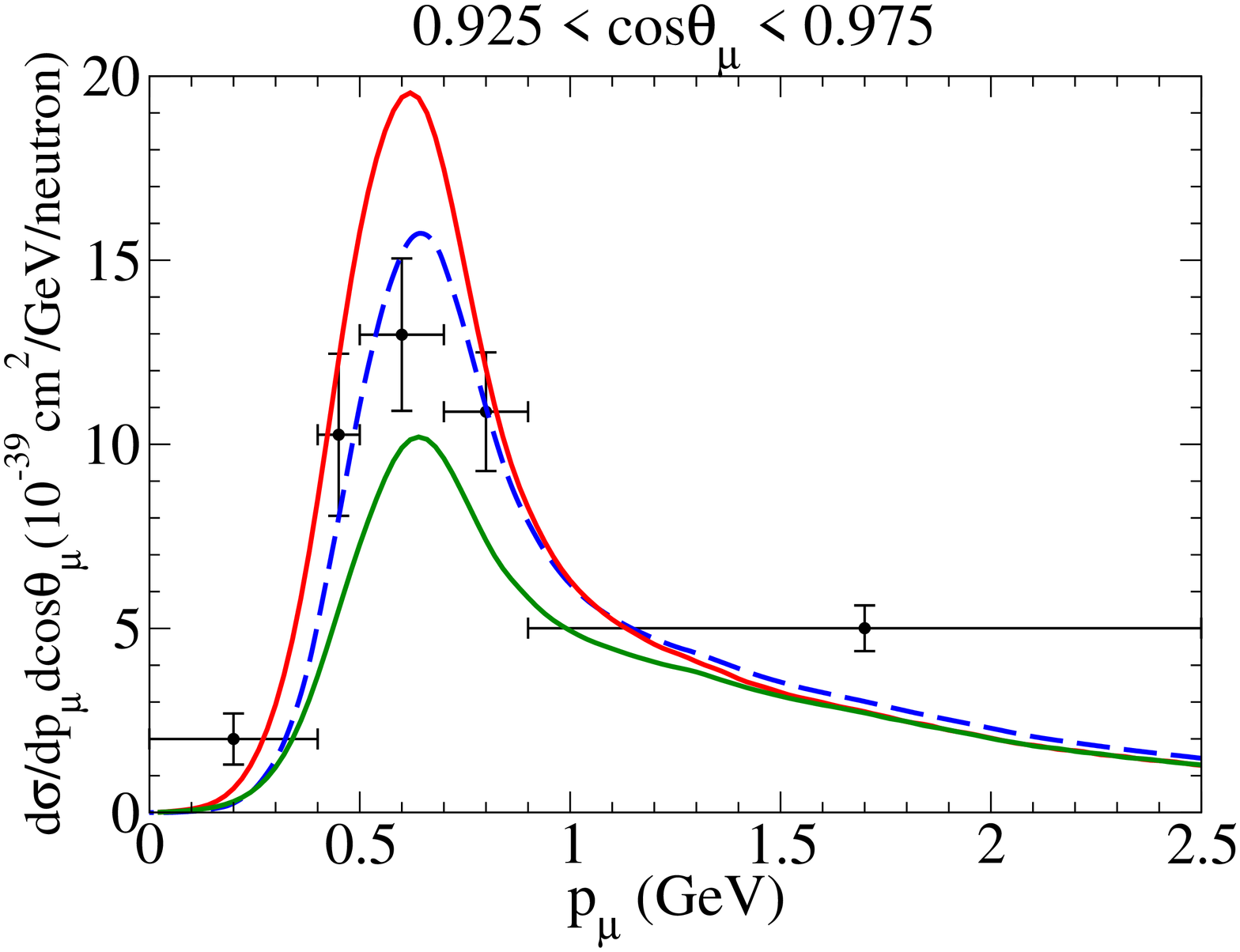}
\includegraphics[width=4.cm,angle=270]{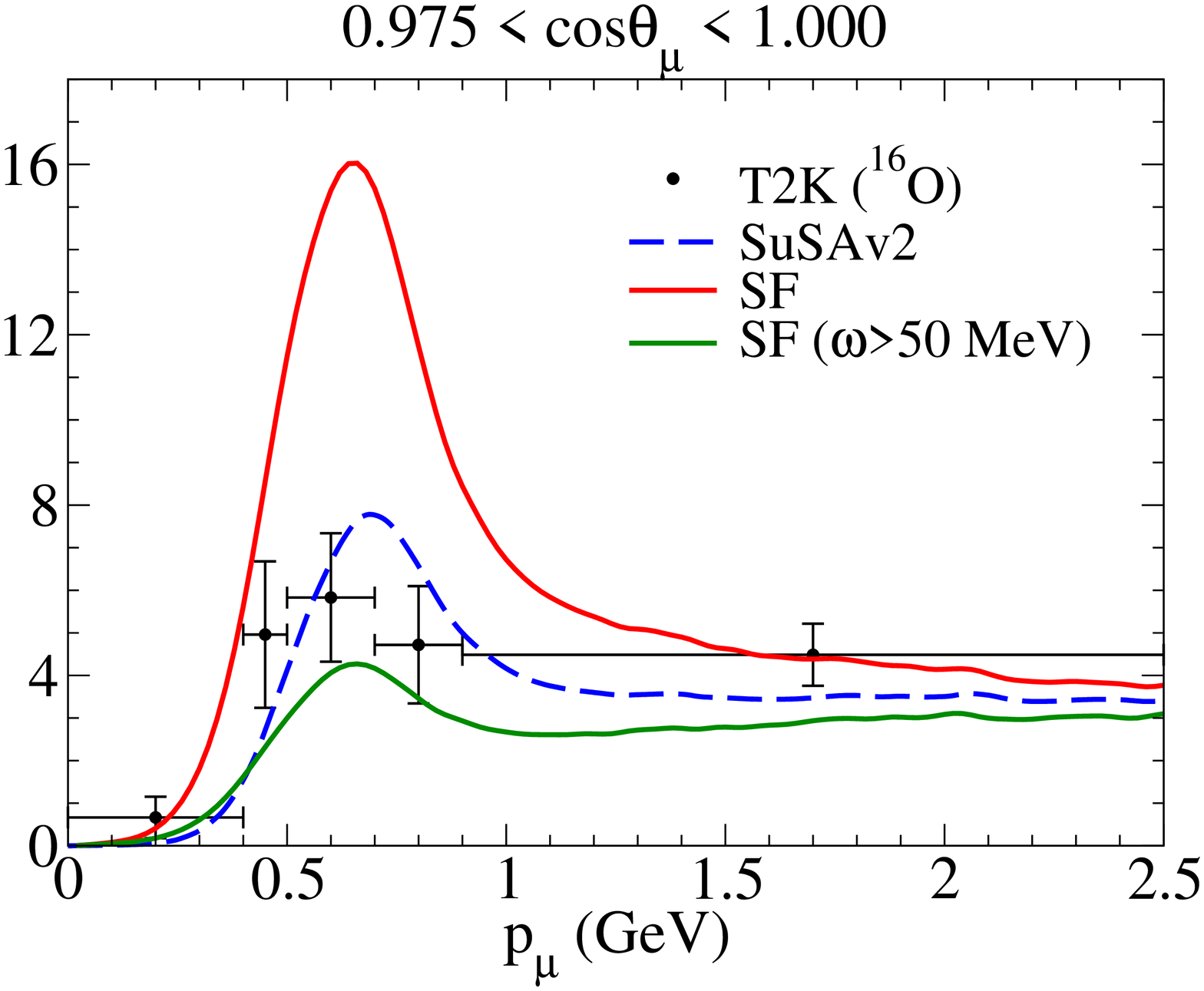}
\end{flushleft}
%\end{center}
\caption{\label{fig:fig2Wally} (Color online)
T2K flux-folded double differential cross section per target neutron for the $\nu_\mu$ CCQE process on \oxygen displayed versus the
muon momentum $p_\mu$ for various bins of $\cos\theta_\mu$ obtained within the SF model and the SuSAv2(QE) model. The contribution above $\omega>50$ MeV is also displayed for the SF model. The data are from~\cite{T2Kwater}.
}
  \end{minipage}
\end{figure}

\subsection{T2K: Oxygen versus Carbon}
\label{sec:T2Kcomp}

To make clear how nuclear effects enter in the analysis of the T2K
experiment, in Fig.~\ref{fig:fig3} we show the predictions provided by
SuSAv2-MEC for the neutrino-averaged double differential cross
sections per neutron in the cases of $^{12}$C (red lines) and
$^{16}$O (blue). Here we show only the total results of adding the QE and MEC contributions, since the latter are essentially equal for carbon and oxygen when scaled by the number of neutrons in the two nuclei; the MEC contributions for carbon are thus proportional to %\xout{essentially} 
those shown for oxygen in Fig.~\ref{fig:fig2}. Although the scaling behaviors of the QE and the 2p2h cross  sections are different --- while the former goes like  $k_F^{-1}$ (scaling of the second kind), the latter increases as  $k_F^2$ --- the results in Fig.~\ref{fig:fig3} are very similar for the  two nuclei in most of the kinematical situations. This is a consequence of the very close values of $k_F$ assumed in both cases, namely, $228$ MeV/c ($230$) for \carbon (\oxygen). Only at forward angles (bottom panels) do some differences between the results for \carbon and \oxygen emerge where the oxygen results are somewhat larger. Also, the amount of this difference increases as the scattering angle approaches zero. This behavior comes essentially 
from the QE response, since, as noted above, the MEC contributions for the two nuclei are very similar. It arises from the fact that at very forward angles the transferred energy in the process is very small and then the different values of $E_{shift}$ for the two nuclei become significant.
It should also be noted that in the last two panels the experimental angular bins are not exactly the same for the two nuclei.
However, in spite of these potential sensitivities to small-$\omega$ dynamics, it is important to point out that the model is capable of reproducing the data for \carbon and \oxygen within their error dispersion. As a test to evaluate the importance of having different $E_{shift}$ values at low energies, these were set equal for the two nuclei and the effect goes away. Again, as stated above, the near-threshold region should be viewed with caution in all existing modeling.
Although not shown here for simplicity, we have analyzed the differential cross section by modifying the values of the Fermi momentum and shift energy by $\pm 10\%$. The relative changes at the maxima of the cross sections are of the order of $\sim 10-15\%$. However, in the case of the most forward angles and larger muon momenta, where the cross section stabilized, these can reach $\sim 30\%$. 
  
We explore the dependence of the C/O differences upon the neutrino energy in a bit more detail by displaying in Fig.~\ref{fig:fig4} the total integrated cross section per neutron with no neutrino flux included versus the neutrino energy.  The results shown here indicate that nuclear effects between these nuclei in the total cross section, that is, including both  the QE and 2p-2h MEC contributions, are very tiny, at most of the  order of $\sim$2-3$\%$.  This minor  difference is also observed for the pure QE response (slightly higher for carbon) and the 2p-2h MEC (larger for oxygen). This is connected with the  differing scaling behavior shown by the QE and 2p-2h MEC responses  with the Fermi momentum, and the very close values of $k_F$ selected  for the two nuclei. 
Note however that the relative amount of 2p-2h in the CCQE-like sample, which has important consequences for the neutrino energy reconstruction, increases as $k_F^2$ and is therefore more important for oxygen that for carbon.
Upon including both  the QE and 2p-2h MEC contributions, one observes that nuclear effects in the total cross section are very tiny. We also show the effect of making a ``cut'' at $\omega = 50$ MeV, namely, setting any contribution from below this point to zero. This has been used in past work as a crude sensitivity test to ascertain the relative importance of the near-threshold region. If significant differences are observed when making the cut, then one should have some doubts about the ability of the present modeling (indeed, likely of all existing modeling) to successfully represent the cross section in this region. What we observe for the total cross section shown in the figure are relatively modest effects from near-threshold contributions, although one should be aware that this is not so for differential cross sections at very forward angles where small-$\omega$ contributions can be relatively important, as discussed above.

Although not shown in Fig.~\ref{fig:fig4} for  simplicity, the use of a smaller value of $k_F$ for $^{16}$O, as the  one $k_F=216$ MeV/c considered in some previous work~\cite{Caballero:2005sj,Caballero:2006wi,Amaro:2005sr}, leads  to more significant differences in the QE (being larger) and 2p-2h  MEC (smaller) 
contributions, but the total response remains rather  similar to the result for $^{12}$C. It is important to point out  that the use of different $k_F$-values only leads to significant  discrepancies for low transferred energies, {\it i.e.}, $\omega\leq 50$  MeV, a kinematical region where other ingredients, not included in the SuSAv2-MEC model, can be important. Moreover, the smaller the  neutrino energy is the larger are the relative contributions coming from  transfer energies below $50$ MeV. % The analysis of the ratio  \oxygen/\carbon leads to the conclusion that, for neutrino energies above $500$ MeV, the normalized cross  sections for the two nuclei are almost identical ({\it i.e.,} their ratio is equal to 1),  independently of the particular $k_F$-value selected for oxygen. On  the contrary, the situation clearly differs for smaller neutrino  energies, $E_\nu \lesssim  500$ MeV. Here the ratio goes  down very quickly as $E_\nu$ falls.  Another ingredient in the calculation that also affects the results within this low-neutrino energy region concerns the energy shift. A careful analysis of $(e,e')$ data for both nuclei leads to the use of energy shift values that differ by $\sim$4 MeV, the one corresponding to $^{16}$O being smaller (see discussion in previous sections).  

The results for the single-differential cross sections per neutron corresponding to the T2K experiment are presented in Fig.~\ref{fig:fig5}. Here we show the cross sections both versus the scattering angle and against the muon momentum using the values of the Fermi momentum given above, namely, 228 MeV/c (\carbon) and 230 MeV/c (\oxygen). The separate contributions of the pure QE (dot-dashed), 2p-2h MEC (dashed) and the total result (solid) are shown. As noted, the differences seen between the two nuclei are very tiny. Then in Fig.~\ref{fig:fig5bis} we show the same cross sections for oxygen obtained using two different values of $k_F$, namely, 230 and 216 MeV/c, to verify the statement made in Sec.~II B that the results are typically relatively insensitive to variations of this magnitude.

Concerning the analysis of the role played by the parameters $k_F$ and $E_{shift}$, that in this model characterize the different nuclei, the uncertainty is of the order of $\sim$2-3$\%$. The analysis of the ratio  \oxygen/\carbon leads to differences below 3$\%$ except for the low-kinematic region where these figures are larger as a consequence of the high sensitivity to different parameters such as the effects arising from $E_{shift}$ and the mass of the residual nucleus. 

\begin{figure}
  \begin{minipage}{\textwidth}
    \begin{flushleft}
%\begin{center}
\includegraphics[width=4.cm,angle=270]{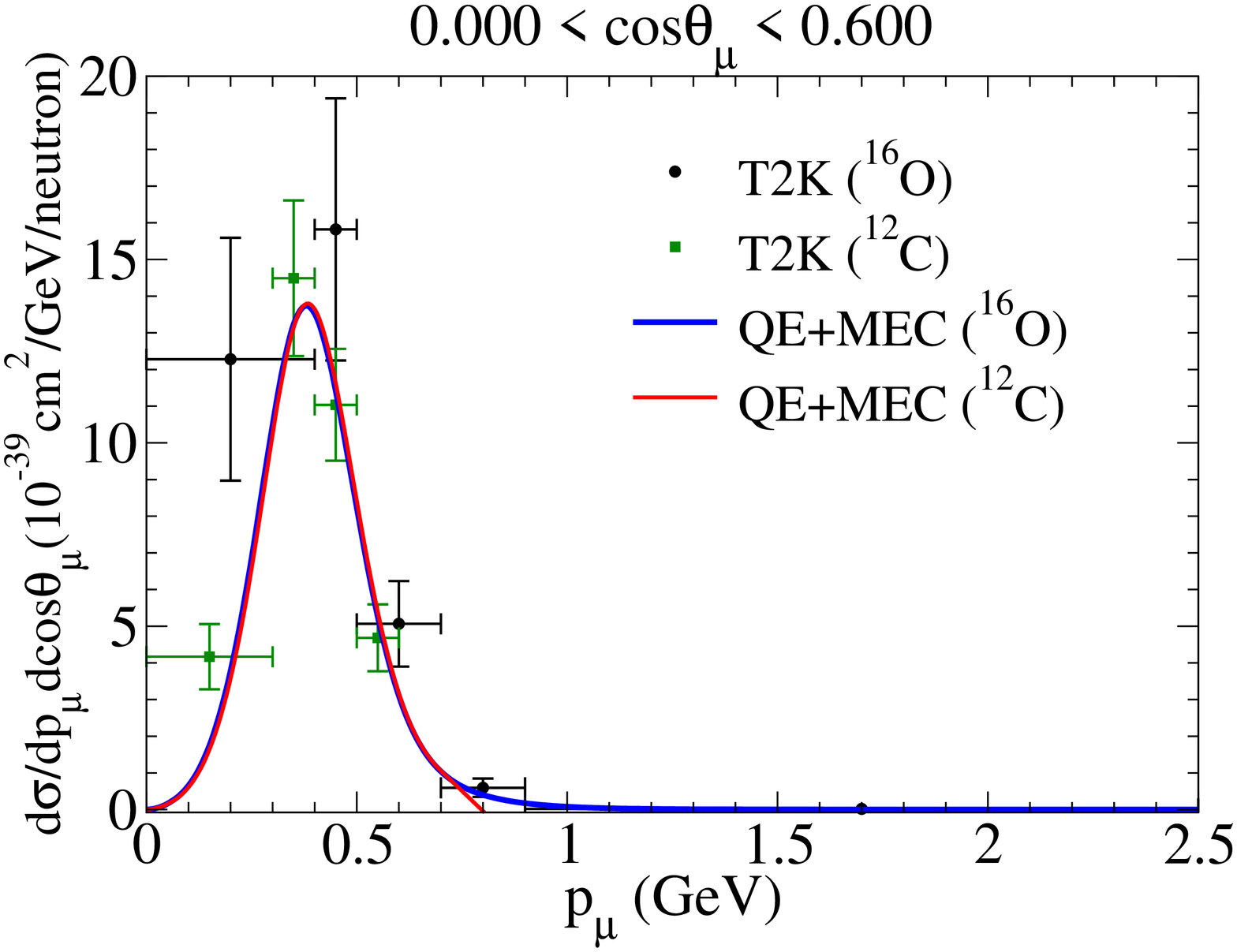}
\includegraphics[width=4.cm,angle=270]{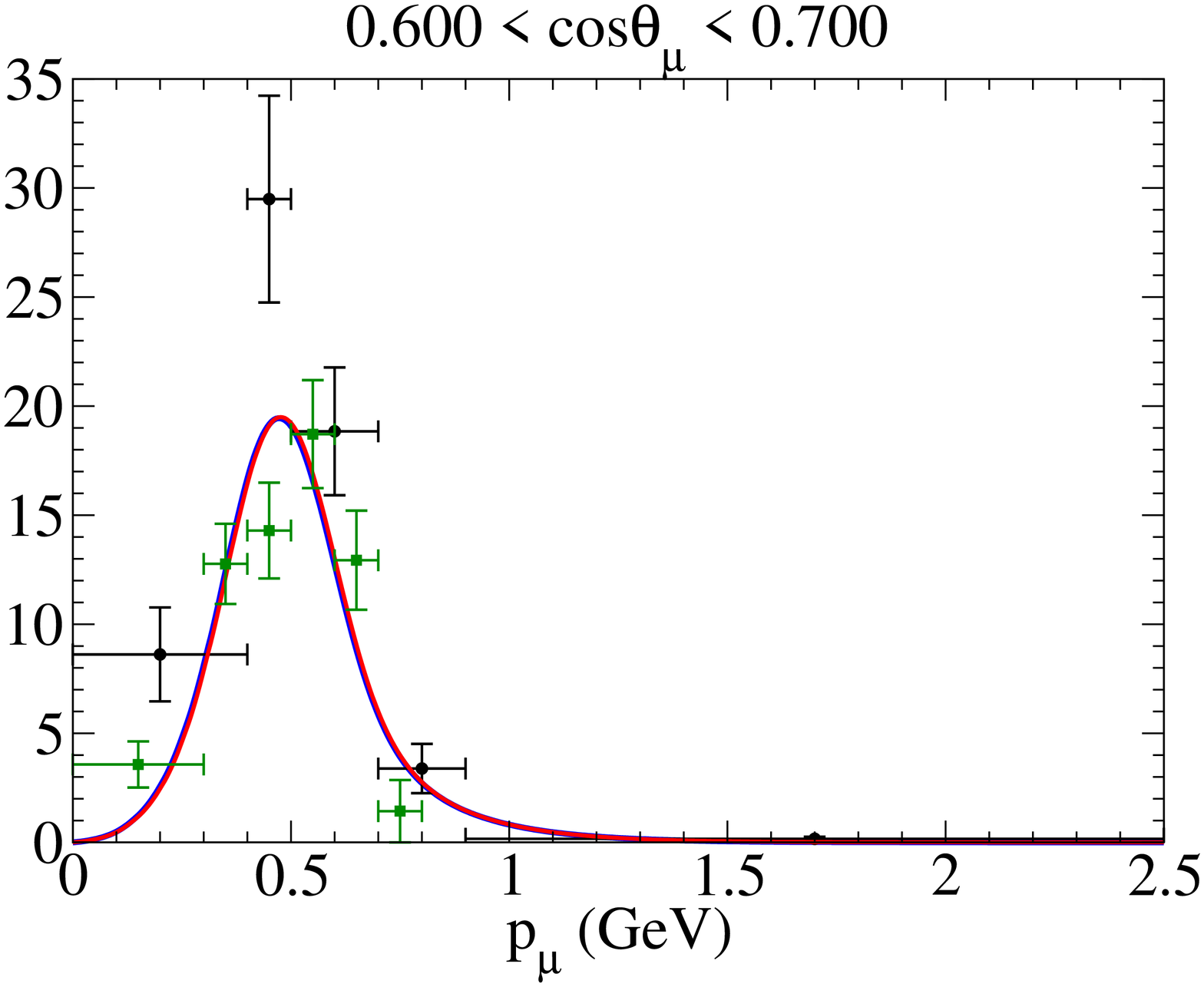}
\includegraphics[width=4.cm,angle=270]{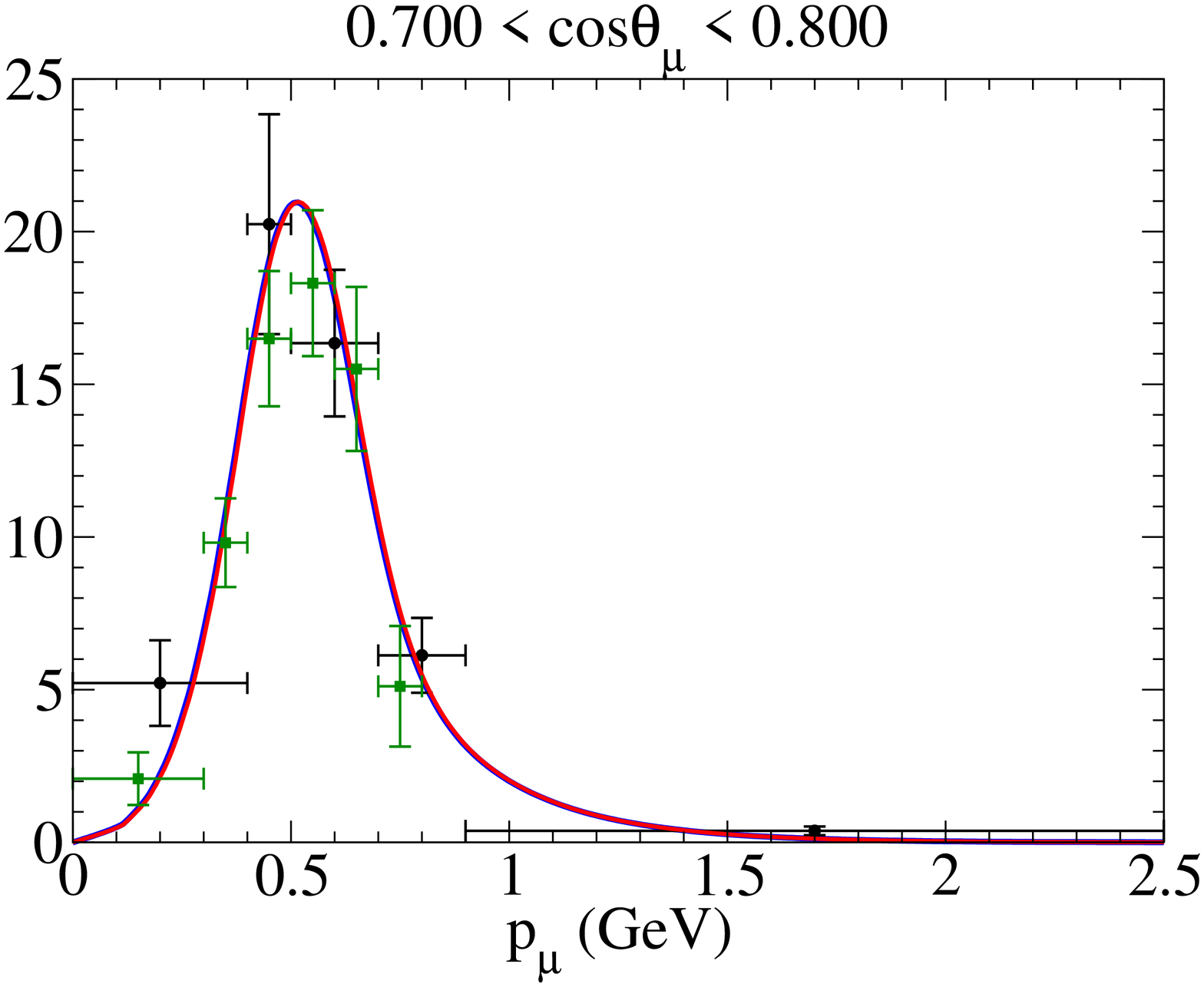}
\\
\includegraphics[width=4.cm,angle=270]{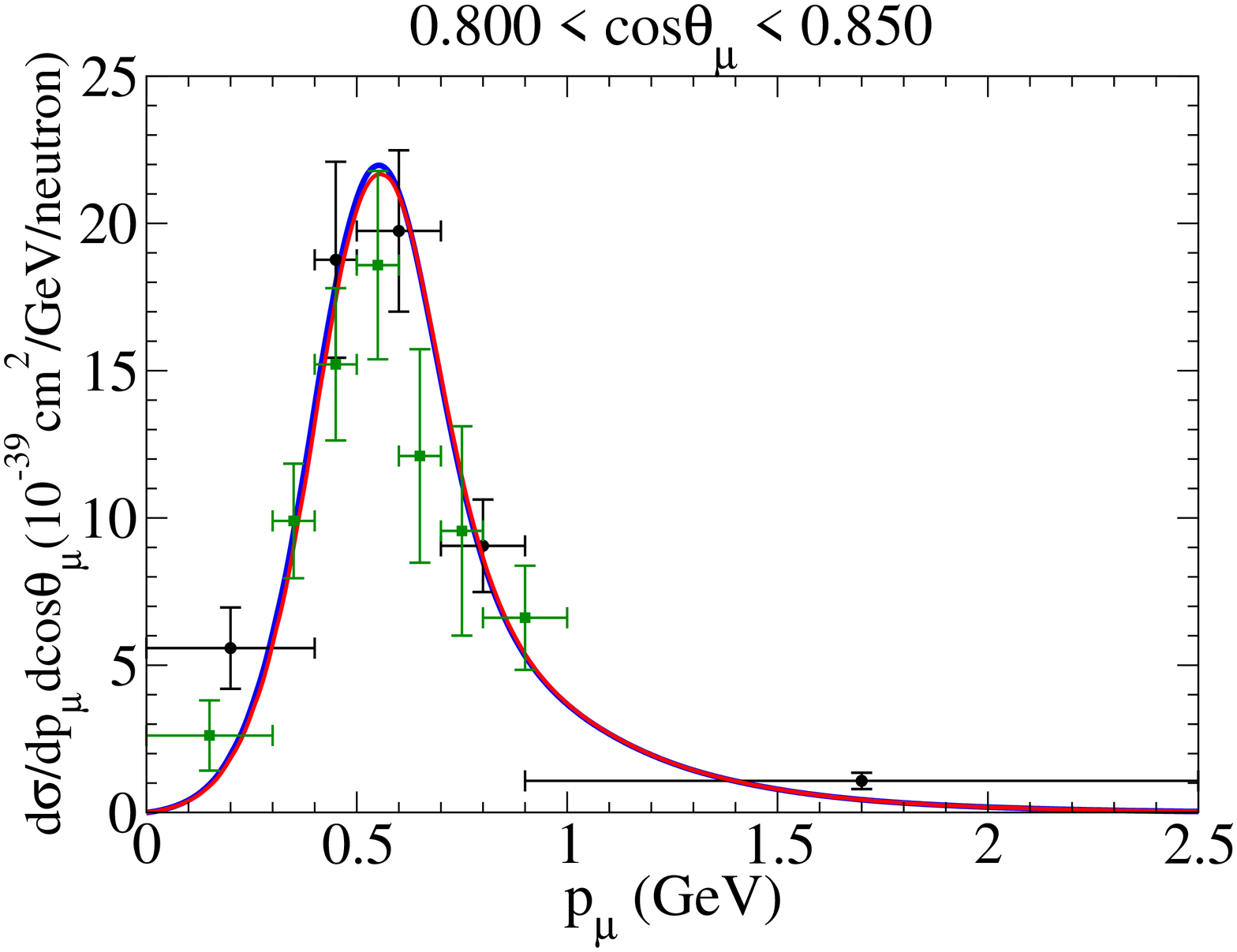}
\includegraphics[width=4.cm,angle=270]{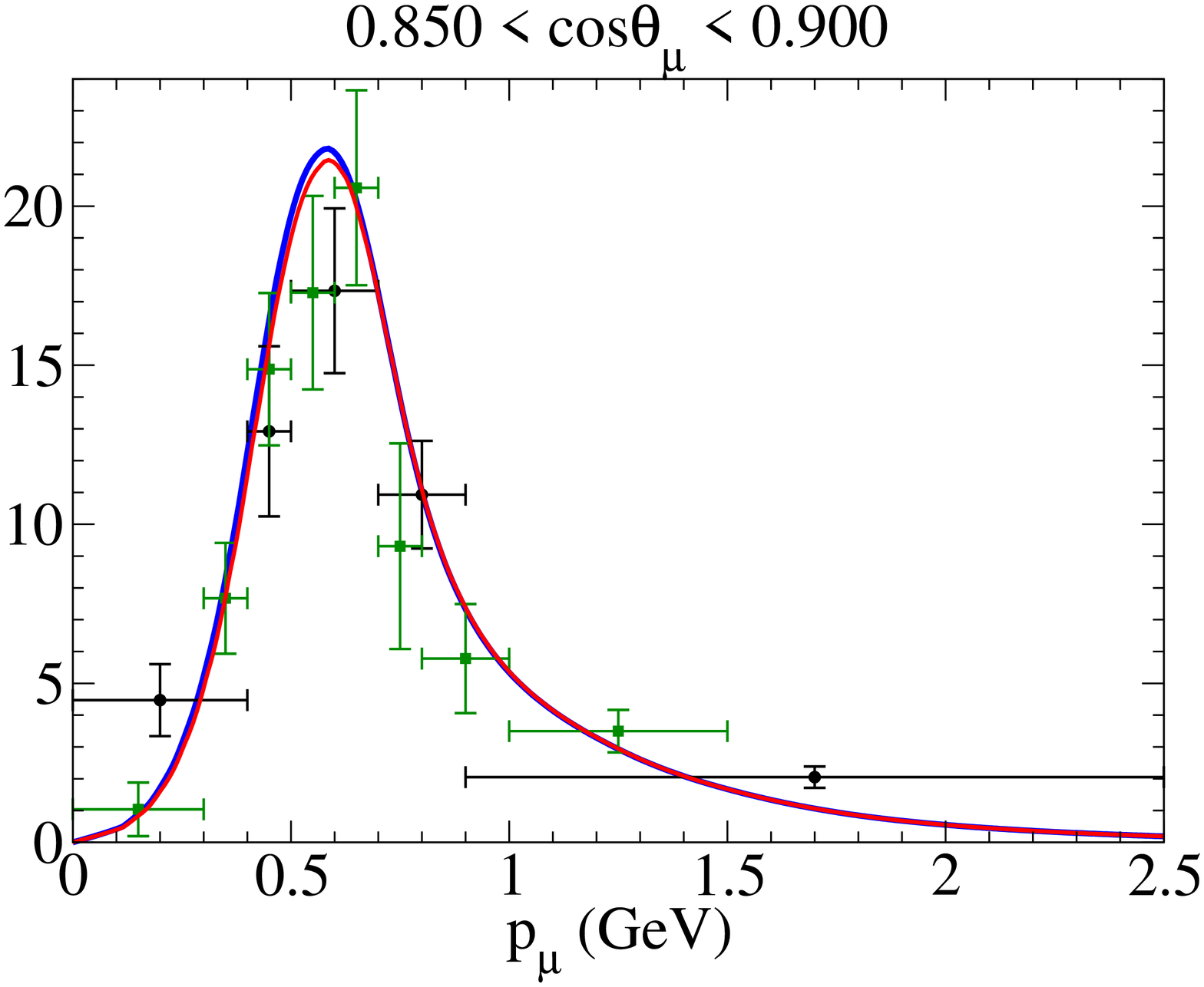}
\includegraphics[width=4.cm,angle=270]{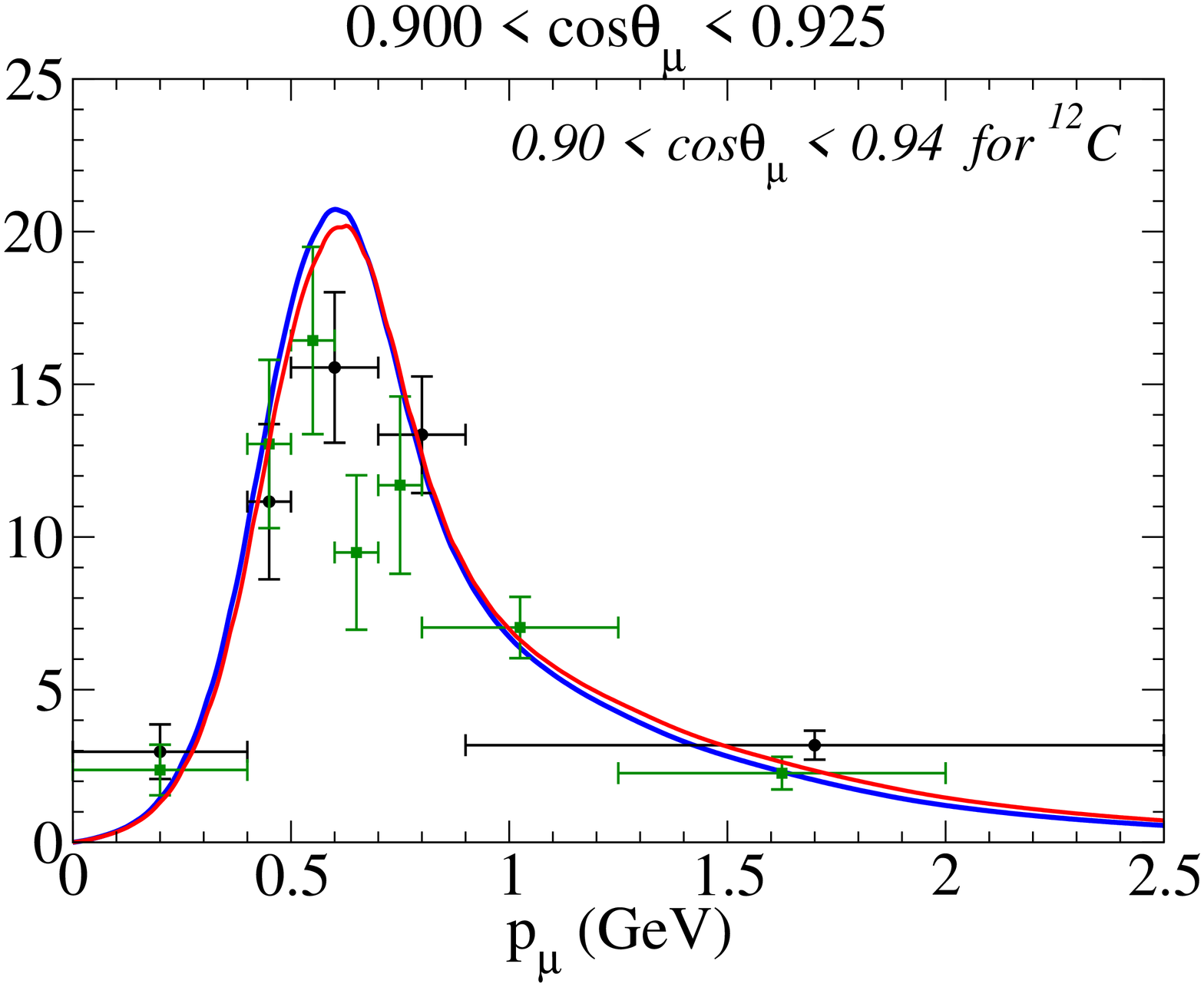}
\\
\includegraphics[width=4.cm,angle=270]{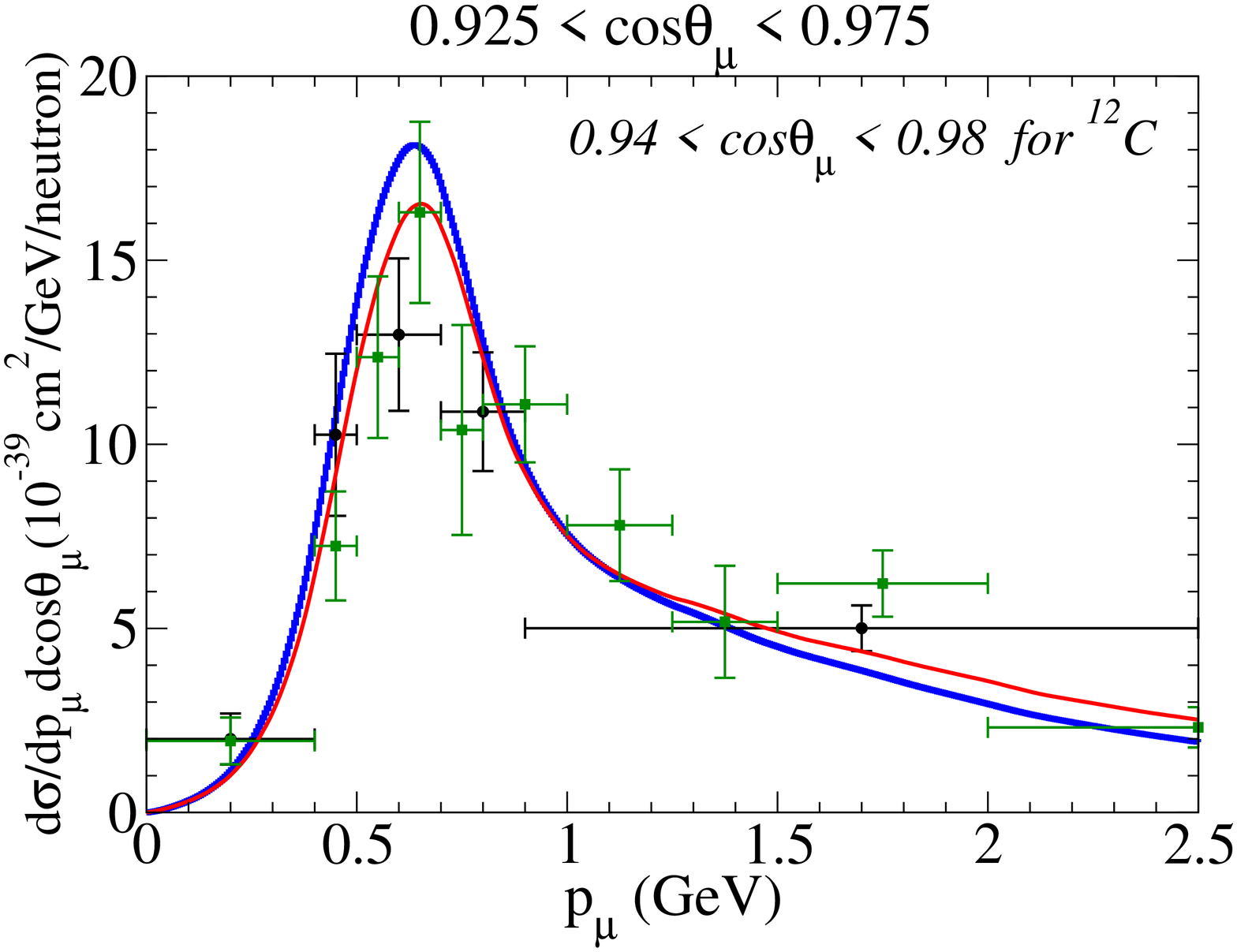}
\includegraphics[width=4.cm,angle=270]{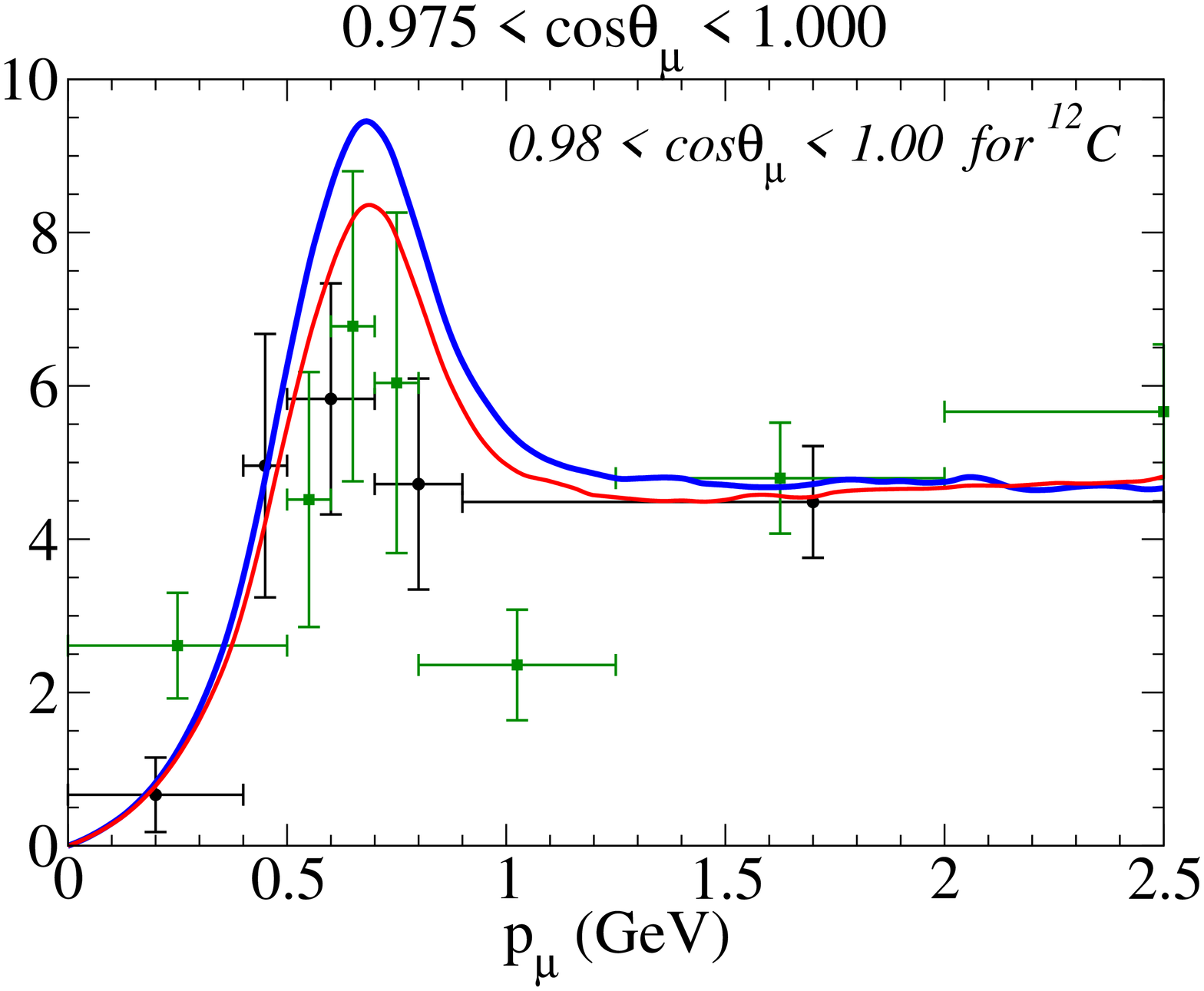}
\end{flushleft}
%\end{center}
\caption{\label{fig:fig3} (Color online)
  Similar to Fig.~\ref{fig:fig2}, but now including also the results corresponding to the T2K-$\nu_\mu$ CCQE process on $^{12}$C. The data are from \cite{T2Kwater,T2Kcc0pi}
}
  \end{minipage}
\end{figure}

\begin{figure}
  \begin{minipage}{\textwidth}
%    \begin{flushleft}
      \begin{center}
\includegraphics[width=9.cm,angle=270]{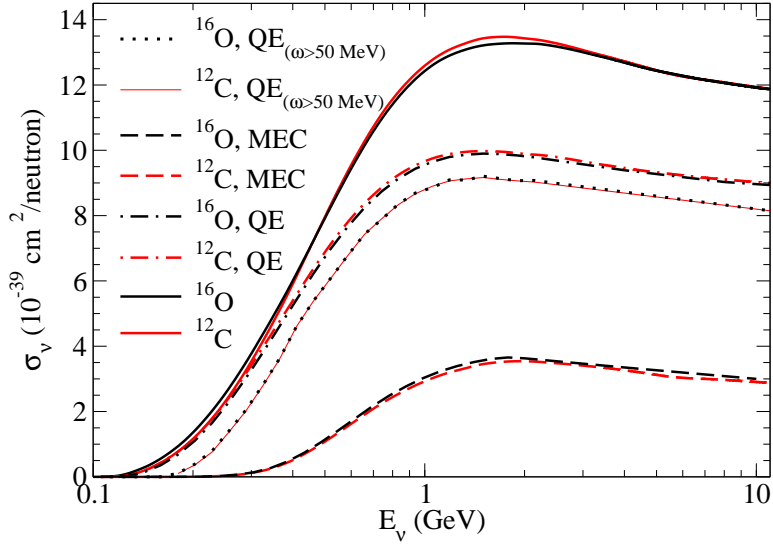}
% \includegraphics[width=6.cm,angle=270]{TOTAL_12C_16O.ps}
%\\
%\includegraphics[width=6.cm,angle=270]{TOTAL_12C_16O_kf230_ratio.ps}
%\includegraphics[width=6.cm,angle=270]{TOTAL_12C_16O_kf230_ratio2.ps}
%\end{flushleft}
\end{center}

\caption{\label{fig:fig4} (Color online) Total $\nu_\mu$
  cross section per nucleon as a function of the neutrino energy
  evaluated for \carbon and \oxygen nuclei. Separate contributions of
  the pure QE (dot-dashed) and 2p-2h MEC (dashed). The effect of making a cut in $\omega$ below 50 MeV for the QE contribution is also shown.
}

  \end{minipage}
\end{figure}

\begin{figure}
  \begin{minipage}{\textwidth}
    \begin{flushleft}
      %\begin{center}
\includegraphics[width=6.cm,angle=270]{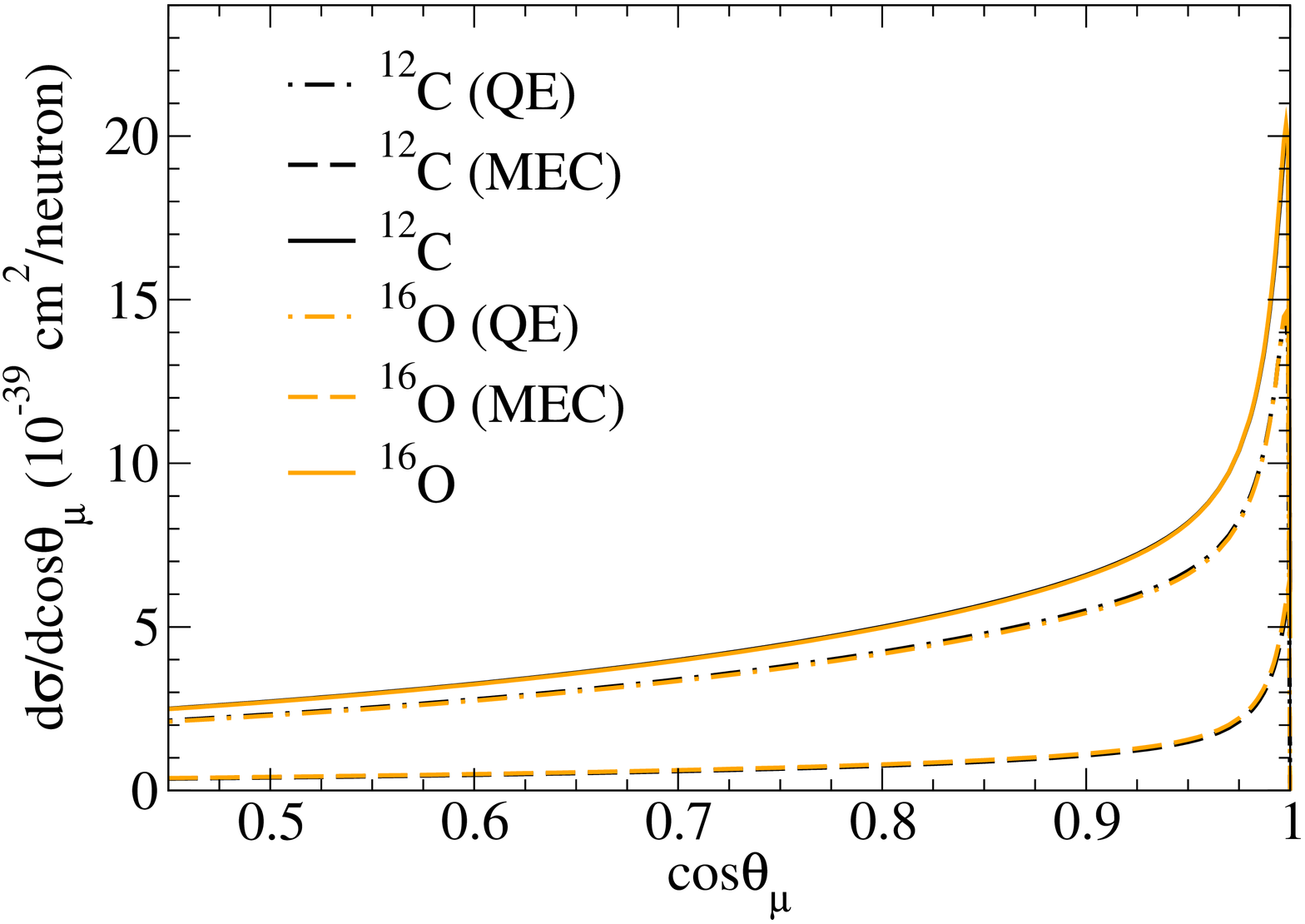}
\includegraphics[width=6.cm,angle=270]{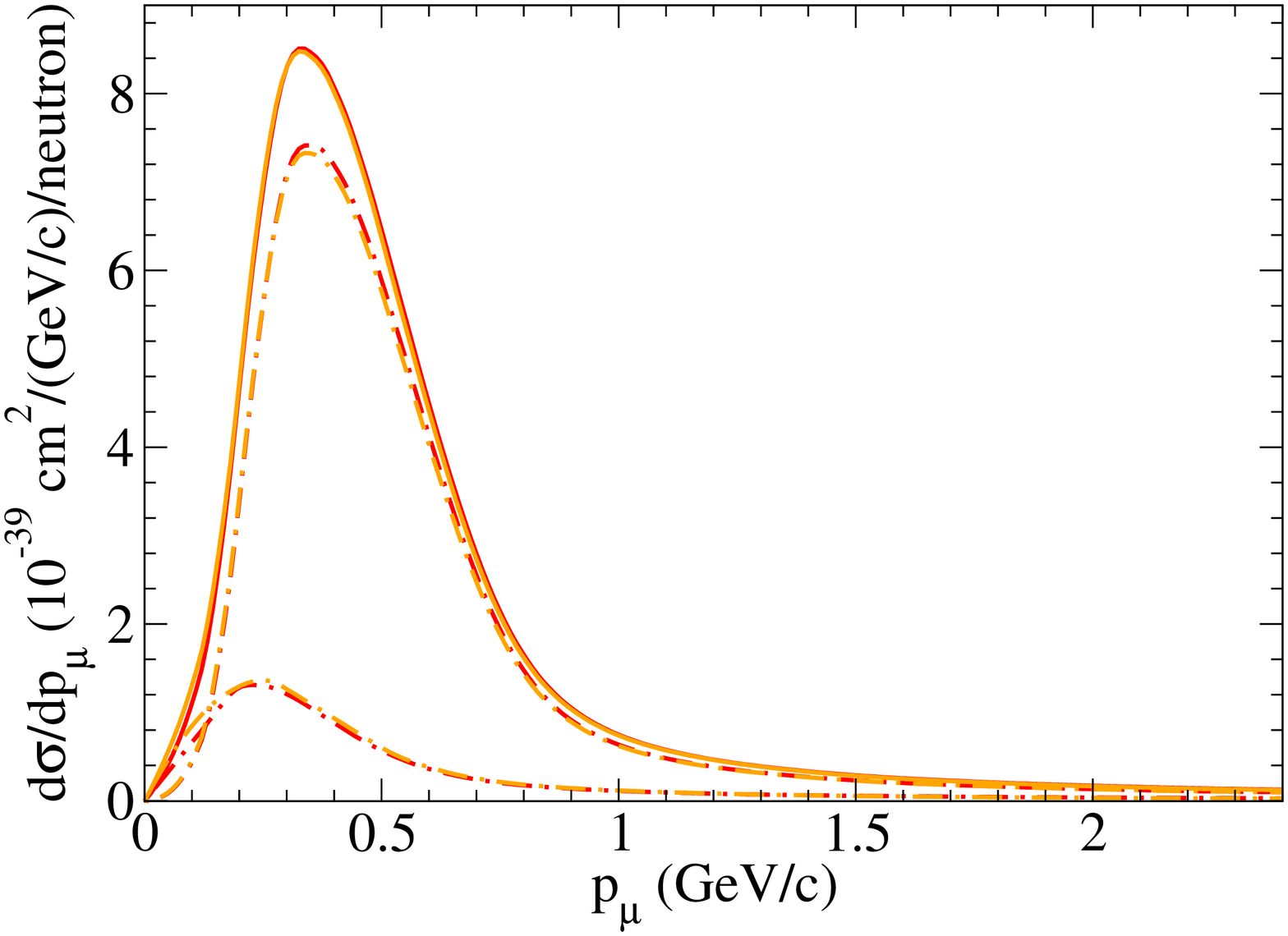}
%\\
%\includegraphics[width=6.cm,angle=270]{T2K_16O_costhetamu_ratio230.ps}
%\includegraphics[width=6.cm,angle=270]{T2K_16O_pmu_ratio230.ps}
\end{flushleft}
%\end{center}
    \caption{\label{fig:fig5} (Color online). %Left panels:
      T2K flux-averaged CCQE neutrino differential cross sections per neutron for \carbon and \oxygen as functions of the muon scattering angle (left panel) and of the muon momentum (right panel). The Fermi momenta here are 228 MeV/c for \carbon and 230 MeV/c for \oxygen. 
}
  \end{minipage}
\end{figure}

\begin{figure}
  \begin{minipage}{\textwidth}
    \begin{flushleft}
      %\begin{center}
\includegraphics[width=6.cm,angle=270]{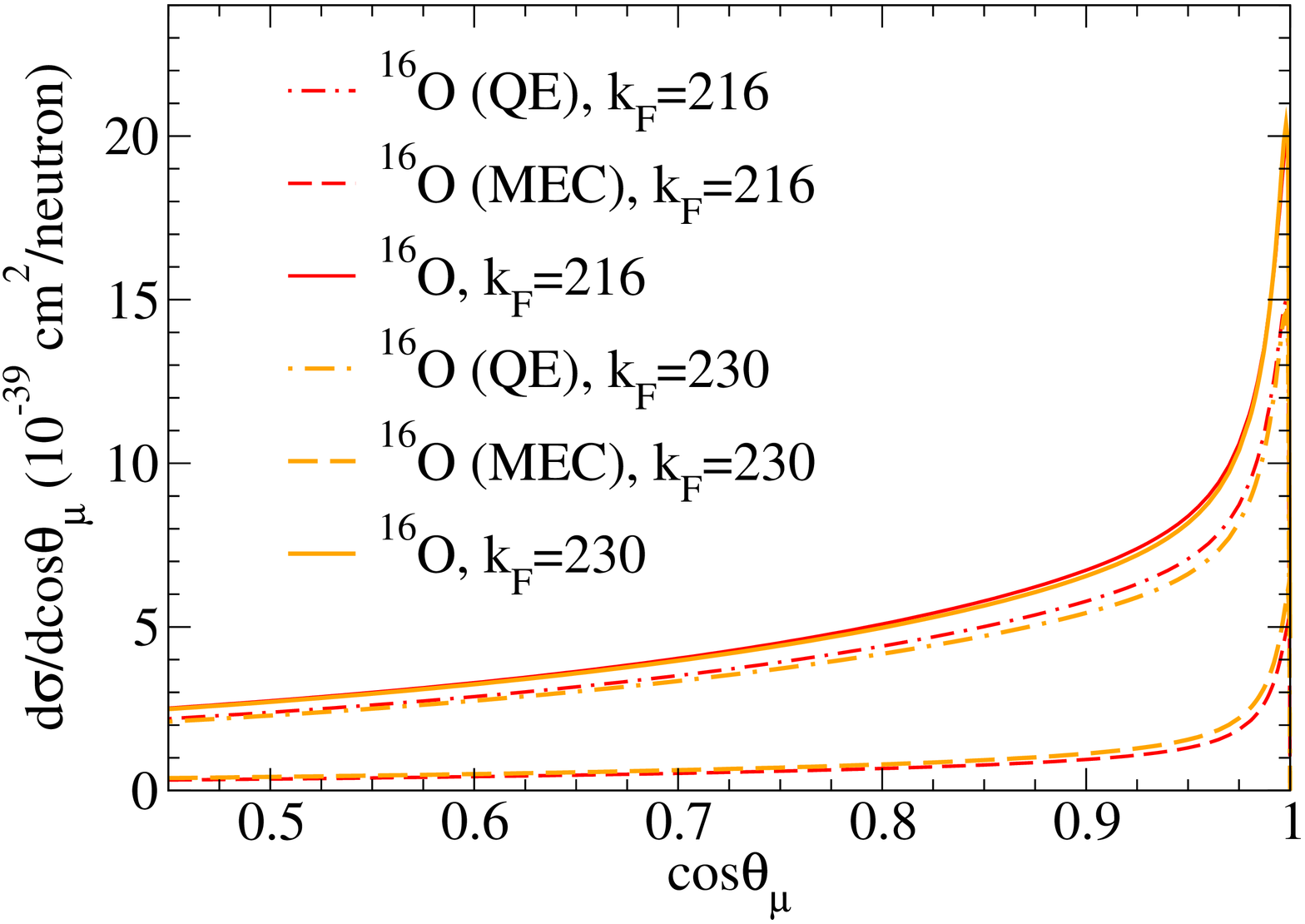}
\includegraphics[width=6.cm,angle=270]{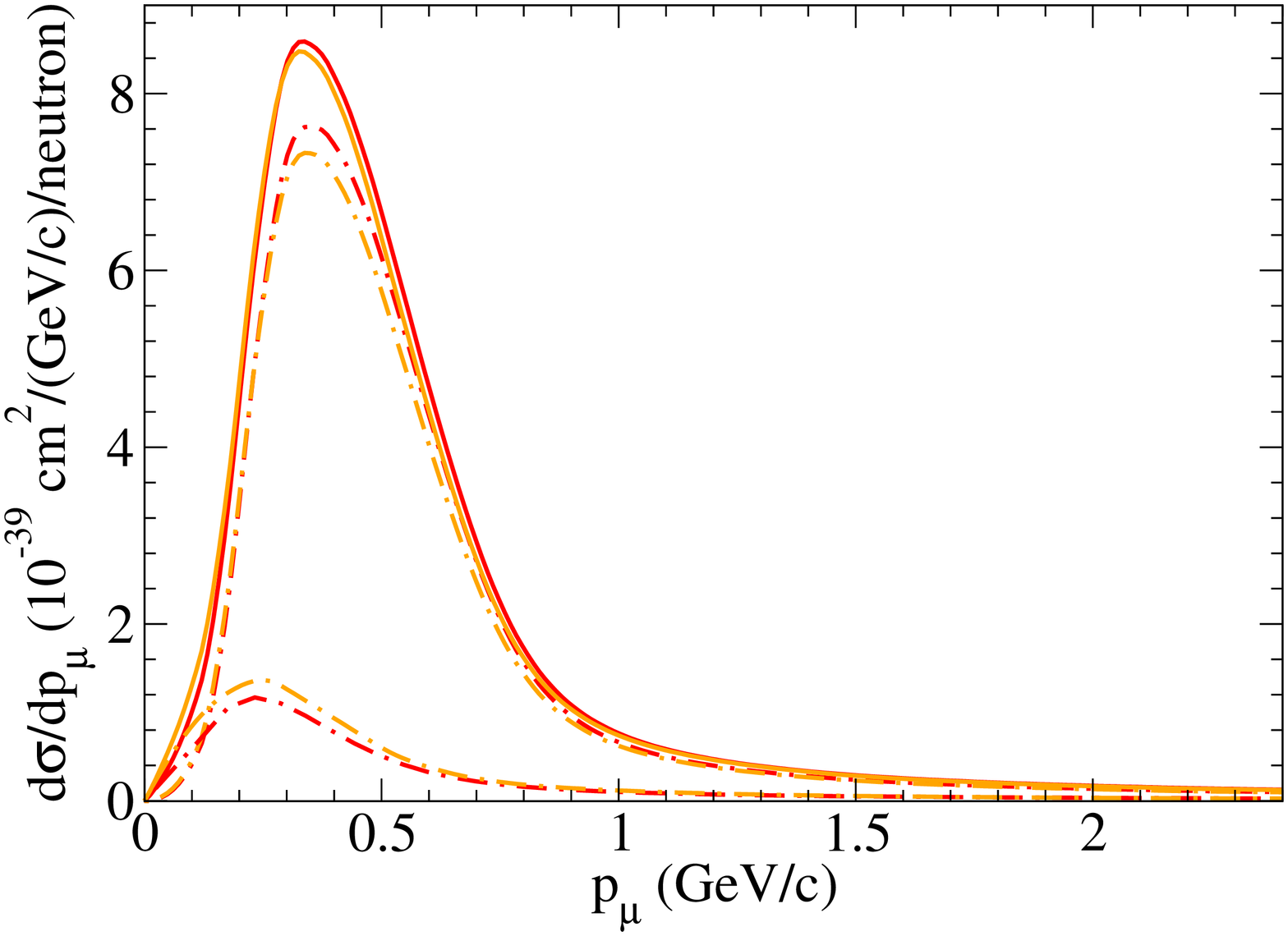}
%\\
%\includegraphics[width=6.cm,angle=270]{T2K_16O_costhetamu_ratio230.ps}
%\includegraphics[width=6.cm,angle=270]{T2K_16O_pmu_ratio230.ps}
\end{flushleft}
%\end{center}
    \caption{\label{fig:fig5bis} (Color online). %Left panels:
      T2K flux-averaged CCQE neutrino differential cross sections per neutron for \oxygen as functions of the muon scattering angle (left panel) and of the muon momentum (right panel), showing the effect of choosing two values for $k_F$ (in MeV/c) and $E_{shift}$, being the latter $16$ MeV (20 MeV) for \oxygen (\carbon).  
}
  \end{minipage}
\end{figure}

\subsection{T2K antineutrino-water scattering}
\label{sec:anti}

Finally, for completeness, in Fig.~\ref{fig:fig2anti} we show the antineutrino-oxygen ({\it i.e.,} with no hydrogen contribution) and in Fig.~\ref{fig:fig2anti2} the antineutrino-water ({\it i.e.,} with the hydrogen contribution) CC double differential cross sections computed using the same model employed above for the neutrino-oxygen case. As observed, the relative contribution of the 2p-2h MEC contribution compared with the pure QE one is very  similar to the case of neutrinos, also showing the same general  shape versus the muon momentum. The SF antineutrino results using the same model as employed for neutrino reactions are also shown in the figures and display similar behavior to what was observed above.

\begin{figure}
  \begin{minipage}{\textwidth}
    \begin{flushleft}
      %\begin{center}
\includegraphics[width=4.cm,angle=270]{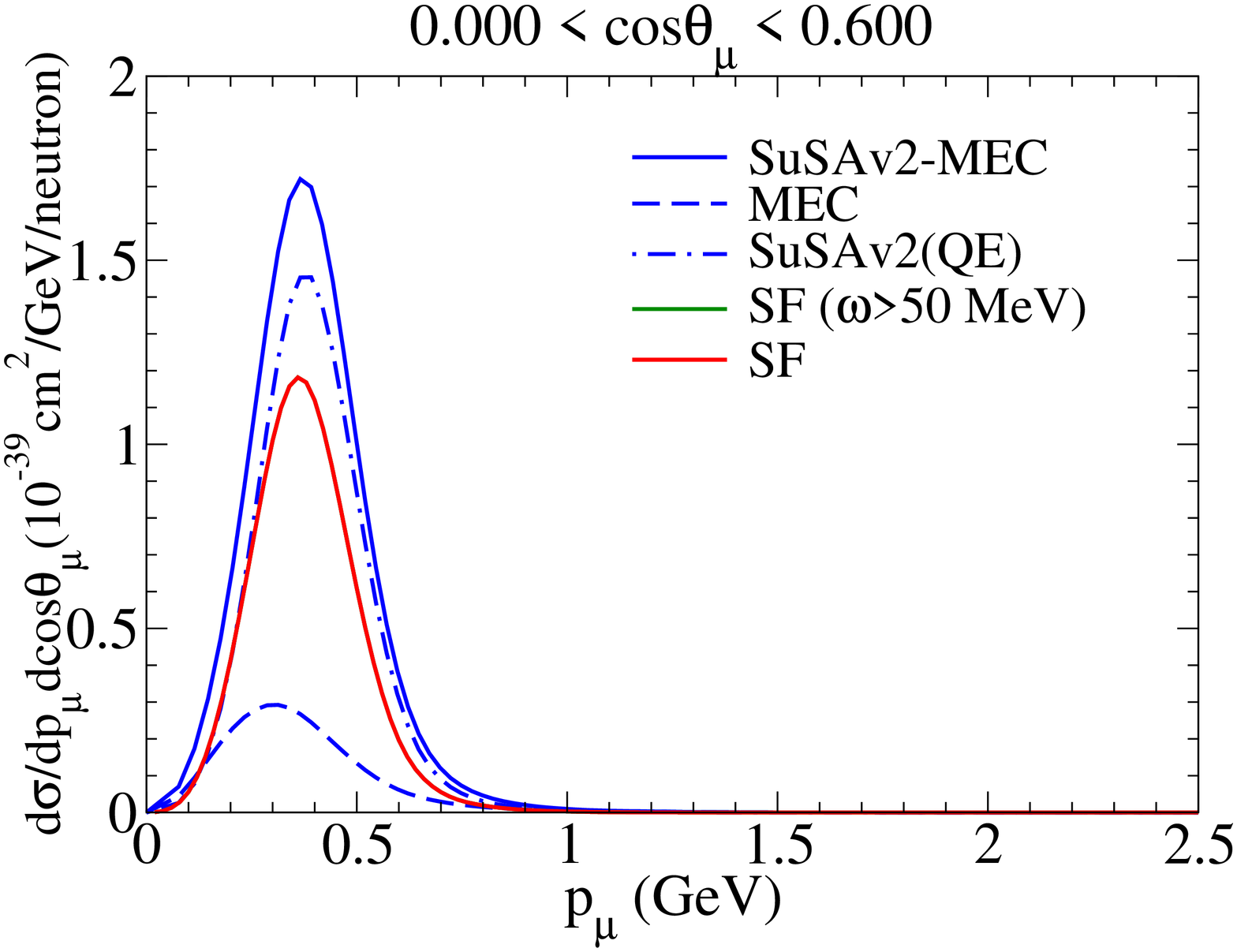}
\includegraphics[width=4.cm,angle=270]{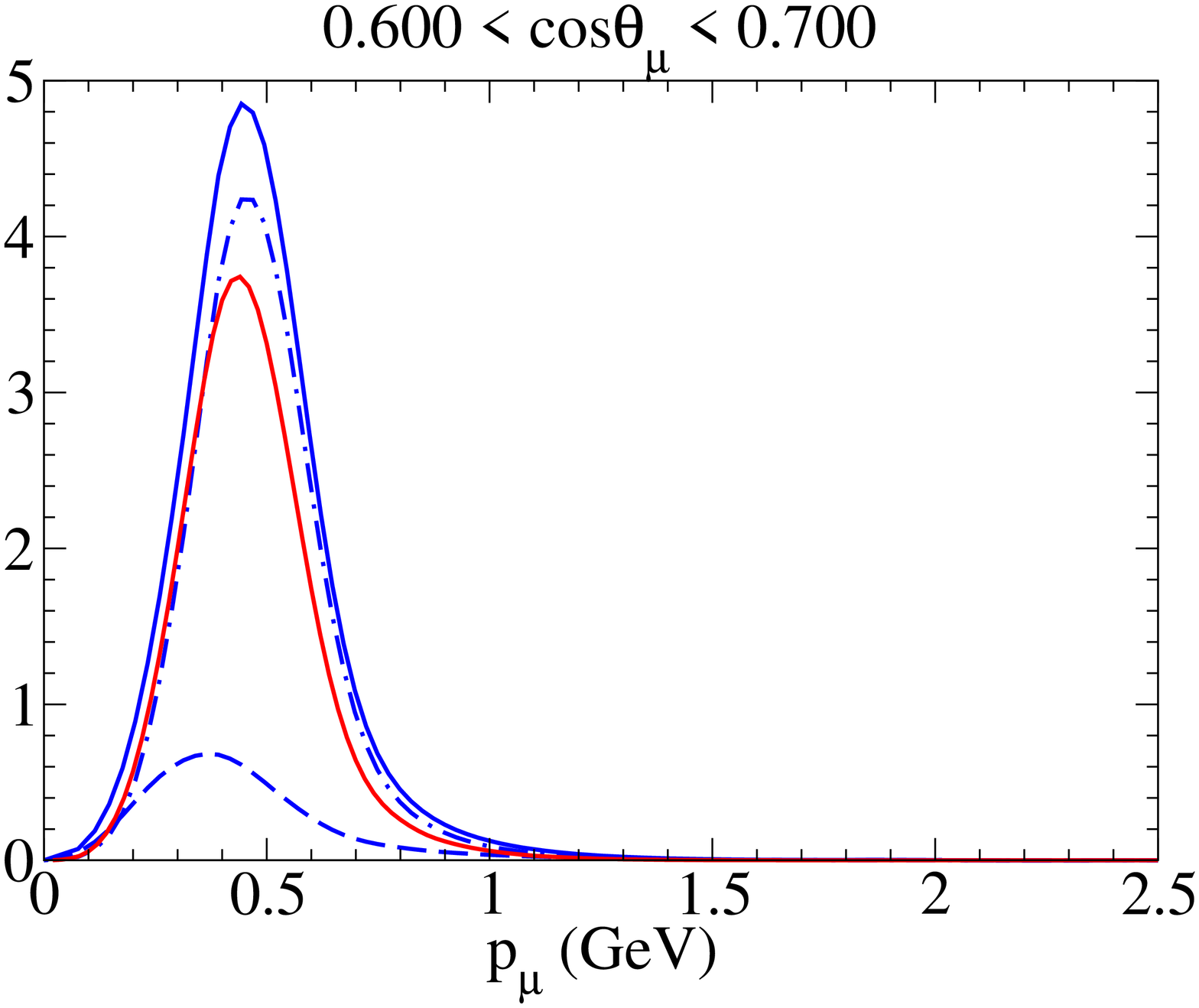}
\includegraphics[width=4.cm,angle=270]{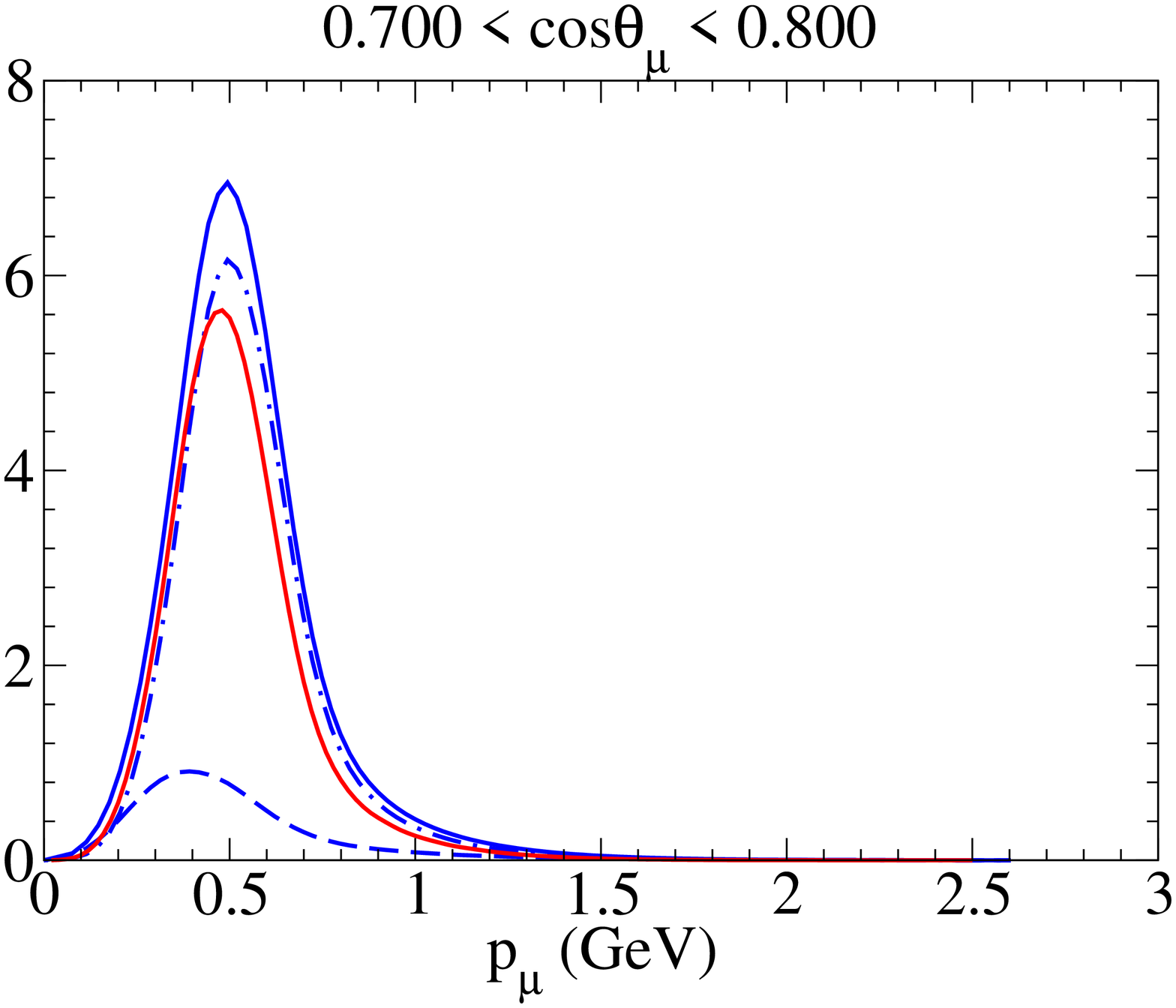}
\\
\includegraphics[width=4.cm,angle=270]{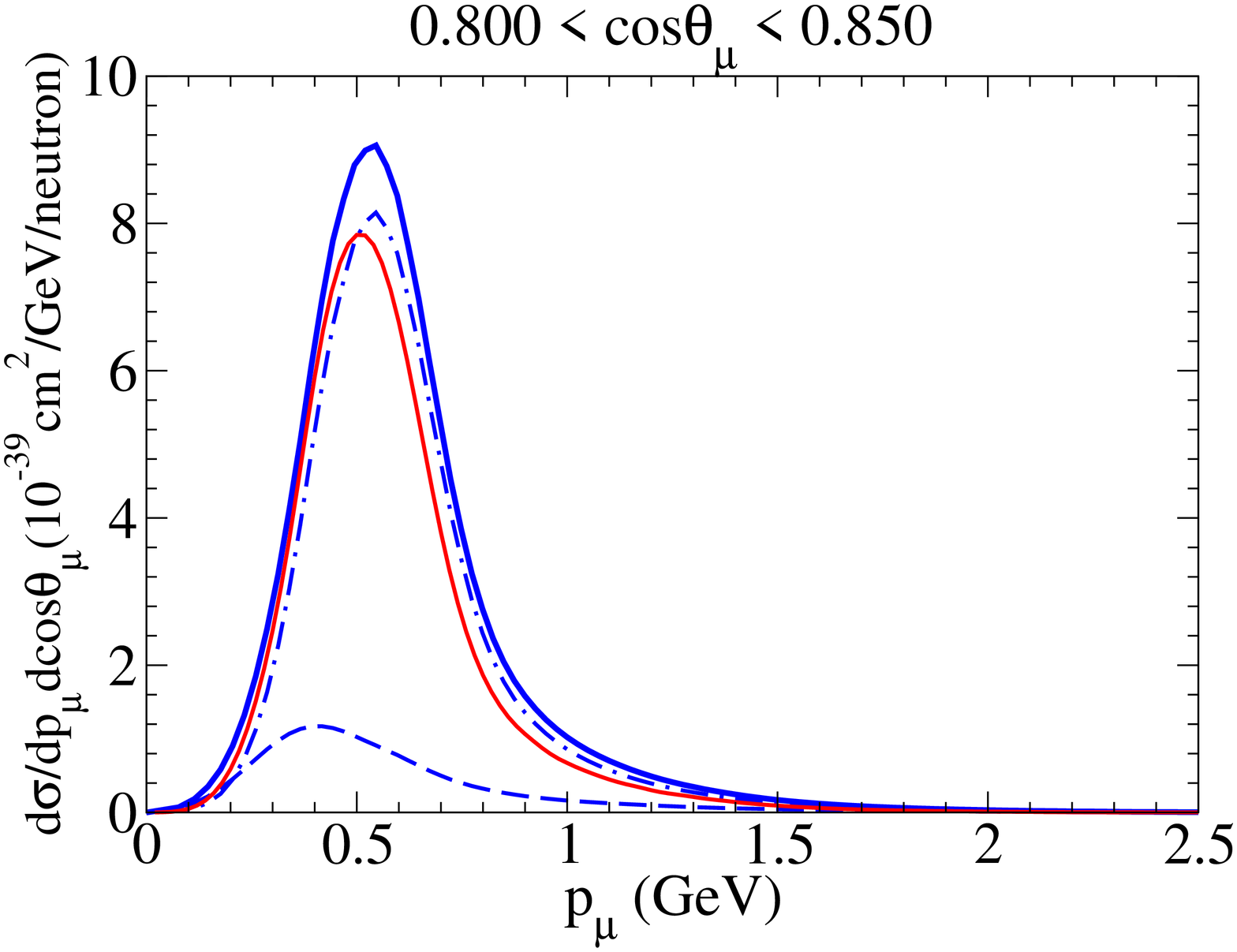}
\includegraphics[width=4.cm,angle=270]{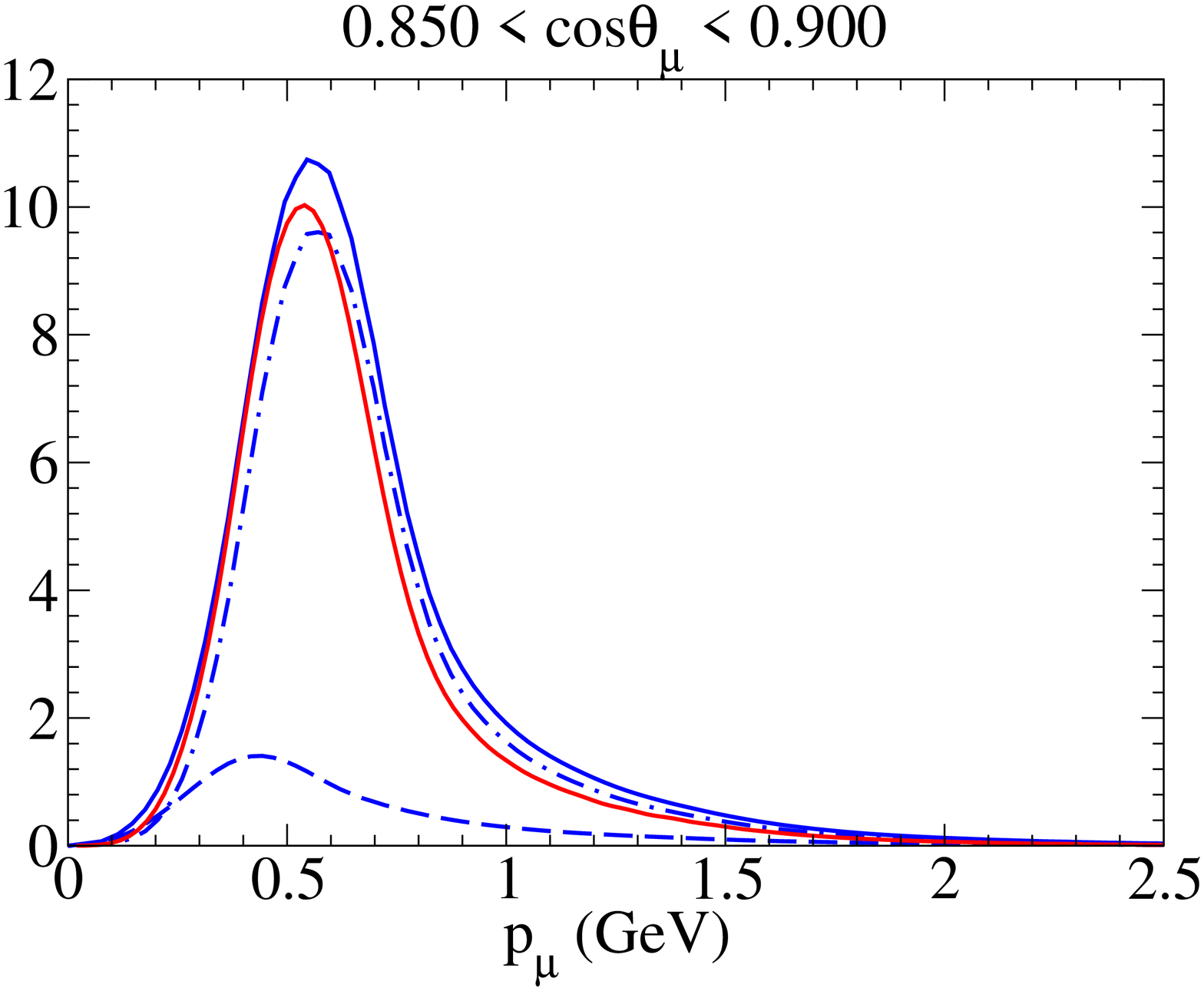}
\includegraphics[width=4.cm,angle=270]{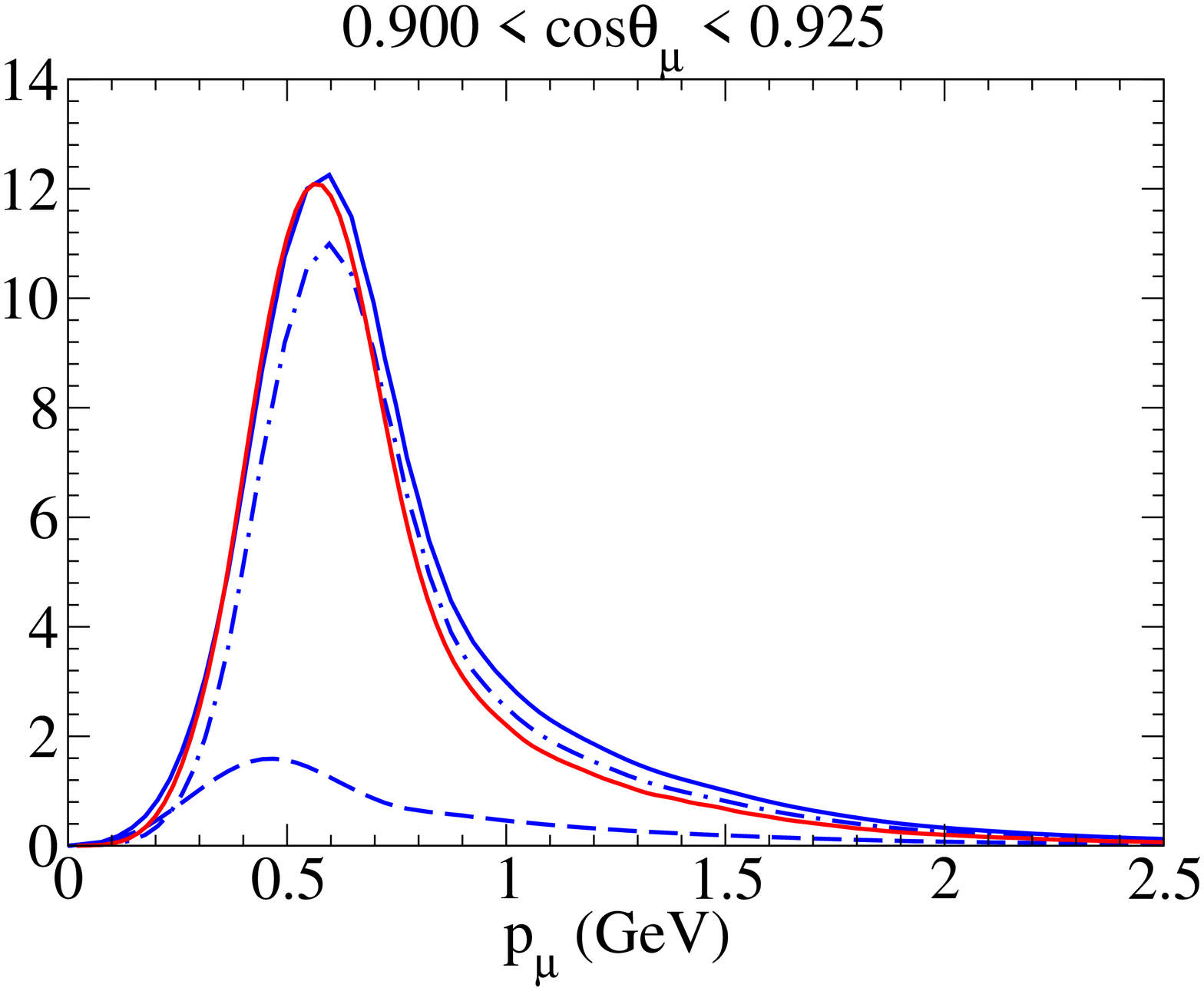}
\\
\includegraphics[width=4.cm,angle=270]{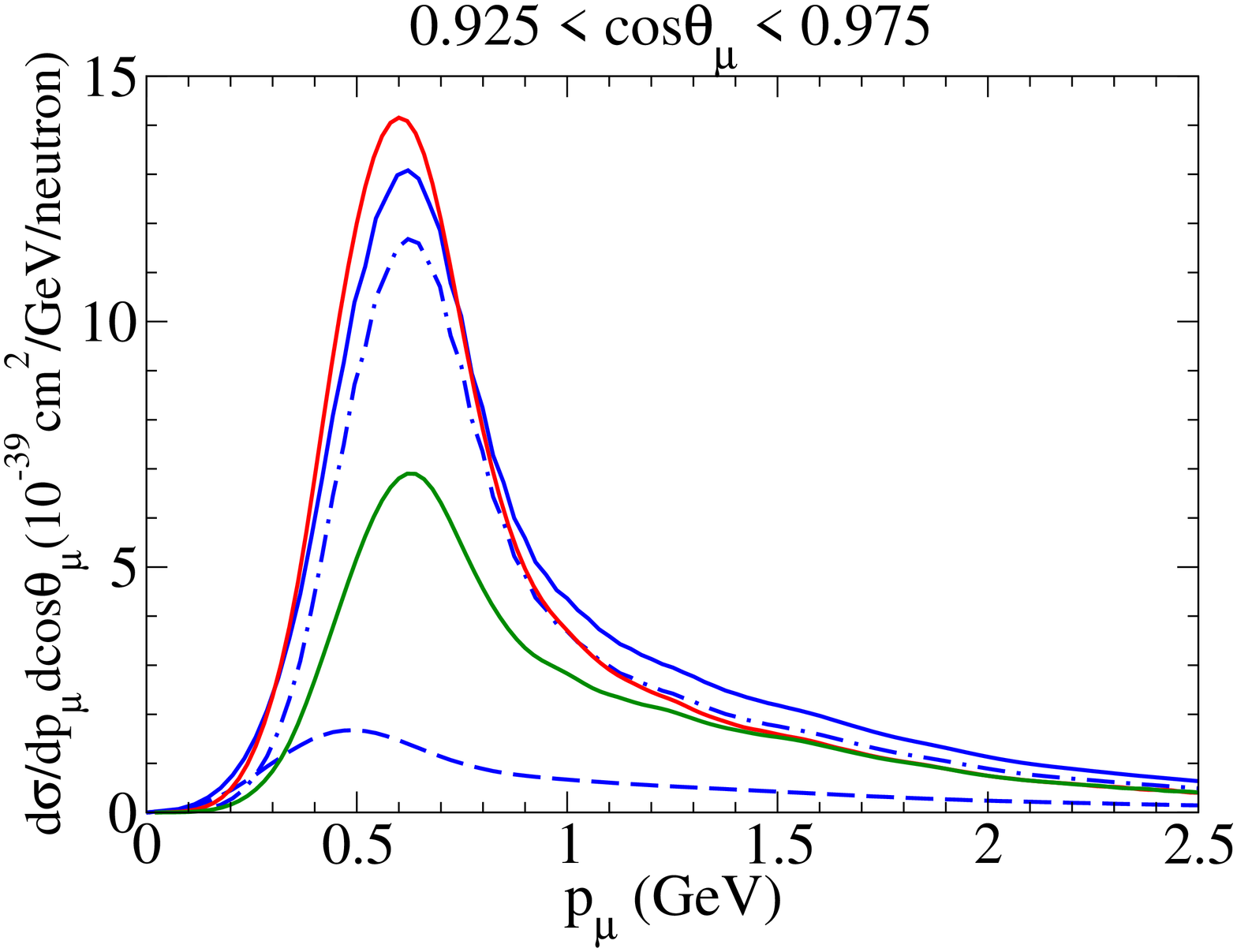}
\includegraphics[width=4.cm,angle=270]{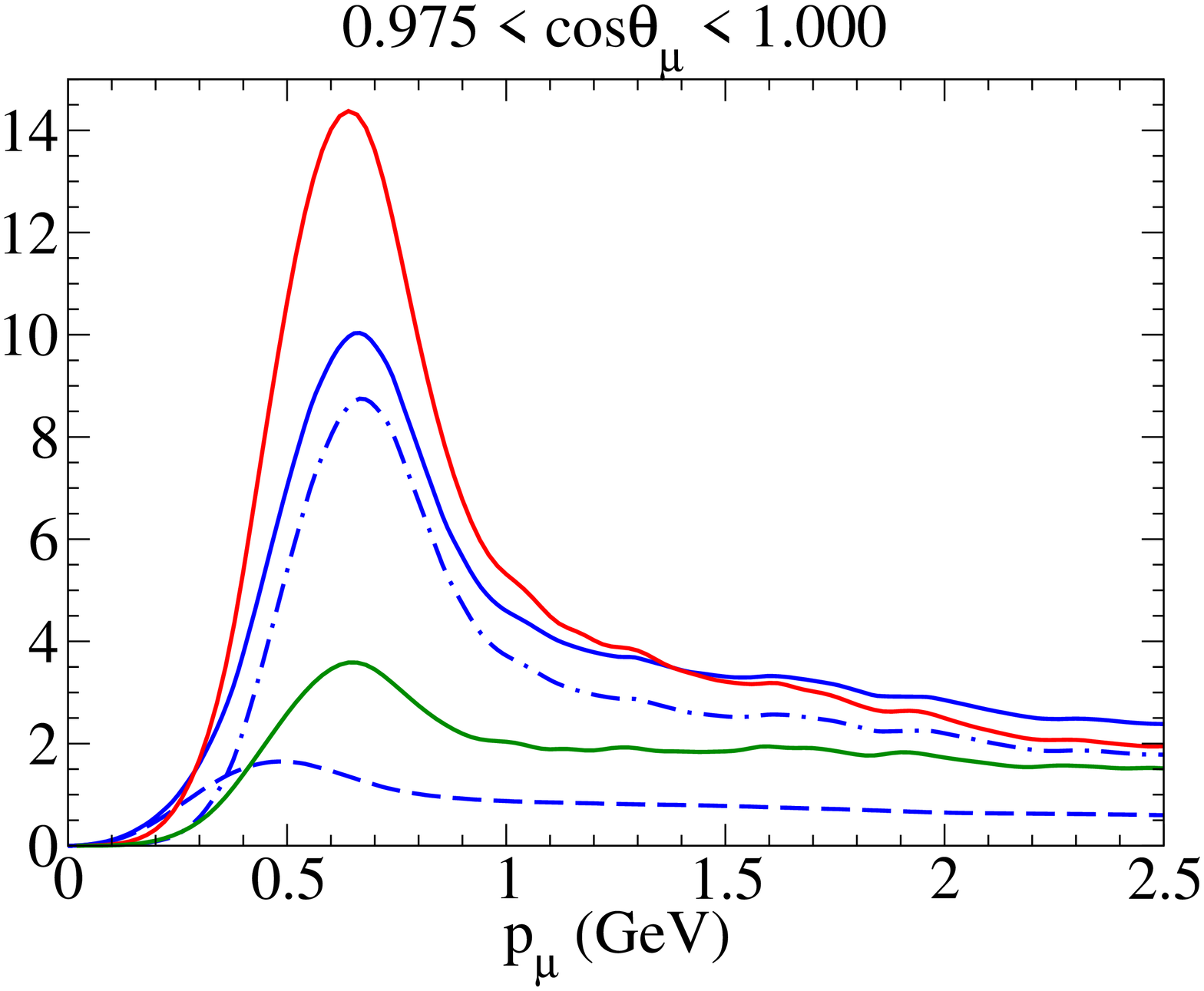}
\end{flushleft}
%\end{center}
\caption{\label{fig:fig2anti} (Color online) T2K flux-folded double differential cross section per target proton for the $\bar\nu_\mu$ CCQE process on oxygen. The SF results are also displayed. In the last two panels the SF result corresponding to $\omega>50$ MeV is also shown (green curve).
}
  \end{minipage}
\end{figure}

\begin{figure}
  \begin{minipage}{\textwidth}
    \begin{flushleft}
      %\begin{center}
\includegraphics[width=4.cm,angle=270]{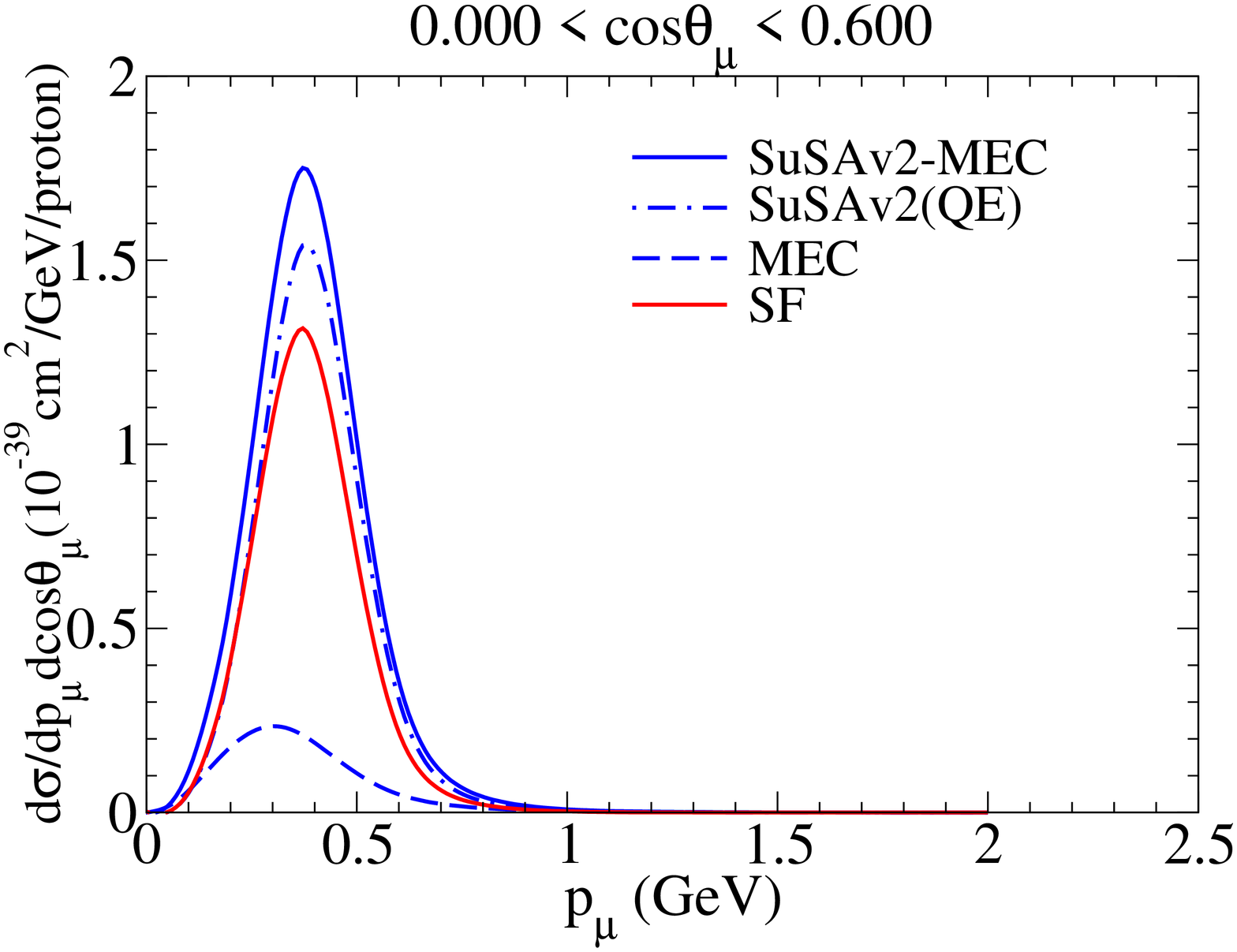}
\includegraphics[width=4.cm,angle=270]{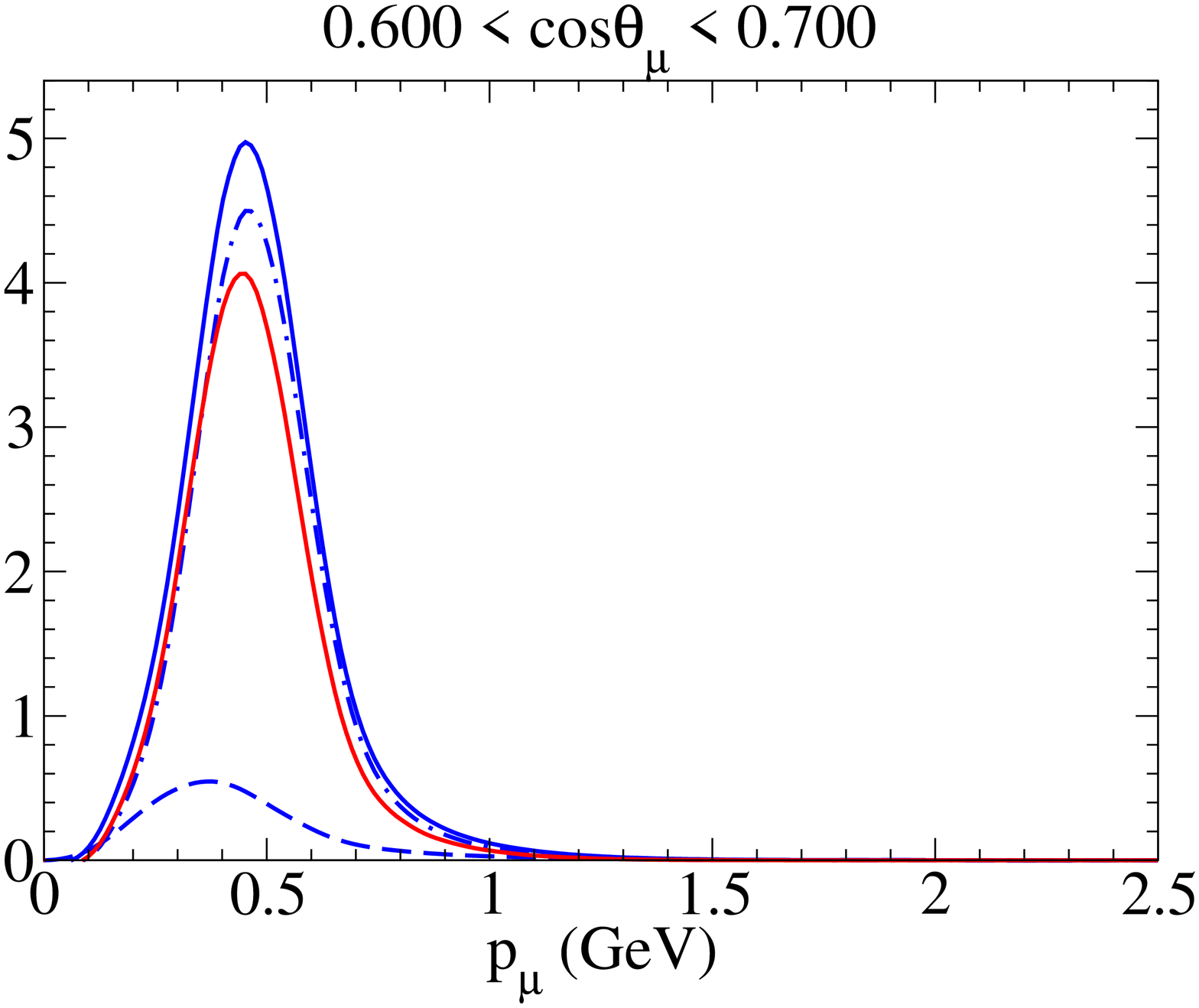}
\includegraphics[width=4.cm,angle=270]{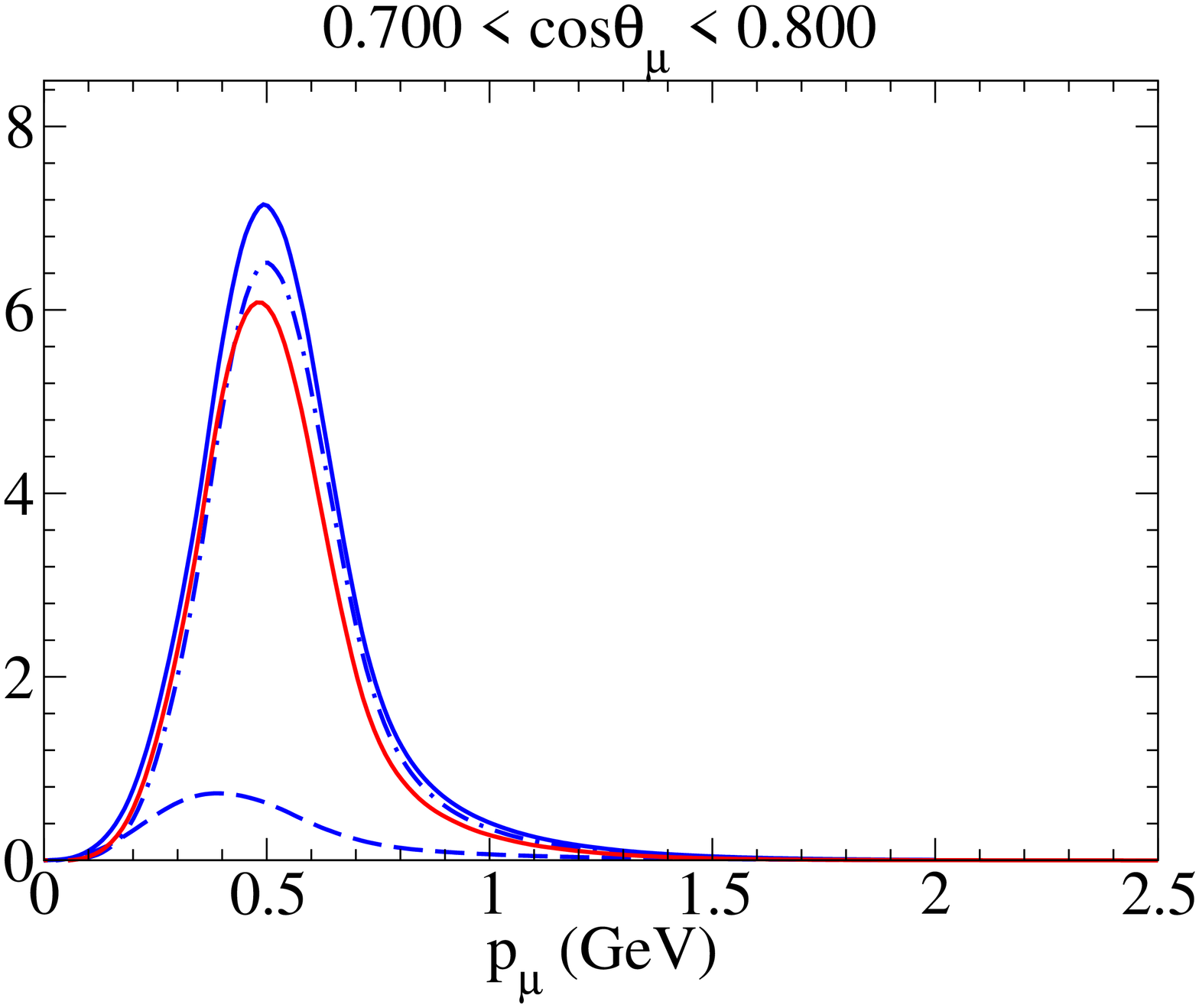}
\\
\includegraphics[width=4.cm,angle=270]{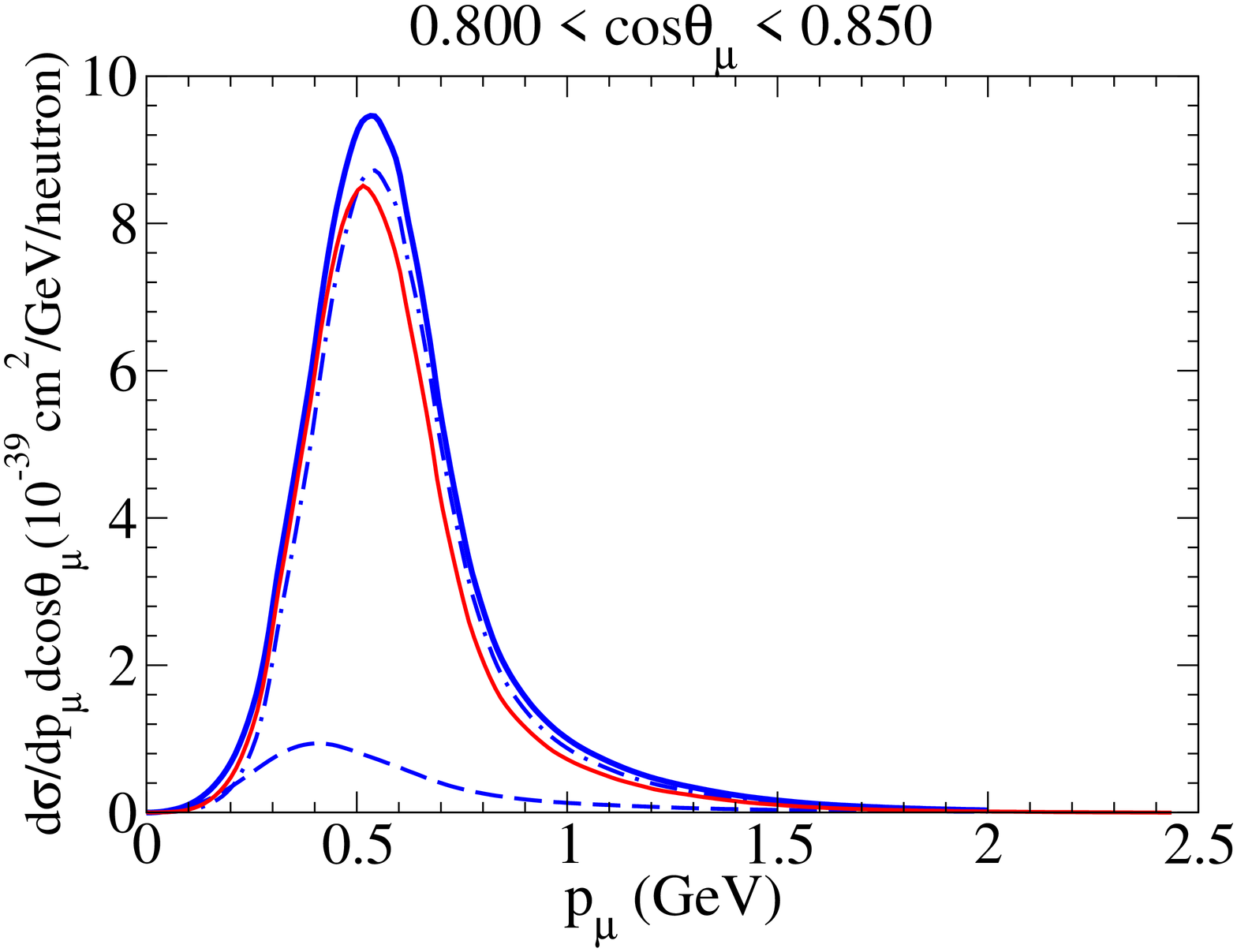}
\includegraphics[width=4.cm,angle=270]{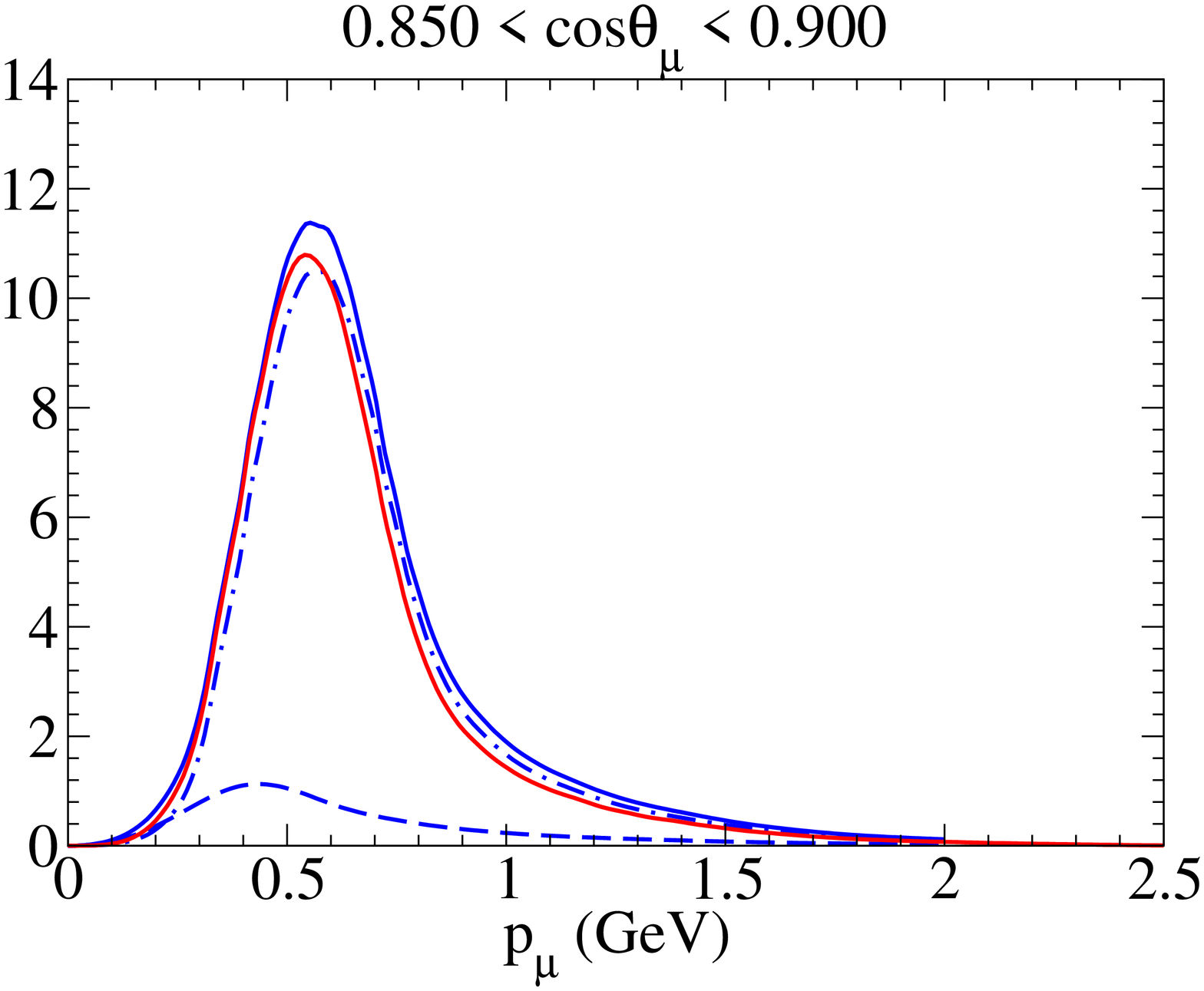}
\includegraphics[width=4.cm,angle=270]{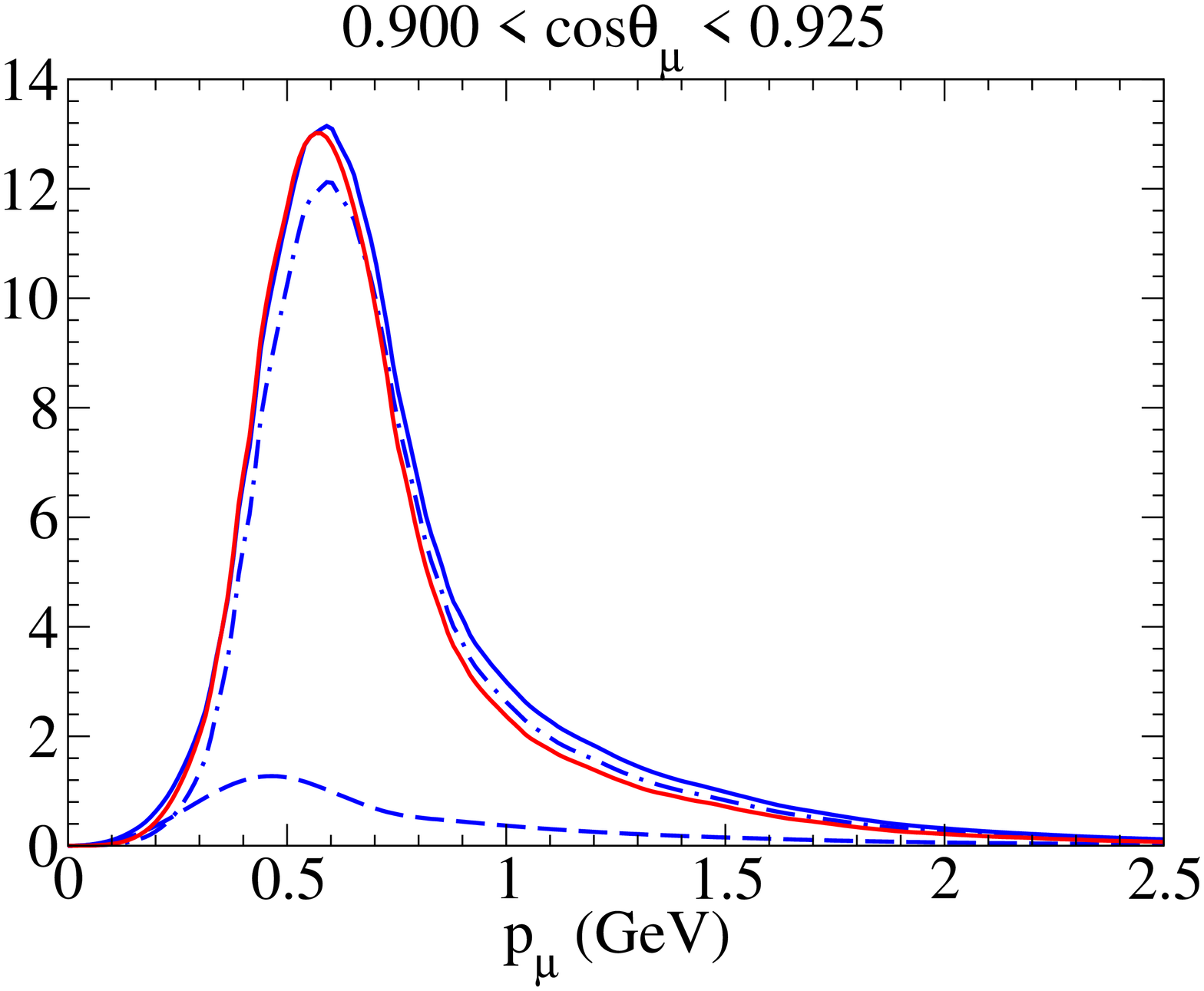}
\\
\includegraphics[width=4.cm,angle=270]{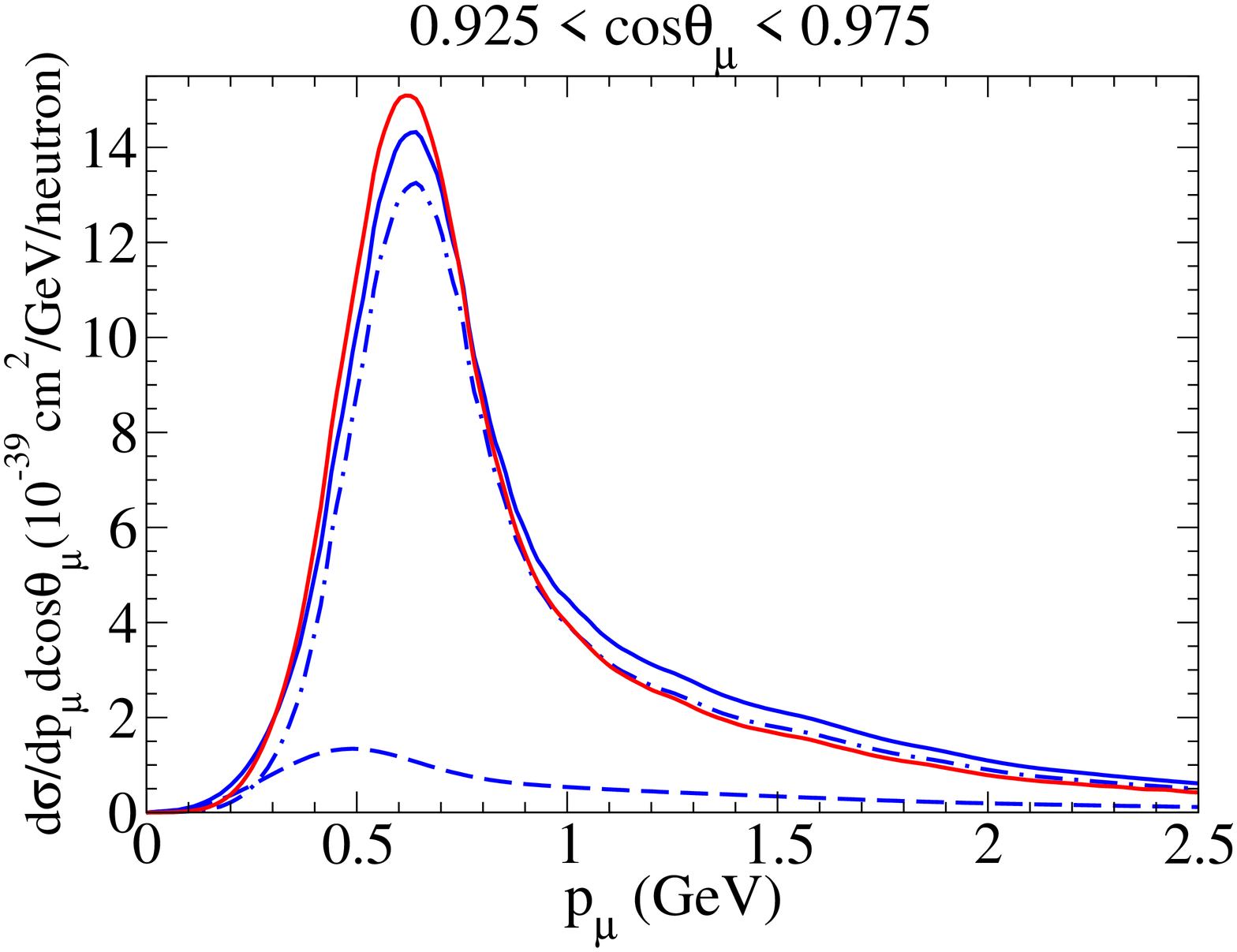}
\includegraphics[width=4.cm,angle=270]{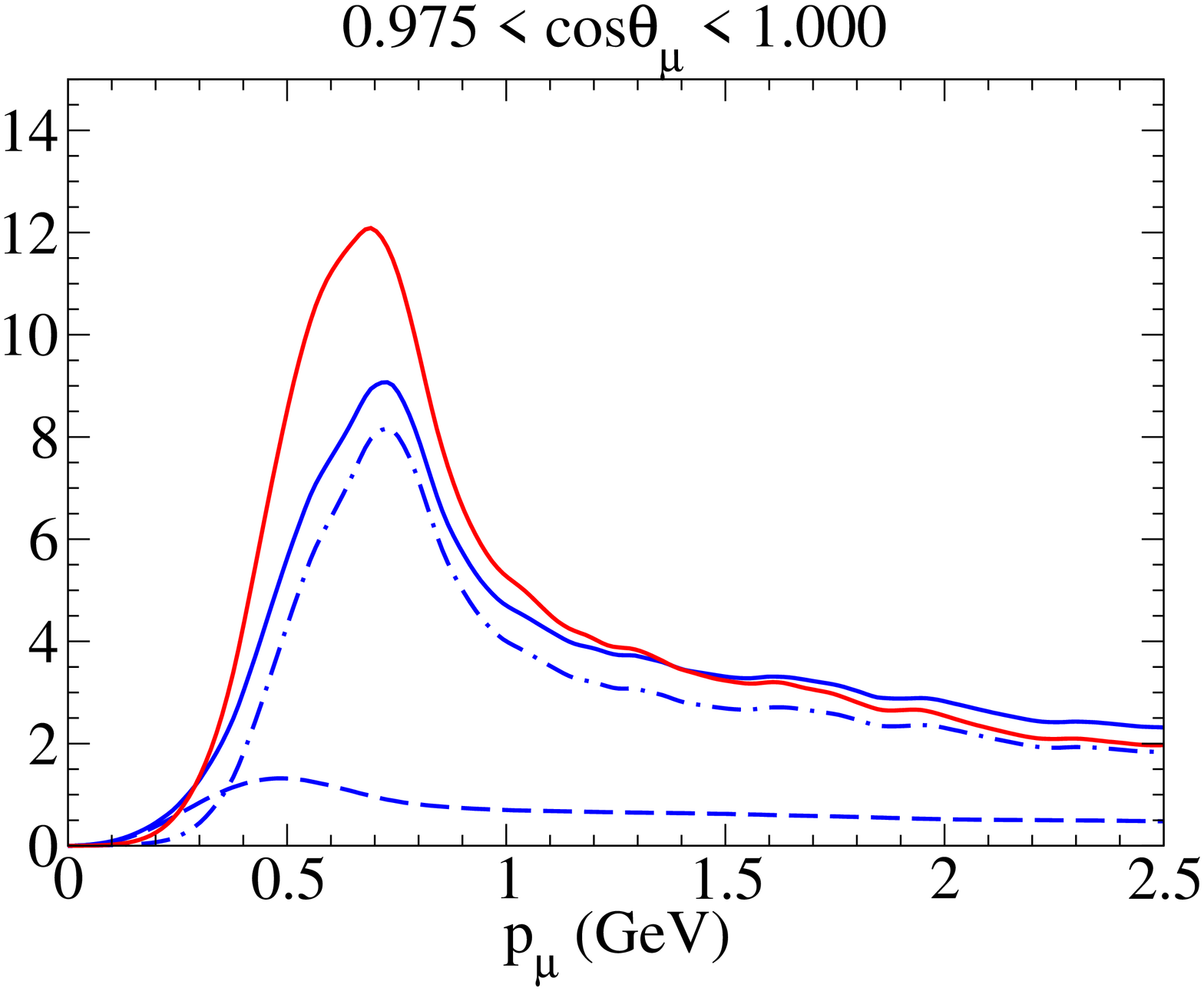}
\end{flushleft}
%\end{center}
    \caption{\label{fig:fig2anti2} (Color online) T2K flux-folded double differential cross section per target proton for the $\bar\nu_\mu$ CCQE process on water. The SF results are also displayed.
    }
  \end{minipage}
\end{figure}

\section{Conclusions}
\label{sec:conclusions}

In the light of new results from the T2K collaboration on neutrino-oxygen CC$\nu$ cross sections we have employed our previous SuSAv2+MEC approach that had been developed for studies of neutrino-carbon CC$\nu$ cross sections to study the relative importance of the various ingredients in the model with respect to how they impact interpretations of the cross sections. This required two basic steps: (1) we first studied the rather limited database of results for inclusive electron scattering from oxygen to determine the (few) parameters in the SuSAv2+MEC model, and (2) we extended the approach from studies of inclusive $(e,e')$ reactions to inclusive CC$\nu$ reactions in exactly the same way used in our previous analyses of carbon. Given some ambiguity in the choice of parameters we also explored the consequences of making different choices, for instance, of the parameter $E_{shift}$ used in our approach. Additionally we inter-compared CC$\nu$ results for oxygen and carbon to explore the robustness of attempts to deduce the cross sections for one from the other. Moreover, we have provided predictions for antineutrino-oxygen and antineutrino-water cross sections in advance of their being available from the T2K collaboration. Finally, we have also included QE inclusive electron scattering and CC$\nu$ (neutrino and antineutrino) results using a spectral function for oxygen together with a factorized PWIA model for the reactions.

The results are very satisfying. 
We see that the SuSAv2+MEC approach agrees reasonably 
%\xout{quite} 
well with the data.
%\xout{, having no significant disagreements given the uncertainties in the data.} 
The SF model used in the present work provides results only for the QE contribution and, when one does a theory-to-theory comparison between this model and the SuSAv2 model for the QE contribution, one sees generally good agreement with the former lying somewhat lower than the latter. 
Such is expected, since the SuSAv2 model contains intrinsic transverse enhancement effects that are absent in most models, certainly in the PWIA SF model. 
Nevertheless, the current experimental uncertainties and the large error bars do not allow us to draw more definite conclusions on the quality of the data comparison.

The main place where disagreements are observed is at very forward angles, namely at rather low excitation energies. To test the sensitivity to this near-threshold region we do as we have in previous work and cut out all contributions from $\omega <$ 50 MeV: when nothing significant occurs one can conclude that these contributions are unimportant. However, when large changes are observed, we need to exercise caution in believing the modeling. For the SF model this forward-angle region shows very large effects, indicating, as should be expected, that the PWIA fails in the near-threshold region. In contrast, the SuSAv2 model contains an extension of what is usually called ``Pauli blocking'' and appears to do much better. Nevertheless, even for the latter approach some caution should be exercised.

Given the success of our modeling for inclusive $(e,e')$ and CC$\nu$ reactions now on two different nuclei we have increased confidence in employing the approach for heavier nuclei. New features are likely to emerge in these cases and presently we are beginning to explore their consequences. Finally, and this was part of the motivation for including the SF modeling in the present study, we are engaged in extending the scope of our studies to include semi-inclusive CC$\nu$ reactions, and being able to ascertain the capabilities of the SF approach for inclusive scattering provides a benchmark for the semi-inclusive studies.

%%%%%%%%%%%%%%%%%%%%%%%%%%%%%%%%%%%%%%%%%%%%%%%%%%%%%%%%%%%%%%%%%%%%%
\section*{Acknowledgments}

%%%%%%%%%%%%%%%%%%%%%%%%%%%%%%%%%%%%%%%%%%%%%%%%%%%%%%%%%%%%%%%%%%%%

This work has been partially supported by the Spanish Ministerio de
Economia y Competitividad and ERDF (European Regional Development
Fund) under contracts FIS2014-59386-P, FIS2017-88410-P, by the Junta de
Andalucia (grants No. FQM-225, FQM160), by the INFN under project
MANYBODY, by the University of Turin under contract BARM-RILO-17, and part (TWD) by the U.S. Department of Energy under
cooperative agreement DE-FC02-94ER40818. IRS acknowledges support from
a Juan de la Cierva fellowship from MINECO (Spain). GDM acknowledges
support from a Junta de Andalucia fellowship (FQM7632, Proyectos de
Excelencia 2011). JWVO acknowledges support by the US Department of Energy under Contract No. DE-AC05-06OR23177, and by the U.S. Department of Energy cooperative research agreement DE-AC05-84ER40150. We thank Omar Benhar for providing the oxygen spectral function. We acknowledge useful discussions during the ``Two-body current
contributions in neutrino-nucleus scattering'' ESNT workshop at
CEA-Saclay, April 2016.
We thank Sara Bolognesi (IRFU, SPP, CEA-Saclay) for her active
participation on discussions of experimental issues.

% \section*{References}
%\bibliography{iopart-num}
\bibliography{PHD}  

\end{document}